\documentclass[preprint,floats,aps,12pt]{revtex4}
\usepackage{times} 
\usepackage{amssymb}
\usepackage{amsmath}
\usepackage{prelim2e}\usepackage[none,bottom]{draftcopy}
\draftcopyName{preprint}{1.2}

\usepackage{graphicx} 
\ifx\pdfoutput\undefined
\usepackage{graphicx}     
\else
\fi
\ifx\pdfoutput\undefined
\DeclareGraphicsExtensions{.eps}
\else
\fi
\usepackage{color}
\graphicspath{{.},{./fig/}} 
\newcommand{\mylab}[1]{\label{#1}{\color{blue}\it \hspace*{.5cm} #1}}
\newcommand{\myfig}[1]{Fig.~\ref{fig:#1}}
\newcommand{\mytab}[1]{Table~\ref{tab:#1}}
\newcommand{\myeq}[1]{Eq.~\ref{eq:#1}}
\newcommand{\mysec}[1]{Section~\ref{sec:#1}}
\newcounter{countfil}

\begin{document}
%
\title{Deterministic transport of particles in a micro-pump}
\author{Philippe Beltrame}
\email{philippe.beltrame@uni-avignon.fr}
\affiliation{Dept. Physique, UMR EmmaH 1114, Universit\'e d'Avignon, Avignon, France}
\author{Peter Talkner}
\email{peter.talkner@physik.uni-augsburg.de}
\author{Peter H\"anggi}
\email{peter.haenggi@physik.uni-augsburg.de}
\affiliation{Institut f\"ur Physik, Universit\"at Augsburg, D-86135 Augsburg, Germany}
\begin{abstract}
We study the drift of suspended micro-particles in a viscous liquid pumped back and forth through a periodic lattice of pores (drift ratchet). In order to explain the particle drift observed in such an experiment, we present an one-dimensional deterministic model of Stokes' drag. We show that the stability of oscillations of particle is related to their amplitude. Under appropriate conditions, particles may drift and two mechanisms of transport are pointed out. The first one is due to an spatio-temporal synchronization between the fluid and particle motions. As results the velocity is locked by the ratio of the space periodicity over the time periodicity.  The direction of the transport may switch by tuning the parameters. Noteworthy, its emergence is related to a lattice of 2-periodic orbits but not necessary to chaotic dynamics.  In this point of view it appears as a generalization of the Stokes' drift. The second mechanism is due to an intermittent bifurcation and leads to a  slow transport composed by long time oscillations following by a relative short transport to the next pore. Both steps repeat in a quasi-periodic manner. The direction of this last transport is strongly dependent on the pore geometry.
\end{abstract}
%
%
\maketitle
%
%
%
\section{Introduction}
\mylab{sec:intro}\\
%
The transport of micro-particles through pores in a viscous fluid in absence of mean force gradient finds its motivation in many biological applications as the molecular motor or molecular pump.
In the last decade, the literature shows that a periodical pore lattice without the symmetry $x\rightarrow -x$ can lead to the so-called ratchet effect allowing an transport in one direction $x$ or $-x$. A review can be found in \cite{HaMa99}.
In order to better understand the transport of particles in a pore lattice, the Max-Planck Institute in Halle (Germany) has build macroporous silicon wafer which is connected at both ends to basins \cite{KRHM00}. The basins and the pores are filled with liquid and micrometer-sized particles pumped back and forth. Thus, a backward and forward motion of liquid occurs draging the particles.
The experiment shows the existence of an effective transport in a certain range of parameter values. By tuning them, the direction of the effective transport may change.
We focus on a simple deterministic model displaying these phenomena but not necessary for the experiment parameter values.\\
The drift of suspended particles in a viscous fluid under an oscillating force of zero average is often explained by the well-known Stokes' drift \cite{Stok1847}. It is a deterministic mechanism involving the existence of a propagating wave of the fluid velocity field. The particle may drift by synchronization with the phase velocity of the traveling wave by a so-called phase locking phenomenon \cite{PNA98}. The drift velocity is the phase velocity. One may use the analogy with a surfer on a wave: the particle rides the traveling wave \cite{HMN05}. This drift remains when  noise is added,  however the efficiency of the Stokes' drift decreases \cite{HMN05}. In contrast for other systems the presence of noise plays a crucial role for the transport of inertial particles. It is noteworthy that the drift dynamic is during a time range close to the Stokes' drift: the drift velocity corresponds to the ratio between the spatial periodicity and the temporal periodicity of the forcing.  Such a transport arises as well in Brownian particles \cite{BHK94,Machal07} as for electric current in a SQUID device \cite{ZBSH96}. Note, that this drift may exist without the presence of traveling wave of the force field. More surprising, if the average of the force is non-zero the drift may be in the opposite direction, it is the absolute negative mobility \cite{Machal07}.\par
For the transport of milli- and micro-particles in a viscous fluid, the role of the thermal noise is debatable. 
The fluid motion without particle is described by the Stokes equations (low Reynolds number) leading to an oscillating creeping motion. If the noise and the inertia are negligible the particles advect with the fluid getting an oscillating motion \cite{KSPH06}. Therefore, in order to explain particle transport,  another phenomenon has to be take into account: either thermal noise or inertia, or both. 
Kettner {\it et al.} \cite{KRHM00} neglect the inertia and consider thermal fluctuations. Stochastic simulations display transport motions and also the direction may change by varying the frequency of pressure oscillations.
However, even the micrometric size of particles the inertia may be larger than the thermal noise. In this paper, we propose a complementary approach where the noise is neglected while the small inertia is taken into account because of a finite mass, then finite drag. Thus, it  is a deterministic approach of inertial particle in a viscous fluid. The question of transport solutions is not trivial since the sinusoidal pumping leads to a flow without the presence of a  traveling wave and then the system is different from the Stokes' drift.\\
In order to clearly identify  the mechanisms of the drift, we focus on the simple  case where the competition between the flow drag and the  particle inertia governs the motion.  Thus, we propose an one-dimensional deterministic problem. 
The  inertia in deterministic ratchet problem showed chaotic dynamics with possible transport and reversal current \cite{BHK94}. Mateos \cite{Mate00} shows that the current reversal for a biased oscillated force corresponds to a transition from a chaotic to a periodic regime  or chaotic crisis \cite{GOY82}. It is a phase synchronization of chaotic oscillators described in the ratchet framework by \cite{JuPa00,PRK01,OPK02}. This phenomenon is generic when a  periodic external forcing is applied. It is observed in several experiments ranging. The resulting transport solution was thoroughly studied in \cite{SER07} in deterministic  for unbiased force. They found transport solutions with locked velocity $c=n/m$ with $n$ and $m$ integers. These solutions may exist for symmetric ratchets where transport are possible in both direction. A current reversal phenomenon is obtained by applying a non-zero biais because of different existence domains of the transport to the left or to the right.  The emergence of the synchronized transport  appears after evolving through a period-doubling route to chaos: attractor crisis \cite{GOY82}. Another studies shows the same mechanism of the emergence of synchronized dynamics as \cite{VKNA05}. However, it seems that chaotic transitions is not required for the existence of synchronized transport rather the coexistence of periodic orbits \cite{BaSa00}.\\
As for \cite{SER07}, we will study the transport solutions for the symmetric case in order to point out the role of pore asymmetry underlying the ratchet effect. The synchronized transport solutions are tracked in the comoving frame using path-following method. Contrary to a direct time integration as it is carried out in the literature, one gets the stability of the dynamics the kind of transition and the domain of existence. For instance it is possible to observe unstable branches even if a stable branch attracts all the dynamics. This approach allows us to give a new interpretation of synchronized transport solution as a generalized Stokes' drift related to 2-periodic orbits. Moreover, we are able to interpret the different transitions found in literature. Nevertheless the parameter domain for which the synchronized transport solution occurs is too small for micro-sized particles.
For large drags there is still transport solution resulting from an intermittent bifurcation. This transport is slow and in contrast to the previous case is strongly related to the asymmetry. This transport does not seem appear in the literature. 

%
%
%
\section{Model}
\mylab{sec:mod}\\
%

\subsection{Stokes approximation}
The Stokes equation is a model of viscous fluid for which the viscous dissipation is compensated with the pressure force work.
Thus, it is a quasi-static fluid motion known as the creeping flow. The main parameter which is relevant for such an approximation is the Reynolds number which measures the mean fluid velocity scaled by the cinematic velocity.
In order to determine the Reynolds number, let us give the magnitude order of the physical parameters of the experiment in \mytab{expval}. The fluid chosen in the water of density $\rho_0=10^3 kg/m^3$ and $\mu=10^{-3} Pa.s$. The time is scaled by the period of the pumping and varies between $[0.1 ;1] ms.$ The length is scaled by the pore length about $L=8.4\mu m$.  Finally, the velocity is scaled by the spatial mean velocity taken at the time where the flux pumping is maximal. One calls it the characteristic velocity $U_m$.
%
\begin{table}
\begin{tabular}{ccll}
$\mu$ & : & Dynamic viscosity of the water & $10^{-3}Pa.s$\tabularnewline
$r_p$ & : & Particle radius & $1.8\cdot10^{-6}m$\tabularnewline
$\rho_{0}$ & : & water density & $10^{3}kg/m^{3}$\tabularnewline
$\rho_{p}$ & : & particle density & $\simeq\rho_{0}$\tabularnewline
$U_m$ & : & Characteristic velocity of the fluid ($>0$) &$2.1\cdot 10^{-2} - 10^{-1}$  m/s\tabularnewline
$T$ & : & Period of the pumping & $0.01--10^{-4}s$\tabularnewline
$L$ & : & pore length & $8.4\cdot10^{-6}m$\tabularnewline
$\Gamma$ & : & friction constant per unit of mass & $10^{6} s^{-1}$\tabularnewline
\end{tabular}
\caption{Dimensional physical parameters of the drift ratchet experiment.}\mylab{tab:expval}
\end{table}
Using these parameters to scale the Navier-Stokes equation, one obtains:
\begin{eqnarray}
\frac{\rho U_m L}{\mu}\frac{\partial v}{\partial t}+\frac{\rho U_m^2 T}{\mu}(v.\nabla)v&=&-\frac{\rho TL}{\mu}\nabla p +\frac{UT}{L}\Delta^2 v\\
Re_1\frac{\partial v}{\partial t}+Re_2(v.\nabla)v&=&-C_1\nabla p +C_2\Delta^2 v
\end{eqnarray}
The first two pre-factors $Re_1$ and $Re_2$ are two possible expressions of the Reynolds number. According to the \mytab{expval}, they are about $10^{-2}$ while both dimensionless parameters $C_1$ and $C_2$ are about 1. Then, it is relevant to neglect the terms on the left side and one obtains the quasi-static approximation of the linear Stokes equation with no-slip boundary
\begin{eqnarray}
\Delta v_0(\bf{r},t) &=& \nabla p(\bf{r},t), \mylab{eq:stokes}\\
v_0(\bf{r},t) & =&0, \mbox{ on the pore boundary} \nonumber
\end{eqnarray}
where $v_0$ and $p_0$ are, respectively, the velocity and the pressure of the fluid without particles.
Because of the fluid is in equilibrium at each time $t$, the sinusoidal variation of the pressure implies a  sinusoidal temporal dependence of the velocity field $v_0:$
\begin{equation}
 {v}_0(r,t)=u_0(r)\sin({2\pi}t),\mylab{eq:vxt}
\end{equation}
where $u_0(r)$ is a velocity field obtained by applying a constant pressure difference back and forth of pore lattice. The lattice being periodic, then the function $u_0(\bf{r})$ gets its periodicity.
The adimensional characteristic velocity $u_m$  is the mean velocity $u_0(\bf{r})$ in one pore. (All dimensionless parameters are noted using small letters).\\
The simulation of the full non-linear Navier-Stokes equation with these parameters corroborates the relevance of the Stockes' approximation (\myeq{stokes}).
\subsection{Governing equation of the particle motion}
The velocity and acceleration of  the particle motion being smaller than those of the flow motion, the Stokes approximation \myeq{stokes} remains valid.  The flow motion is described by the Stokes equation but only the boundary conditions of \myeq{stokes} have to be added in order to fit the no-slip condition on the particle surface:
\begin{eqnarray}
\Delta \vec{v}_f &=& \nabla p_f, \mylab{eq:stokespart}\\
\vec{v}_f(\vec{r},t) & =&0, \mbox{ on the pore boundary},\mylab{eq:stokespart2}\\ 
\vec{v}_f(\vec{r},t) & =&\vec{v}(\textbf{r},t), \mbox{ on the pore surface},\mylab{eq:stokespart3}
\end{eqnarray}
where $v_f(\vec{r},t)$ is the fluid velocity with the presence of the particles and $v(\vec{r},t)$ is the particle velocity at the point $\vec{r}$ and the time $t$. At each time $t$ the stress tensor on the particle surface may be estimated using boundary equations (see e.g. \cite{PoPa94,Bonn95}). By adding the Newton's and the momentum equations of the solid particle, the particle motion can be completely described. The motion can be quite complicated if the particle does not stay on the symmetry $x$ axis of the pore because, on one hand, the particle may rotate and, on another hand, the interaction between the particles and the wall has to be taken into account \cite{Brenn64}.
Such a study is out of the scope of the paper since we aim at finding the most simple transport mechanisms in a viscous fluid.\\
Therefore, we focus on the axisymmetric case: the particle is centered on the axis and does not rotate. With this assumption, the problem becomes 1D. Let us note $s(t)$ the position $x$ of the particle center at the time $t$. Each point of the particle has the same velocity note $v(t)$:
\begin{equation}
 m_p \frac{dv(t)}{dt} = F(s(t),t),
\end{equation}
where $m_p$ is the particle mass and  $F(s(t),t)$ the drag force on the particle. This last one is determined by the system \ref{eq:stokespart}, \ref{eq:stokespart2} and \ref{eq:stokespart3}.
Now one assumes that the drag force ${F}$ is determined by the Stokes' drag of a spherical particle in an uniform flow, i.e. the drag force $F$ is proportional to the relative velocity of the particle. Then 
\begin{equation}
 F(s(t),t)=-\beta\left( v(t) - v_0(s(t),t) \right) \mylab{eq:linear-drag}
\end{equation}
The \myeq{linear-drag} is verified only by assuming: firstly, the pore boundaries are neglected in the drag force estimation and secondly, the unperturbated $v_0$ flow velocity is almost uniform in the volume occupied by the particle. Therefore these last assumptions yield only for small particles about 0.1 $\mu m$ in the framework of the experiment since the narrow region of the pore is about $2.5\mu m$. \\
In this case the particle motion is one-dimensional model described by the ODE
\begin{eqnarray}
 \frac{dv(t)}{dt}&=&\frac{\beta}{m_p}\left( v(t) - v_0(s(t),t) \right)\\
 &=&\gamma\left( v(t) - v_0(s(t),t) \right)\mylab{eq:1dim}
\end{eqnarray}
with $\gamma =\frac{\Gamma}{T}$.
The relevant parameter is the dimensionless drag $\gamma$  defined by
\begin{equation}
 \gamma =\frac{\Gamma}{T}= \frac{6\pi\mu r_p}{Tm_p}
\end{equation}
Introducing the mean particle density $\rho_p$ it comes:
\begin{equation}
\gamma = \frac{9}{2}\frac{\mu}{T\rho_p r_p^2}
\end{equation}
According to the value of \mytab{expval}, we get the approximation of $\gamma\simeq10$--$1000$. However, the lower value of $10$ is obtained for particle radius about 5-6 $\mu m$. Then, the Stokes' force can not anymore be determined by the linear relation \myeq{linear-drag} if the pore radius is still about $8-9\mu m$. However, in the paper we do not restrict our study to  the experiment framework, for instance the pore width may be larger in order that the \myeq{linear-drag} remains relevant. The main goal of this analysis is to answer the open questions, is it possible to transport passive particles in a slow viscous flow with zero mean value in time and space? And if the answer is positive, which mechanisms are responsible? In this way, one does not aim at computing the exact velocity field $u_0(x)$ from the experiment set-up but  the velocity field will be rather described by three parameters, for instance the characteristic velocity $u_m$, the velocity contrast $a$ and the spatial asymmetry $d$. These parameters vary around values relevant for Stokes' flow and small particle inertia.%

Finally, one obtains a second order differential equation of the particle motion
\begin{equation}
 \ddot{s}+\gamma \dot{s}=\gamma v_{0}(s,t),\mylab{eq:1dode}
\end{equation}
and according to \myeq{vxt} $v_0(s,t)$ can be decomposed 
\begin{equation}
v_{0}(s,t)=u_0(s)\sin(2\pi t)\mylab{eq:1dode}
\end{equation}
where  $u_0(s)$ is 1-periodic function, i.e. the pore periodicity in dimensionless problem.
\subsection{Dimensionless parameters}
As explained above, the drag may vary from values about 10 to infinite.
The dimensionless characteristic velocity $u_m$ defined by
\begin{equation}
u_m=\int_0^1v_0(x)dx
\end{equation}
is about $1$ to $25$ depending on the pressure difference on both sides of the pore network.\\
Then one introduces the scaled relative variations of the velocity $w_0(x)$ by:
\begin{equation}
u_0(x)=u_m(1+aw_0(x))
\end{equation}
where $w_0$ is a periodic function and smooth enough, i.e. $C^1$ of zero mean value:
\begin{eqnarray}
 w_0(x+1)&=&w_0(x) \mylab{eq:w0per}\\
 \int_0^1w_0(x)dx&=&0.\mylab{eq:w00}
\end{eqnarray}
One assumes that $|w_0(x)|$ and $a$ vary in the range $[0,1]$ so that $a$ is related to the extrema of $u_0$
\begin{equation}
a=\frac{u_{0max}-u_{0min}}{u_{0max}+u_{0min}}.
\end{equation}
Therefore, the parameter $a$ is called the velocity contrast.\par
The particle motion is governed by the second order non-linear ordinary differential equation:
\begin{equation}
\ddot{s}+\gamma \dot{s}=\gamma u_m \left( 1 + a w_0(s) \right) \sin(2\pi t),\mylab{eq:ode}
\end{equation}
Such a equation admits an unique solution $C^2$ for a given position and velocity $(s_i,v_i)$ at a time $t_i$. One can show that the particle acceleration $\ddot{s}$ and its velocity $\dot{s}$ are bound (see Appendix I).
%
%
\section{Basic results and numerical method}
There is no analytical results of the non-linear \myeq{ode} in the general case. In this section, one gives the behavior of the particle in some limit cases. The generic case is numerically solved.
\subsection{Basics}%
\label{sec:basics}
\subsubsection{Uniform flow: $a\ll1$.}
%
If $a=0$ the flow without particles  $v_0(x,t)$ is spatially uniform. The ODE \myeq{ode} becomes second order linear equation with a periodic force and a damping force:
\begin{equation}
\ddot{s}+\gamma \dot{s}=\gamma u_m  \sin(2\pi t),\mylab{eq:ode a=0}
\end{equation}
 The dynamics tends exponentially to an 1-periodic solution with the amplitude
 \begin{equation}
 A=\frac{u_m}{2\pi}\left(1+\frac{4\pi^2}{\gamma^2}\right)^{-1/2}
 \end{equation} 
 The amplitude of the oscillations is maximal for large drag (advection). The characteristic time of relaxation is scaled by the inverse of the drag $1/\gamma$. \par
If $a\neq0$, the invariance in the $x$ direction is broken and the existence of 1-periodic solution is no more guaranteed  even if $a$ is close to zero. Such a scenario occurs in the \mysec{asym} for an  asymmetric pore profile. However, if the geometry of the pore has the parity left-right symmetry, then there is generically solutions which have this symmetry. Because there is two points  of reflection symmetry  --the minimum and the maximum of the flow $v_0(x)$-- there is two 1-periodic solutions which move around each symmetry point (see next \mysec{sym}). The existence of both solutions plays a relevant role for the understanding of the transport dynamics. 
\subsubsection{Large drag $\gamma$.}
In the experience the drag is at least equal to 100. The limit case $\gamma=+\infty$ corresponds to a pure advection of the particle and leads to the first order ODE
\begin{equation}
 \dot{s}_\infty=u_m \left( 1 + a w_0(s_\infty) \right) \sin(2\pi t).
\end{equation}
The solutions of this equations are 1-periodic in time. Then, a drift is not possible if no constant force is present at the gravity. In the literrature it will be found as overdamped ratchet. In the \mysec{aint} we will study the nearly case of overdamped ratchet: $\gamma$ is finite but large.

For large values of $\gamma$ but finite, the continuum of 1-periodic solutions disappears but it may still remain both 1-periodic solutions mentioned in the paragraph above.  The existence of periodic solutions at large drag $\gamma$ will be discussed in the last section \ref{sec:asym}  and it depends strongly on the asymmetry of the flow. \\

Then, for small values of parameters $a$, $u_m$ and large values of $\gamma$, one expects only 1-periodic motions when the flow profile is symmetric or slightly asymmetric.
\subsubsection{linear variation of $v_0(x)$}
Here, we do not consider a pore lattice rather a pore with varying diameter so that the velocity $v_0$  varies linearly
\begin{equation}
 u_0(x)=u_m(1+\frac{x}d),\mylab{eq:w0lin}
\end{equation}
with $a>0$ and $u_m>0$. For a dynamic close to $x=0$, for instance far from the singular point $x=-d$, the time integration of the ODE (\ref{eq:ode}) shows  a drift to the negative direction, i.e. to the decreasing values of $v_0(x)$. The consequence for the particle motion in a pore is  that for small enough oscillations, the particles tend to the region where $v_0(x)$ is minimal, i.e. where the pore has the larger diameter. Then, the periodic solution close to the velocity minimum should be stable and the other one close to the maximum should be unstable. Of course, when the  oscillation amplitudes become comparable to the pore length, this argument does not yield any more.
\subsection{Numerical method}
The above analysis shows that it may exist two 1-periodic solutions if the profile is slightly asymmetric and the oscillation amplitude is small. Moreover, one does be stable. Therefore, this scenario does not allow the existence of a drift. We need to explore larger parameter values.
The time integration of the \myeq{ode} allows to find possible transport solution but each simulation depends on the initial values and there is no information on stability. Moreover, it is difficult to understand the mechanism of the transport. The continuation technics is a powerful tool to track periodic solutions in the parameters space. Moreover, one gets information about their stability. This approach allows to better understand the dynamics and to find transport solution. In some cases, the onset of existence of periodic solutions may explain a transport phenomenon.  In order to perform the continuation task, one uses the freeware package AUTO \cite{AUTO00}.  It allows the continuation of periodic orbit and the detection of doubling period and fold bifurcation of a ODE's system.
The ODE system (\ref{eq:ode}) is transformed in a three dimensional system with $(s,v(=\dot{s}),t)$ as variables
\begin{equation}
\left\{
\begin{array}{rcl}
\dot{s}&=&v\\
\dot{v}&=&\gamma\left(u_m[1+aw_0(s)]\sin(2\pi t)-v\right)\\
\dot{t} &=& 1
\end{array}
\right.
\end{equation}
This last system can be formal written
\begin{equation}
\left({\dot{s},\dot{v},\dot{t}}\right)=F(s,v,t)\mylab{eq:ds}
\end{equation}
 Each point of the bifurcations diagrams represents the norm $||.||$ of a $T$-periodic state $s$. Its norm is defined as the $L_2$-norm of the particle velocity:
\begin{equation}
||s||=\frac1{T}\left( \int_0^Tv^2(t) dt \right)^{1/2}.
\end{equation}
Then, the norm does not depend on the particle position and for instance two identical particle motions but shifted by an integer of pore lengths have the same norm.\\
The path-following method requires to know the starting solution. The periodic motions for an uniform flow, i.e. $a=0$, being known, one performs the continuation of the periodic solution branch expanding first by the parameter $a$.
%
%
\section{parity-symmetry of pores} 
\mylab{sec:sym}\\
%
If the geometry of pores lattice has the spatial parity symmetry $x\rightarrow-x$, an effective transport is not possible. However, this fact does not forbid the existence of transport solutions for a given initial condition. Then, the symmetry of the flow velocity implies the existence of a symmetric transport in the opposite direction.
 The motivation to consider a parity symmetry geometry lies in a better understanding of the different mechanisms resulting in a {\em ratchet effect}: dissymmetric flow and  synchronization phenomenon. In this section, we show the existence of transport solutions related to a spontaneous symmetry breaking of symmetric periodic solutions.
In the next \mysec{amtw}, the existence of transport solution for an asymmetric geometry is interpreted as a perturbation of the symmetric case using continuation method.
The asymmetry of the pore geometry ensures only that the mean transport measured over all initial conditions is generically non-zero.
\par

\subsection{Equivariance of the ODE \myeq{ds}}
The parity symmetry means the parity of the incident velocity field at a given time $t$, i.e. 
\begin{eqnarray}
v_0(-x)=v_0(x) &\Longleftrightarrow&w_0(-x)=w_0(x)
\end{eqnarray}
 Then, the function $F$ of the system \myeq{ds} is equivariant by the spatio-temporal transformation
\begin{eqnarray}
{\cal S}_0&:&(s,v,t)\rightarrow(-s,-v,t+1/2) \mylab{eq:equi0}
\end{eqnarray}
what means
\begin{equation}
F(S_0(s,v,t))=S_0(F(s,v,t)).
\end{equation}
The consequence is that for each   solution $(s(t),v(t))$ of  \myeq{ds} then the symmetric dynamic: $(s_{-}(t),v_-(t))=(-s(t+1/2),-v(t+1/2))$ is solution too. 
Let us remark that because of the time and spatial 1-periodicities, the system is also equivariant by the transformation
\begin{eqnarray}
{\cal S}_{m}&:&(s-1/2,v,t)\rightarrow(-(s-1/2),-v,t+1/2), \mylab{eq:equidemi}
\end{eqnarray}
which corresponds to a parity symmetry around the middle of the pore: $x=1/2$.  \\
We say that a periodic solution is {\em symmetric} if it is invariant by one of these transformations ${\cal S}_0$ or ${\cal S}_{m}$. We denote by $s_0$  the solution invariant by ${\cal S}_0$. If $<s>_t$ is the mean position over the temporal period:
\begin{eqnarray}
<s>_t &=& \frac{1}{T}\int_0^Ts(t) dt.
\end{eqnarray}
then the equivariance  show that $<s_0>_t=0$.
In this sense it is centered at zero. In the same way the symmetric solution $s_m$ invariant by ${\cal S}_{m}$ is centered at the pore middle: $<s_m>_t=1/2$. Obviously, the spatial periodicity implies that at each point $x=k/2$, $k\in \mathbb{Z}$ there is a symmetric solution. But there are spatially translated solutions from $s_0$ or $s_m$. Then we focus only on both solutions $s_0$ and $s_m$. \\
Because of the spatio-temporal symmetries, the solutions $s_0$ and $s_m$ generically exist according to the concluding remarks of \mysec{basics}.
In order to find transport solutions, this parity symmetry of the dynamics needs to  be broken. Then by varying the three parameters: the drag $\gamma$, the characteristic velocity $u_m$ and the velocity contrast $a$, we explore the bifurcation diagrams of the periodic solution branches.\\
A sinusoidal dependence is chosen for the velocity field $v_0(x)$ of the driving flow:
\begin{equation}
v_0(x) = u_m\left( 1+a \cos(2\pi x)\right),\mylab{eq:u0sym}
\end{equation}
which corresponds for $a=0.65$ roughly to the velocity profile  on the axis  in the micro-pump experiment in \cite{KRHM00}.  Thus, this value will often used for the different bifurcation diagrams.
\subsection{1-periodic solutions}\label{sec:1per}
\begin{figure}[tb]
\includegraphics[width=0.45\hsize]{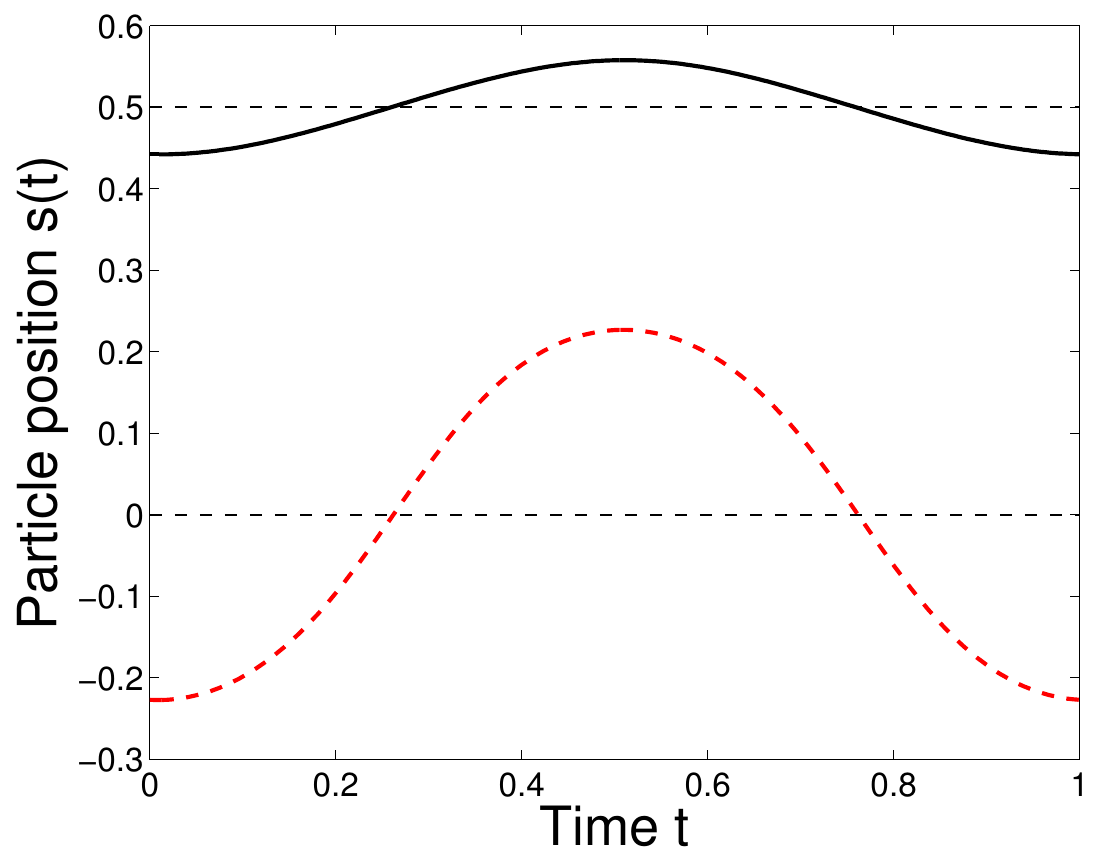}
\caption{Time evolution of the 1-periodic symmetric solutions (red) $s_0$ and (black) $s_m$  for $u_m=1$,  $\gamma=100$ and $a=0.65$ of the velocity field $v_0(x)$ given by \myeq{u0sym}. Solid [Dashed] line indicates stable [unstable] solution.}\mylab{fig:symu1solt}
\end{figure}
%
%
\begin{figure}[tb]
\includegraphics[width=0.45\hsize]{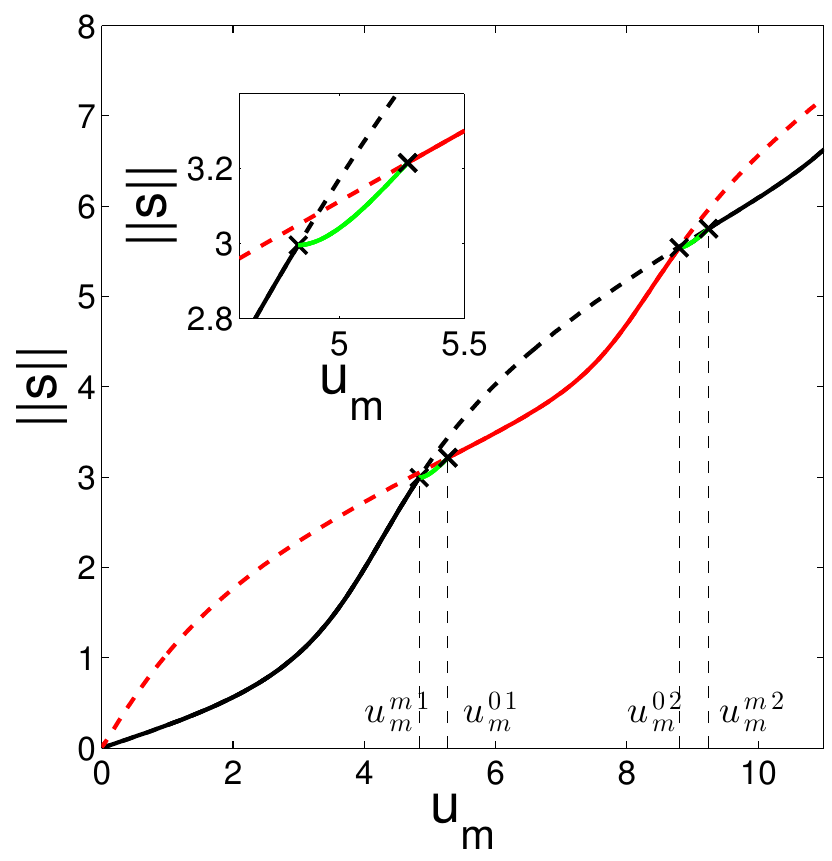}\hfill
\includegraphics[width=0.45\hsize]{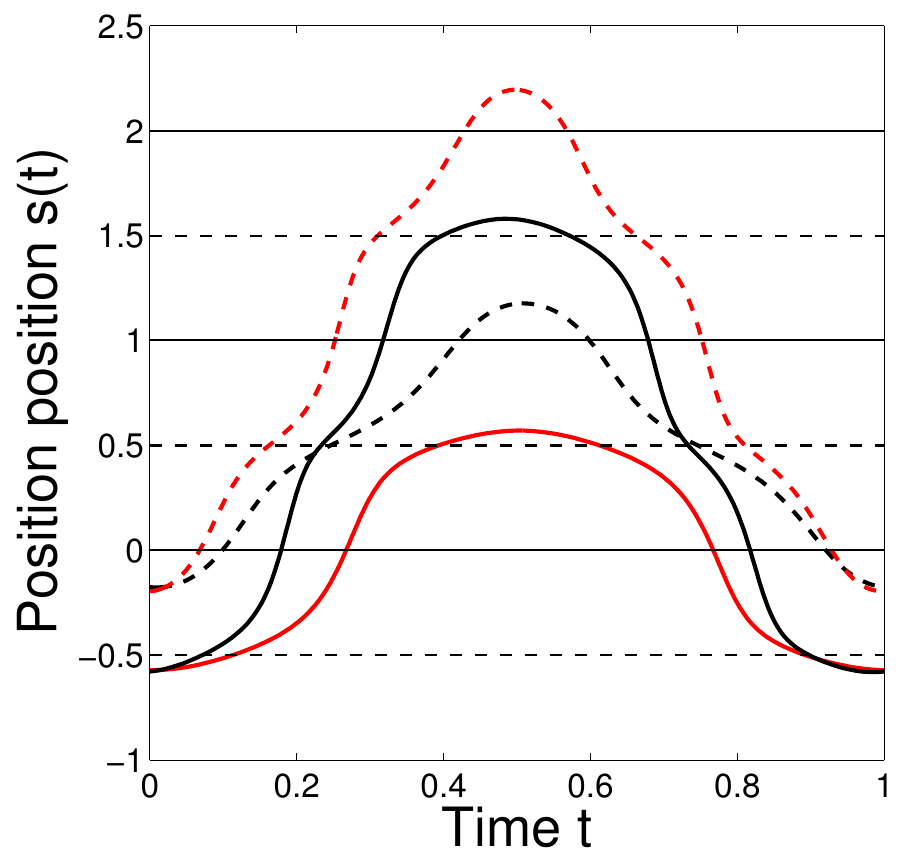}\\
\caption{Continuation of branches of 1-periodic orbits (red line) $s_0$ and (black line) $s_m$  for the symmetric velocity profile $v_0(x)$ (\myeq{u0sym})  by varying $u_m$  at $\gamma=100$ and $a=0.65$. Red color corresponds to $s_0$ and black to $s_m$. Symbol 'x' indicates a pitchfork bifurcation. (b) Time evolution of periodic branches at the pitchfork bifurcations of the panel (a): (plain red) $s_0$ at $u_m^{m_1}=5.27$, (dashed red) $s_0$ at $u_m^{m_2}=9.24$,  (dashed  black) $s_m$ at $u_m^{0_1}=4.84$ and at (plain black) $u_m^{0_2}=8.79$}\mylab{fig:symgia65cu}
\end{figure}
%
\begin{figure}[tb]
\includegraphics[width=0.65\hsize]{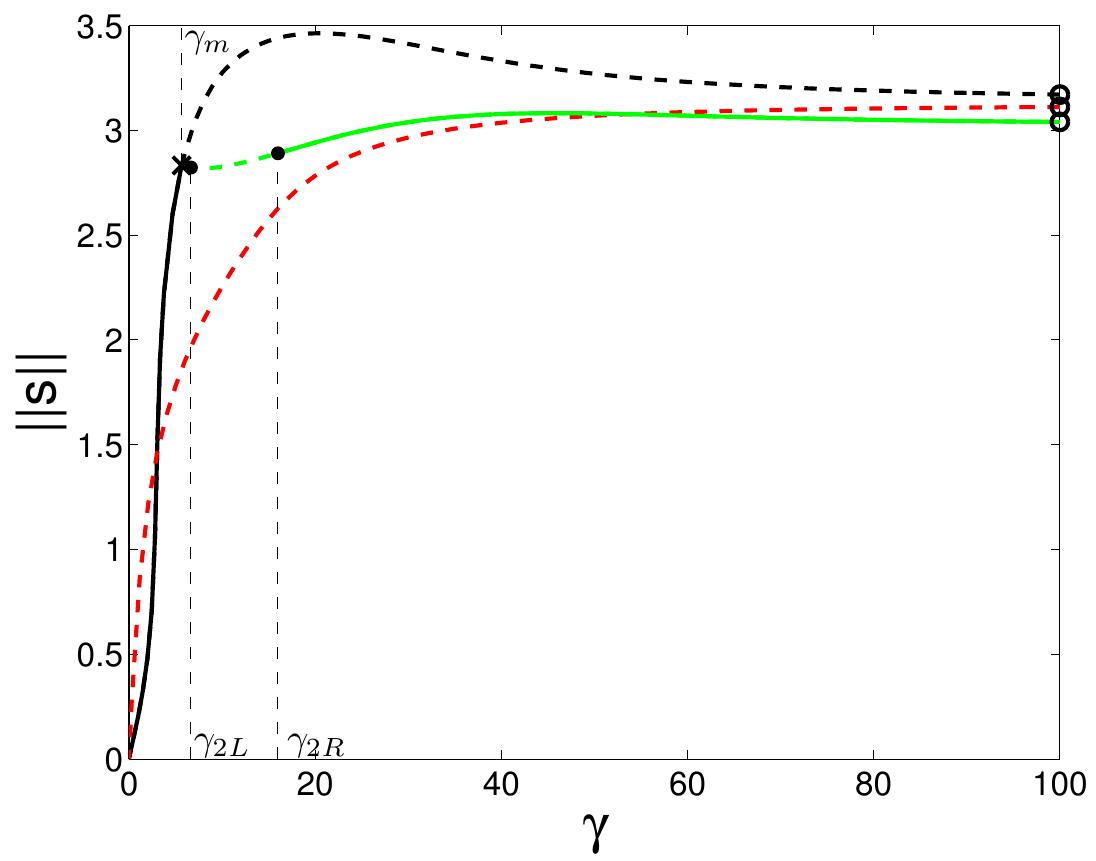}

\caption{Continuation of 1-periodic orbits  by varying the drag $\gamma$ from $100$ to zero. The starting  states at $a=0.65$ are the circles in the bifurcation diagram \myfig{symgia65cu}a. Color code of the branches is the same as  \myfig{symgia65cu}a. Symbol 'x' indicates a pitchfork bifurcation  while a dot indicates a period doubling bifurcation.}\mylab{fig:symu5a65cg}
\end{figure}
%

%
The parity-symmetric periodic solutions $s_0$ and $s_m$ get the periodicity of the forced oscillations, i.e. they are one-periodic. 
The 1-periodic symmetric solution is given for $a=0$ by the linear differential equation \myeq{ode a=0} and it is the starting solution used for the path-following.%
\\
The first bifurcation diagram is performed with a large drag $\gamma=100$. Then, one expects an almost advective motion of the particle of small amplitudes.  When $a$ is non-zero the continuum of periodic solutions breaks into the two solutions $s_0$ and $s_m$ as shows \myfig{symu1solt}.
The solutions $s_0$ and $s_m$ are respectively  oscillations around the maximum and the minimum of the velocity field. As expected in \mysec{basics} the solution $s_0$ is unstable and $s_m$ is stable. 
The continuation of these solutions shows that for any velocity contrast $a$ there is no bifurcation and the stable solution $s_m$ remains stable.
By decreasing the drag, there is no bifurcation too  when $a=0.65$. Therefore, all the dynamics are attracted by the only stable periodic solution $s_m$.%
\\
Now one varies the characteristic flow velocity $u_m$ and the velocity contrast and the drag are fixed: $a=0.65$, $\gamma=100$. Both solutions exchange their stability (\myfig{symgia65cu}). The solution $s_0$ stabilizes at $u_m^{0_1}=5.27$. The \myfig{symgia65cu}-b shows its time evolution during one period: the amplitude ends just beyond the position $x=\pm1/2$, i.e., where the flow velocity is minimal. In contrast, the periodic solution $s_m$ destabilizes at $u_m^{m_1}=4.84$ for which the amplitude of oscillations ends close to $x=1$, i.e. the maximum of the velocity flow (\myfig{symgia65cu}-b). Moreover, by increasing $u_m$, the solution stabilizes again at $u_m^{m_2}=9.24$ when the amplitude of oscillations are just beyond the minimum of velocity $x=1.5$ or $x=-0.5$ (\myfig{symgia65cu}-b).  An analogous destabilization occurs for $s_0$ at $u_m^{0_2}=8.79$: the boundary of the oscillations of the periodic solution $s_0$ are close to $\pm1$ (maxima of flow velocity). Therefore the stability of the one-periodic solutions seems to follow the rule: if the boundary of oscillations are close to the minimum of velocity $v_0$ (i.e. at $x=1/2mod[1]$) the solution is stable and if the boundaries of oscillations are close the maximum of the velocity $v_0$, then the solution is unstable.\\
The exchange of stability implies the emergence of a new branch which breaks the ${\cal S}_0$ reflection symmetry, then it is called the asymmetric branch $s_a$. It is a pitchfork bifurcation of periodic orbit and then there are two solutions $s_{a+}$ and $s_{a_-}$ which emerge from the critical point such $s_{a+}={\cal S}_0 (s_{a-})$. Because they are symmetric, they have the same norm and then they do not appear twice in the bifurcation diagram \myfig{symgia65cu}-a. Thus both solution branches are called  $s_a$. When the solution $s_m$ loses its stability at $u_m^{m_1}$ the branch $s_a$ emerges supercritically and connects the branch $s_0$ at $u_m^{0_1}$ which gains stability. A similar scenario appears between the critical values $u_m^{02}$ and $u_m^{m2}$ (\myfig{symgia65cu}). The branches $s_a$ being stable, there is always a stable periodic solution for all values of $u_m$. The time integration of the ODE \myeq{ode} shows that  the dynamics tend to the stable solution which forbids any transport solution.
\\

To conclude, the symmetric solutions $s_0$ and $s_m$ may change their stability which depends strongly on the amplitude of their oscillations. Roughly, the stability changes when the  oscillation amplitude increases of a half pore length. At the onset of stability,  a new branch $s_a$ of 1-periodic solution bifurcates. This branch is no more invariant by the reflection symmetry ${\cal S}_0$ nor by ${\cal S}_{m}$. For a large drag ($\gamma=100$) this asymmetric branch $s_a$ is stable and then there is always a stable periodic solution which attracts the dynamics. 
In the following, one studies the dynamics into smaller friction term $\gamma$. %
%
\subsection{Doubling period cascade}
\begin{figure}[tb]
\includegraphics[width=0.45\hsize]{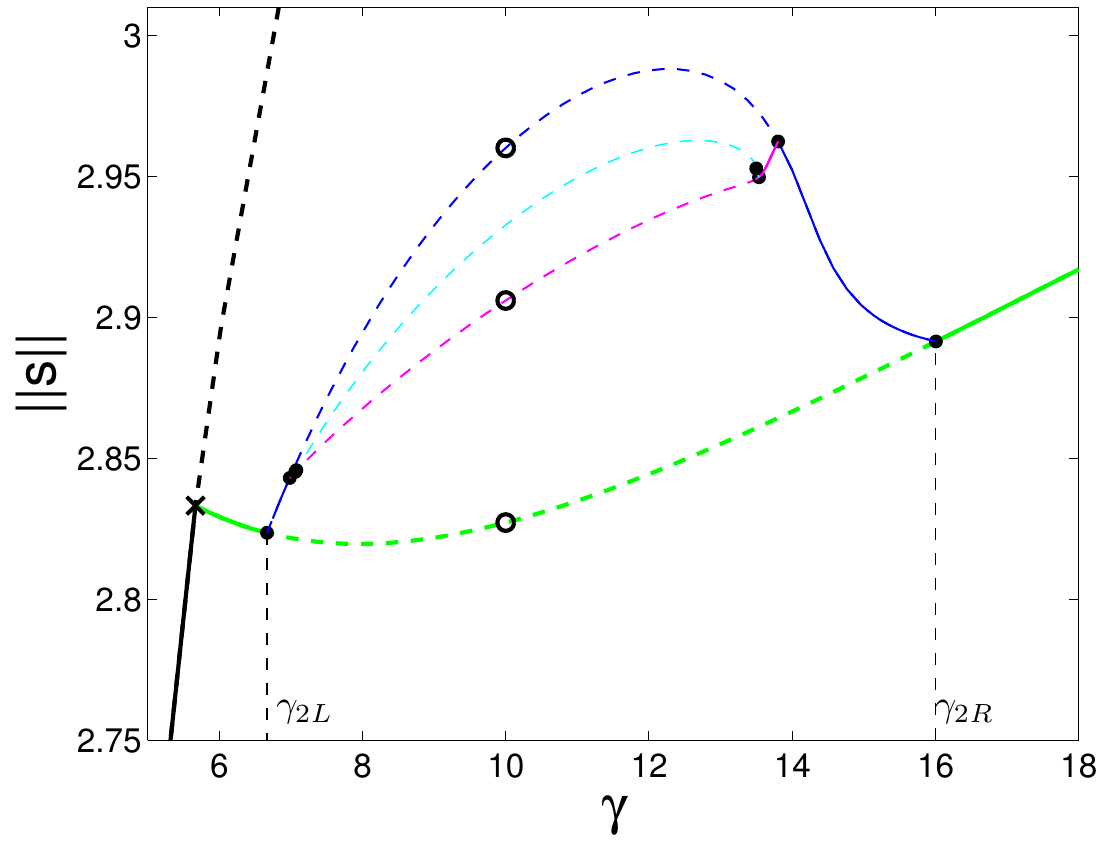}\hfill
\includegraphics[width=0.45\hsize]{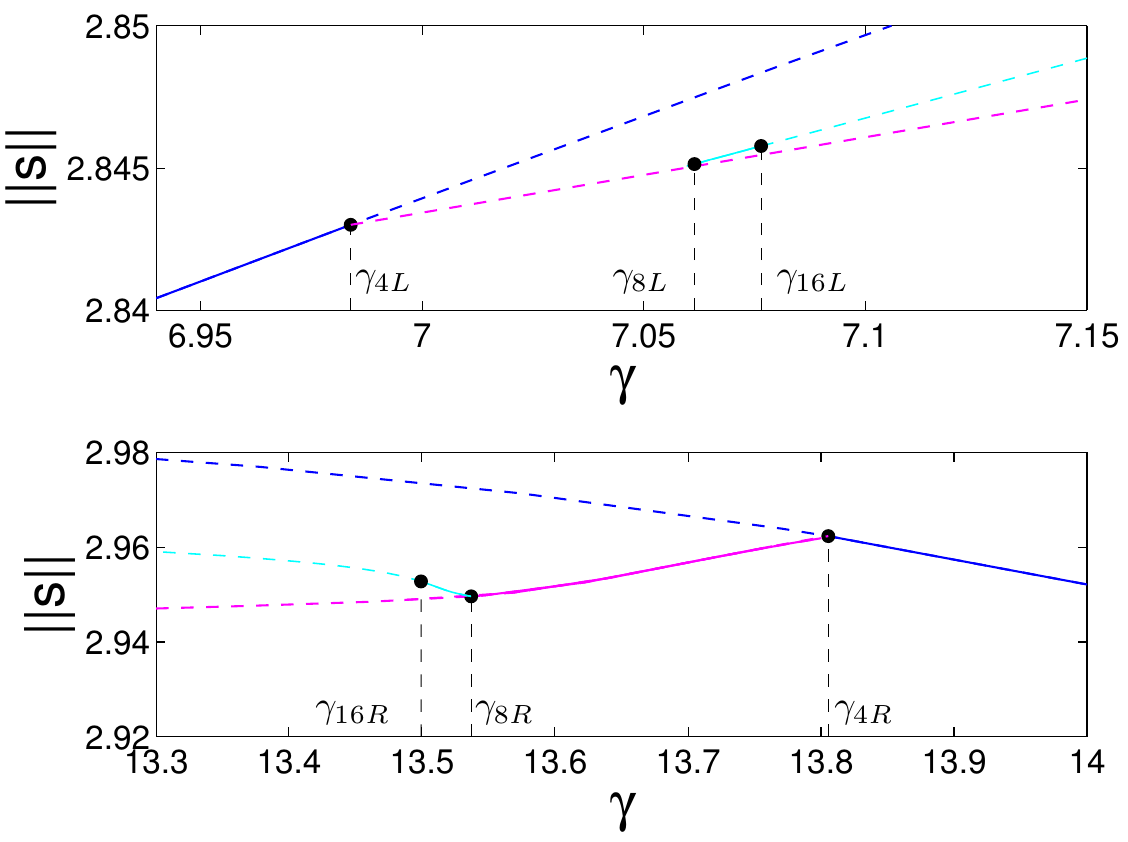}\\
(a)\hfill (b)
\caption{Bifurcated period-doubling branches  emerging from (green line) $s_a$ in  \myfig{symu5a65cg}. Shown is (a) the bifurcation diagram and (b) the magnification near the period doubling bifurcations. For the one-periodic orbits the color code is the one used in \myfig{symu5a65cg}. The 2-, 4- and 8-periodic orbits correspond respectively to the thin blue, cyan and sky-blue lines. Solution profiles at $\gamma=10$ indicated by circles are shown in \myfig{snapd}. }\mylab{fig:symu5pd}
\end{figure}
\begin{figure}[tb]
\includegraphics[width=0.7\hsize]{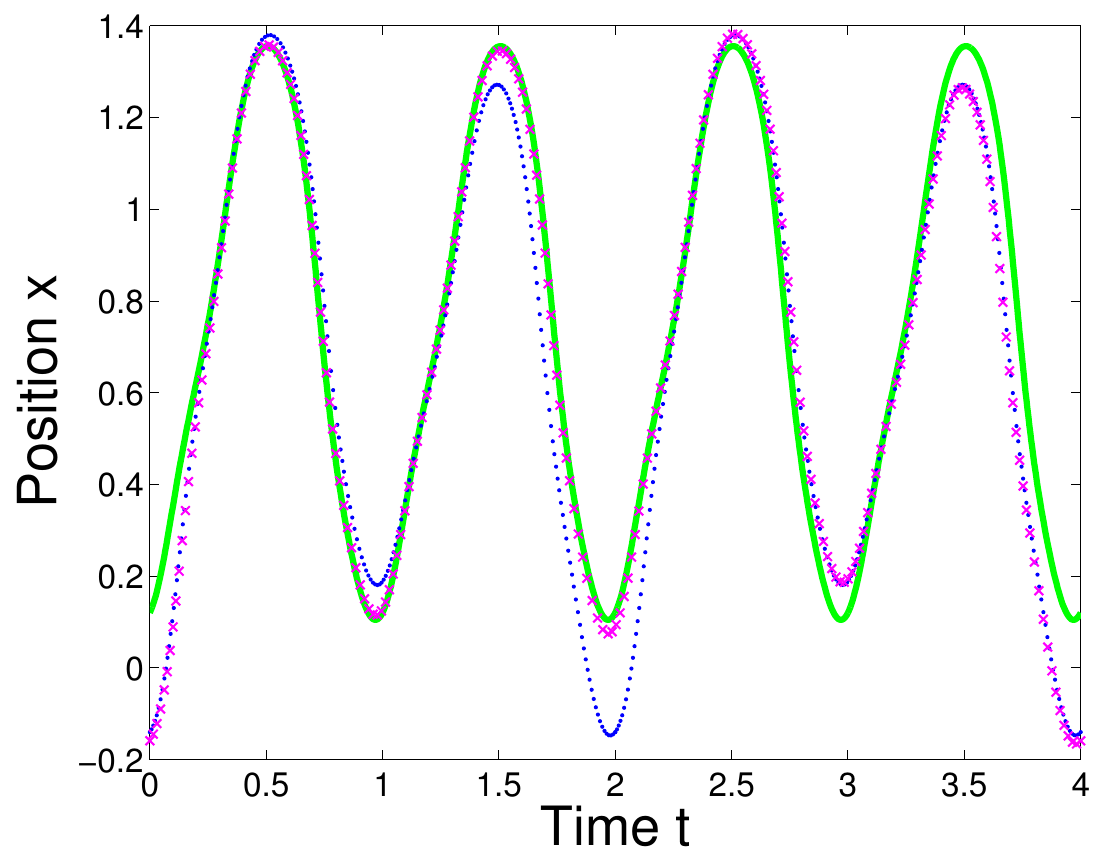}
\caption{Time evolution of  the 1-(plain green line), 2- (blue dots), 4-(magenta 'x' symbols) periodic solutions for $\gamma=10$ at locations indicated by circles in \myfig{symu5pd}. }\mylab{fig:snapd}
\end{figure}
\begin{figure}[tb]
\includegraphics[width=0.75\hsize]{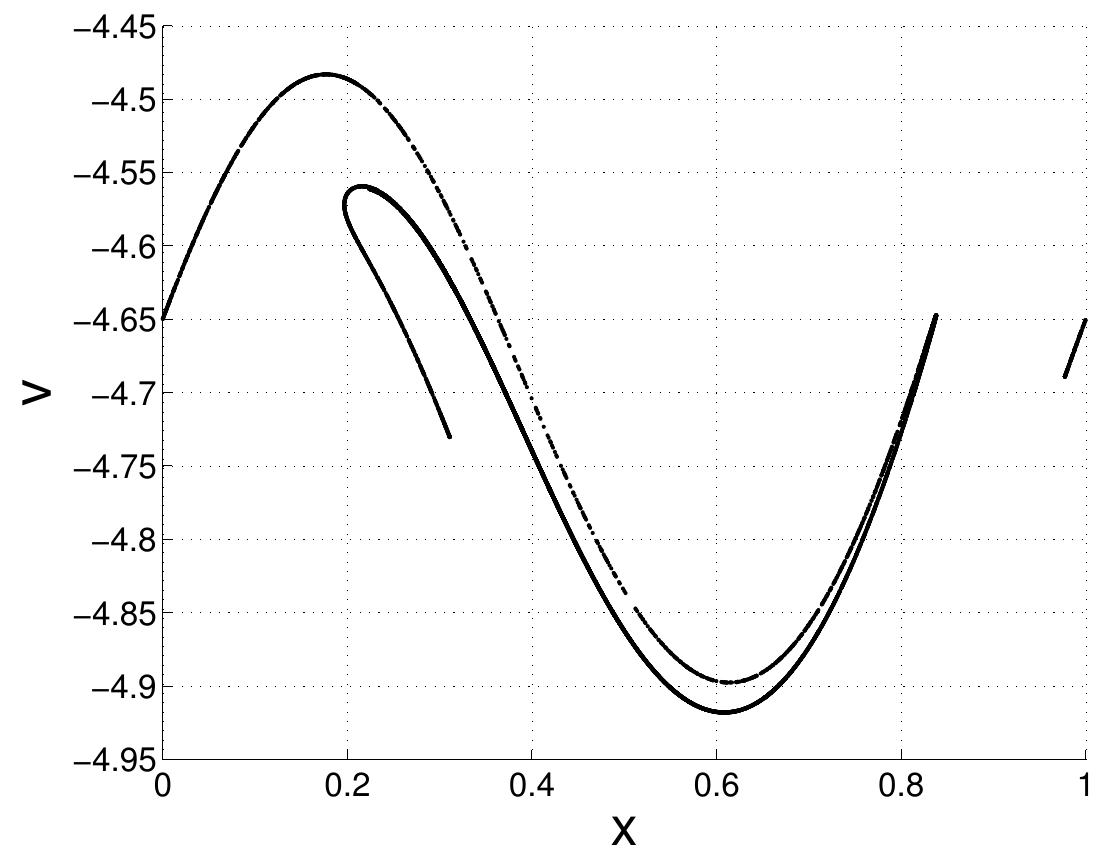}\hfill
\caption{Strange attractor in the $(x,v)$ Poincar\'e map of the dynamics at $\gamma=11$ for the parameters of the bifurcation diagram \myfig{symu5pd} obtained for 100 different initial conditions. Parameters: $a=0.65$, $u_m=5$.}\mylab{fig:symu5quasiper}
\end{figure}
In this section, we show that the $s_a$ branch may be unstable and then a cascade of period doubling bifurcations occurs as it is  generically exhibited by  nonlinear oscillators with dissipation and additional time dependent force \cite{TPJ82}. 
For that let us carry out the continuation of the three 1-periodic solutions at $a=0.65$ and $u_m=5$ (circles in \myfig{symgia65cu}a) by decreasing the drag. The \myfig{symu5a65cg} shows that the branch $s_a$ connects the branch $s_0$ which gains stability at $\gamma_m=5.67$. In contrast to other scenarios of \mysec{1per}, the branch $s_a$ becomes unstable in the range $[\gamma_{1L}=6.67 ; \gamma_{1R}=16.0]$ via two period doubling bifurcations.
Both period-doubling bifurcations are connected by a branch of 2-periodic orbits (see panels (b) of \myfig{symu5pd}). This last kind of solution needs two temporal periods to come back to the same state (position and velocity). In particular, there is two different maxima (\myfig{snapd}). Due to the finite drag, there is a small delay for the direction change of the particle since the extrema are slightly after the times $t=n\pi/2$ with $n$ entire. The 2-periodic branch is stable close to the bifurcations points but loses its stability via a period doubling bifurcation at a slightly larger or smaller $\gamma$. Again a 4-periodic branch relies these points ($\gamma_{4L}=6.98$ and $\gamma_{4R}=13.8$) and it loses its stability via period doubling at $\gamma_{8L}=7.06$ and gains its stability at $\gamma_{8R}=13.5$. Then, there is a range for which the 4-periodic solution is unstable.  The 8-periodic branch is displayed in \myfig{symu5pd}-a and its period-doubling bifurcations are very close from the branches emergence (\myfig{symu5pd}b). This cascade of period doubling continues without end and leads to chaotic dynamics. Moreover, the period doubling bifurcation of the branch $2^{n+1}$-periodic is always closer to the previous bifurcation point of the $2^n$ branch implying that there is a range of non-zero length where every periodic branch is unstable.
 Thus, there is a range for which all periodic orbits are unstable and letting appear quasi-periodic or chaotic motions. The poincar\'e section in the plane $(x,v)$ in \myfig{symu5quasiper}  shows a strange attractor for $\gamma=11$.  The visited domain is included in one pore length thus it is bound. The dynamics may be a quasi-periodic since there is often the case for systems with forced oscillations.
 This route to the chaos  is widely described in deterministic inertia ratchet \cite{Mate00,SKPK03,SER07,MaAl08}. Note however that the route to chaos studied in these papers concerns already drifting solutions which are periodic in the comoving frame.\\
Eventually, one found also $3-$ and $5-$ periodic solutions. These solutions may be stable but they lose their stability by the same period-doubling phenomenon. We will study such a dynamic for non symmetric geometry in \mysec{asym}.\par

The complexity of the dynamics exemplifies the classical results relative to forced systems present in a plethora of nonlinear oscillations topics. Here we do not aim at studying chaotic dynamics since they do not necessary lead to a transport solution.
\subsection{Synchronized transport solutions}
\label{sec:synchro transport}
\begin{figure}[tb]
\centering
\includegraphics[width=0.75\hsize]{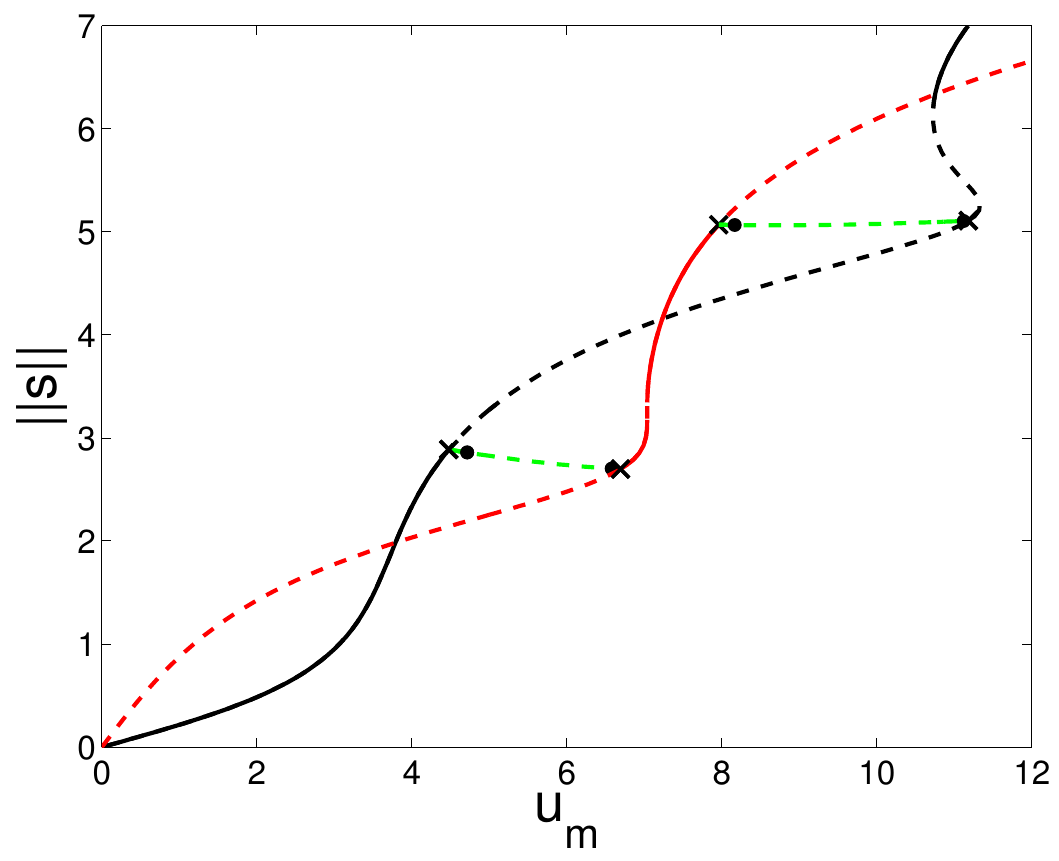}
\caption{Continuation of 1-periodic orbits branches spanned by the characteristic velocity $u_m$ for $a=0.65$ and $\gamma=10$. The velocity profile is given by (\myeq{u0sym}). The same color code and the same bifurcation symbols as in \myfig{symu5a65cg} are used. }\mylab{fig:symcu}
\end{figure}
\begin{figure}[tb]
\includegraphics[width=0.7\hsize]{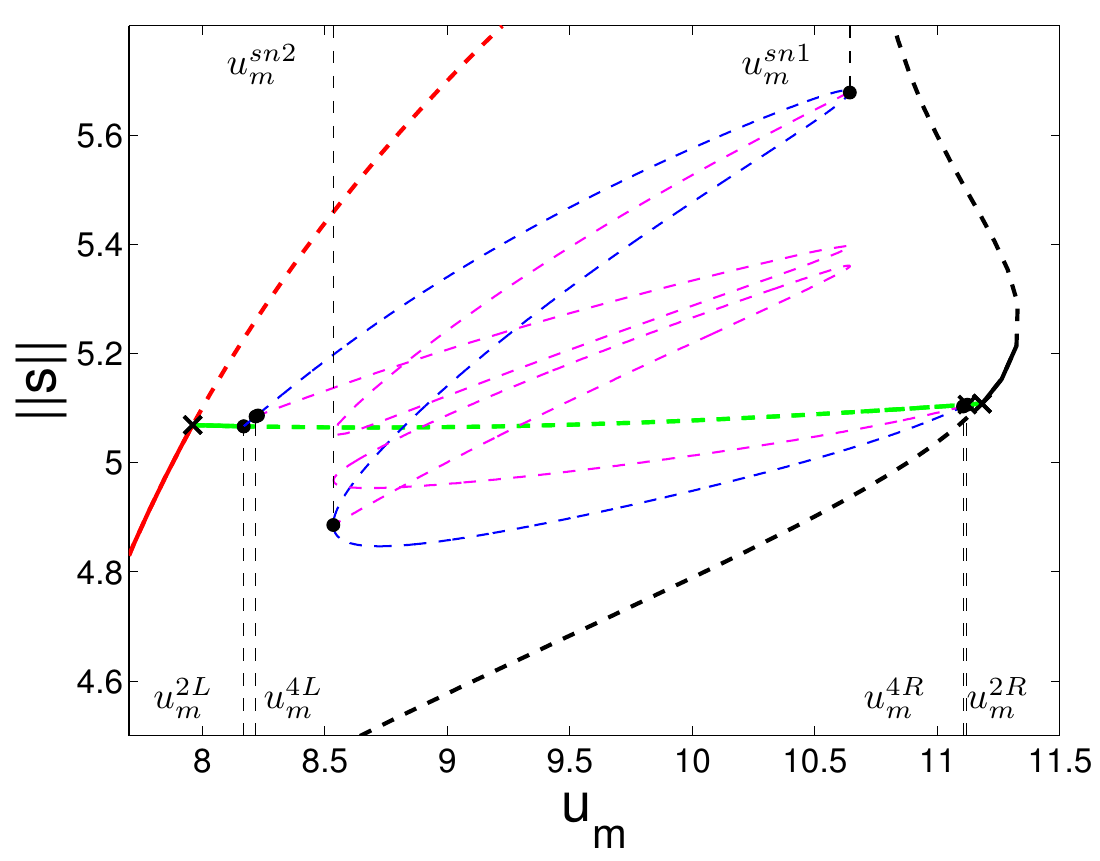}
\caption{Magnification in the range $u_m\in[8;11]$ of the bifurcation diagram \myfig{symcu} (parameters: $a=0.65$, $\gamma=10$). Shown are the (dark blue) 2- and (magenta) 4-periodic branches emerging from the (green lines) $s_a$ branch.  Unstable branch is represented by dashed line otherwise the line is plain. The color code and the bifurcation symbols are the same as used in \myfig{symu5pd}.}\mylab{fig:symg10cu}
\end{figure}
%
%
\begin{figure}[tb]
\includegraphics[width=0.45\hsize]{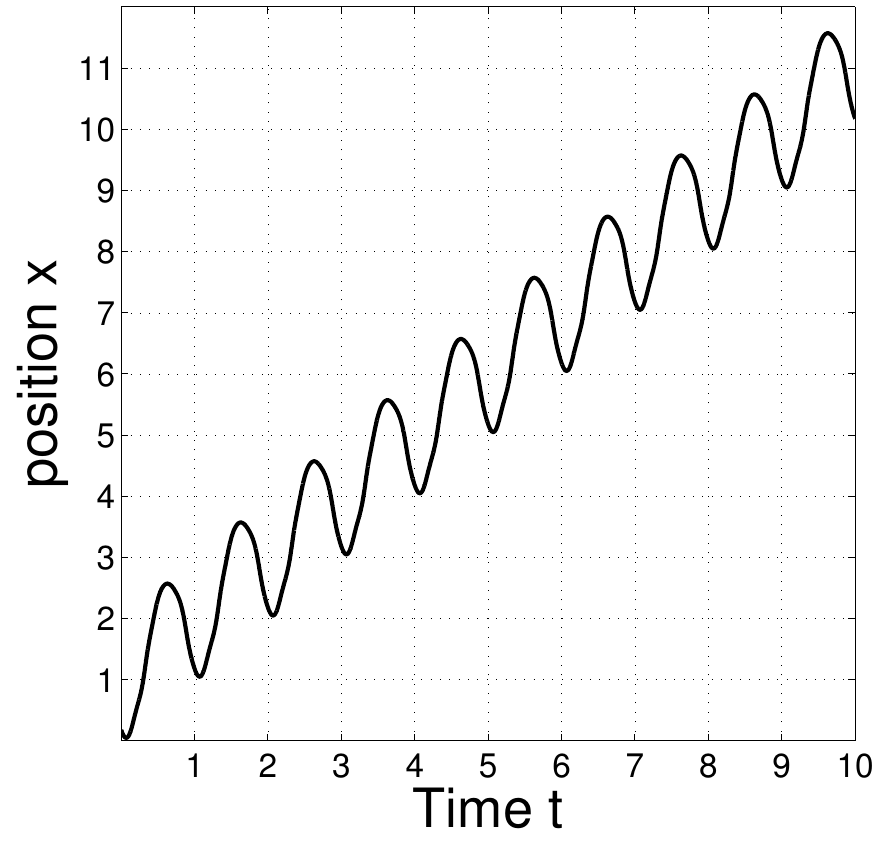}
\caption{Time evolution of the synchronized transport solution $s_t$ at $u_m=9$. The remaining parameters are as in \myfig{symg10cu} ($\gamma=10$, $a=0.65$).}\mylab{fig:symtw}
\end{figure}
\begin{figure}[tb]
\includegraphics[width=0.7\hsize]{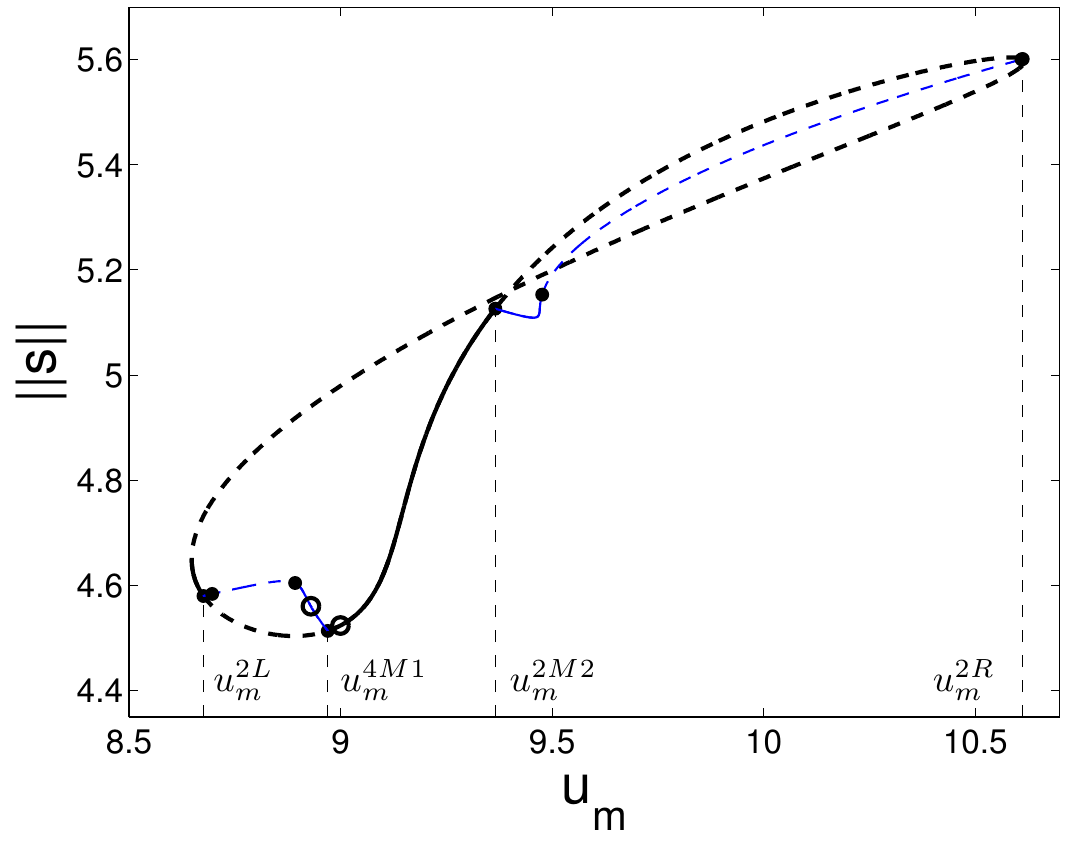}\hfill
\caption{Continuation of the periodic orbits $s_p$ (\myeq{tw}) associated to the synchronized transport solutions $s_t$ with drift velocity $c=\pm1$. Black dots are period doubling bifurcations. The circles indicate the locations of solutions shown in \myfig{symtwpd}. Blue lines are 2-periodic transport solutions. The remaining parameters are as in \myfig{symg10cu} ($\gamma=10$, $a=0.65$).}\mylab{fig:symctw}
\end{figure}
\begin{figure}[tb]
\includegraphics[width=0.45\hsize]{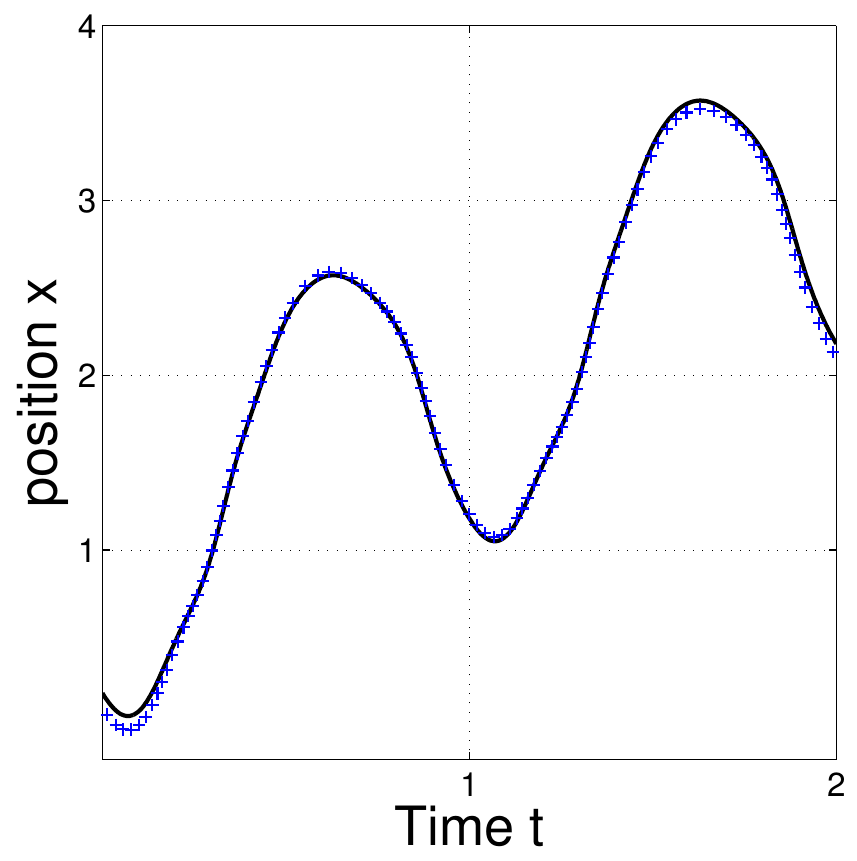}
\caption{Time evolution of the synchronized transport solution $s_t$ at location indicated by circles in \myfig{symctw}. Plain black line corresponds to the solution at $u_m=9$ and the blue $+$ correspond to the 2-periodic transport solution $s_t$ at $u_m=8.93$. The remaining parameters are as in \myfig{symg10cu} ($\gamma=10$, $a=0.65$).}\mylab{fig:symtwpd}
\end{figure}
\begin{figure}[tb]
\includegraphics[width=0.45\hsize]{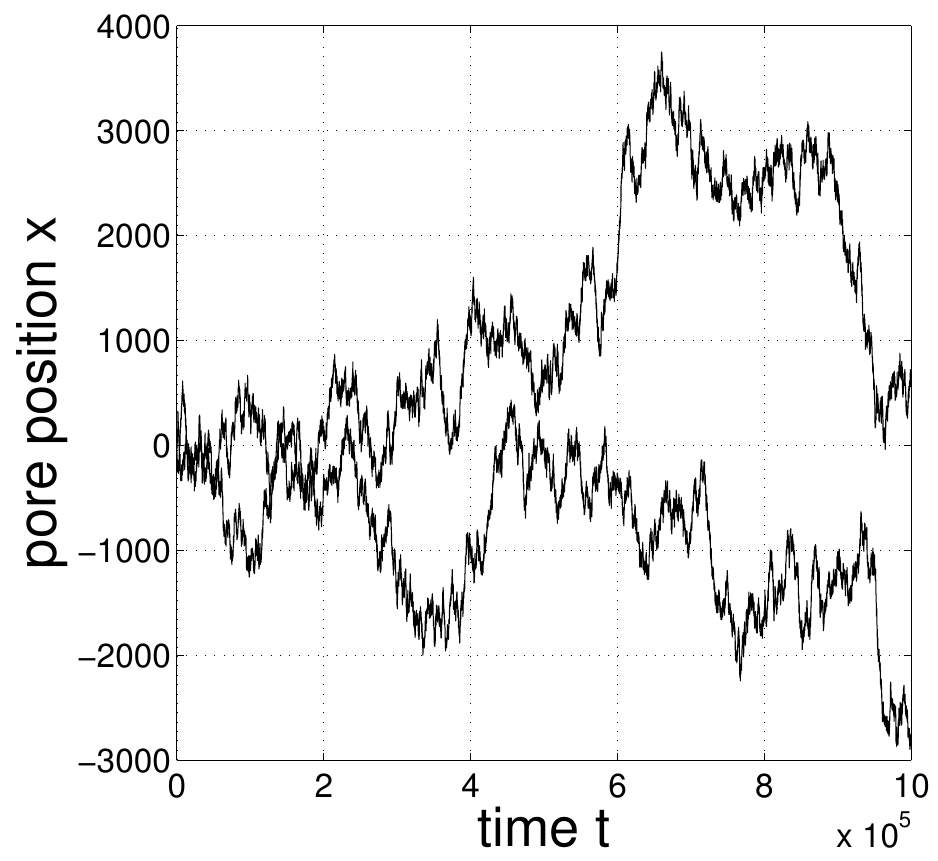}
\includegraphics[width=0.45\hsize]{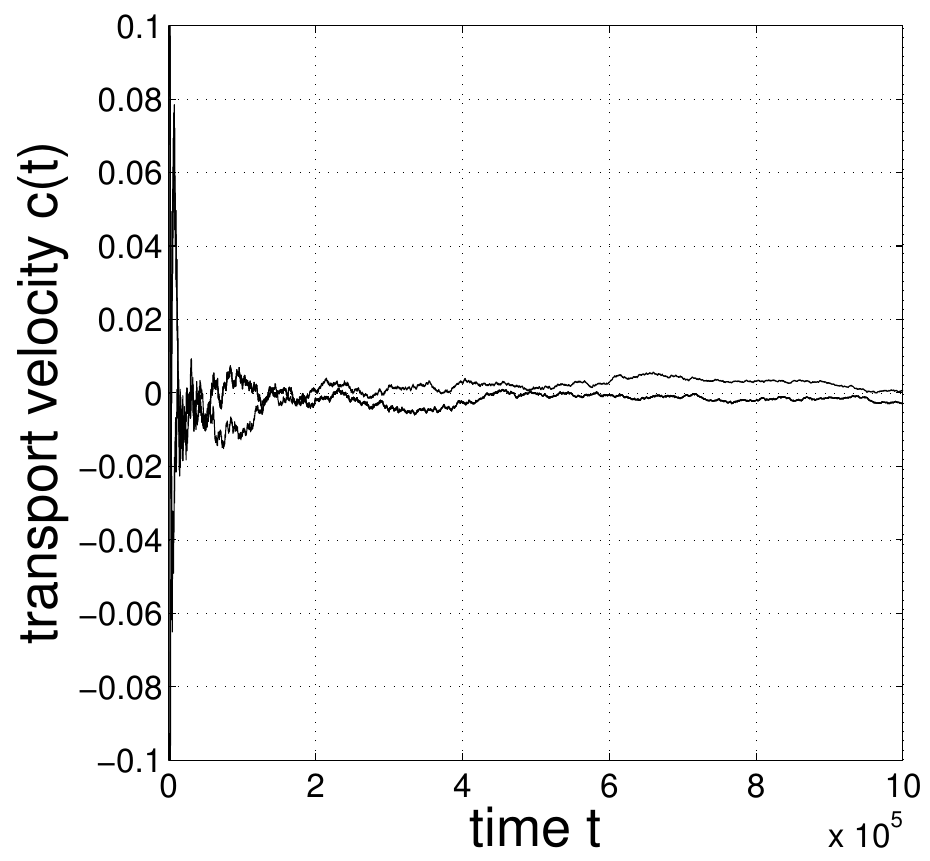}
\caption{(a) Particle dynamics in the presence of two unstable opposite transport solutions at $u_m=8.8$ for two different initial conditions. The remaining parameters are the same as in \myfig{symg10cu} ($\gamma=10$, $a=0.65$). (b) Convergence of the transport velocity of the dynamics of panel (a).}\mylab{fig:symtwt}
\end{figure}
\begin{figure}[tb]
\includegraphics[width=0.45\hsize]{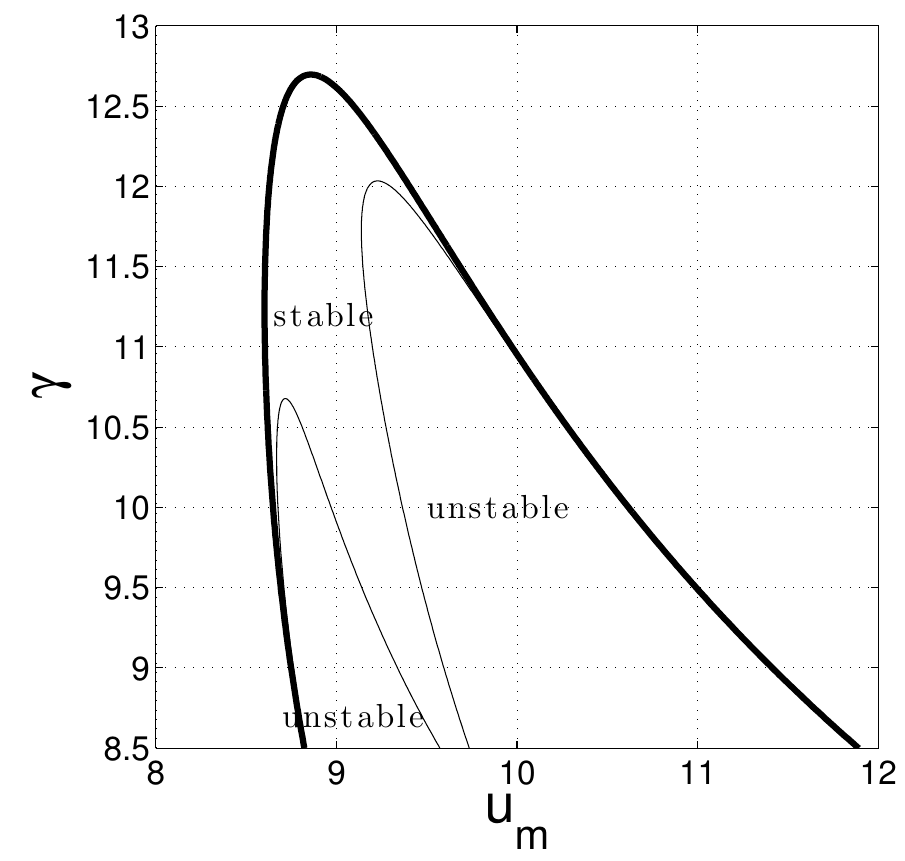}
\caption{Existence domain of the synchronized transport velocity with $c=\pm1$ demarcated by the bold line. Both subdomains (fine lines)  bound the unstable solution and the remaining domain between bold and fine lines corresponds to stable solution.}\mylab{fig:twrange}
\end{figure}
\begin{figure}[tb]
\includegraphics[width=0.8\hsize]{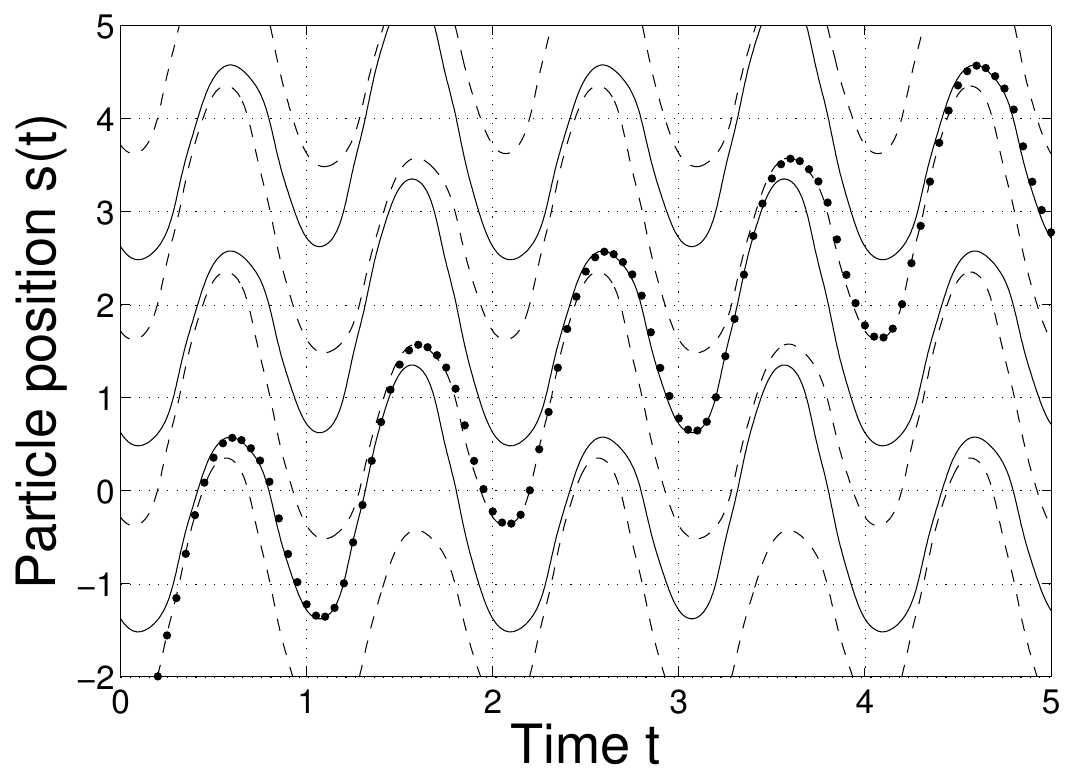}
\caption{Time evolutions of (dots) synchronized  transport solution and (plain and dashed lines) and 2-periodic solutions $s_2$ at $u_m=u_m^{2R}$ in \myfig{symg10cu}. Each solution $s_2'$ (dashed line) is deduced from the solution $s_2$ by spatial and temporal translation: $s_2'(t)=s_2(t+1)+1$.  Parameters $\gamma=10$, $a=0.65$.}\mylab{fig:gluing}
\end{figure}
%
%
\begin{figure}[tb]
\includegraphics[width=0.8\hsize]{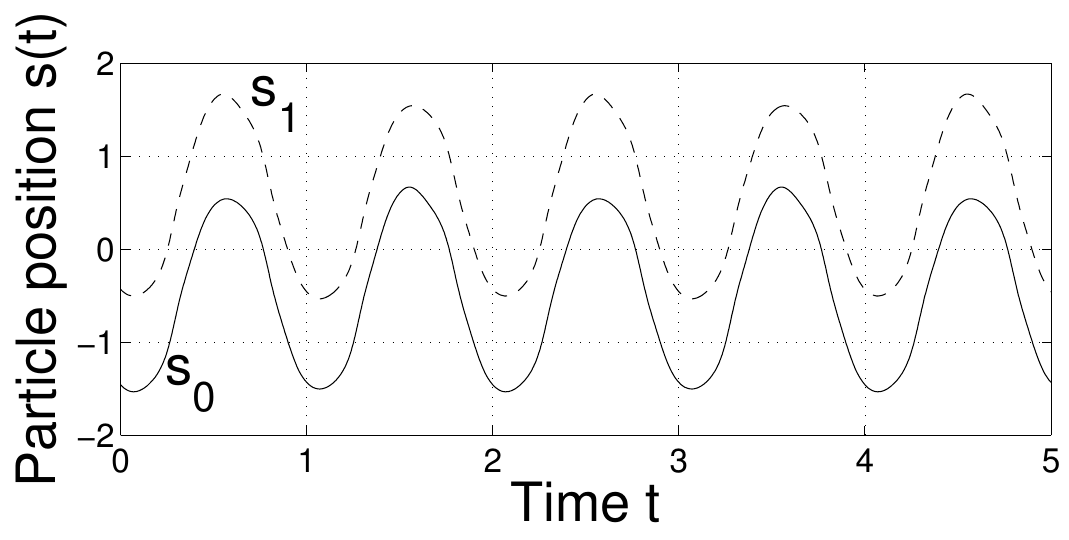}
\caption{Time evolutions of 2-periodic orbits at $u_m=10.7$ (circle in \myfig{symg10cu} $\gamma=10$, $a=0.65$). The orbit is shifted from $s_0$ such $s_1(t)=s_0(t+1)+1$.}\mylab{fig:noglu}
\end{figure}
\begin{figure}[tb]
\centering
\includegraphics[width=0.45\hsize]{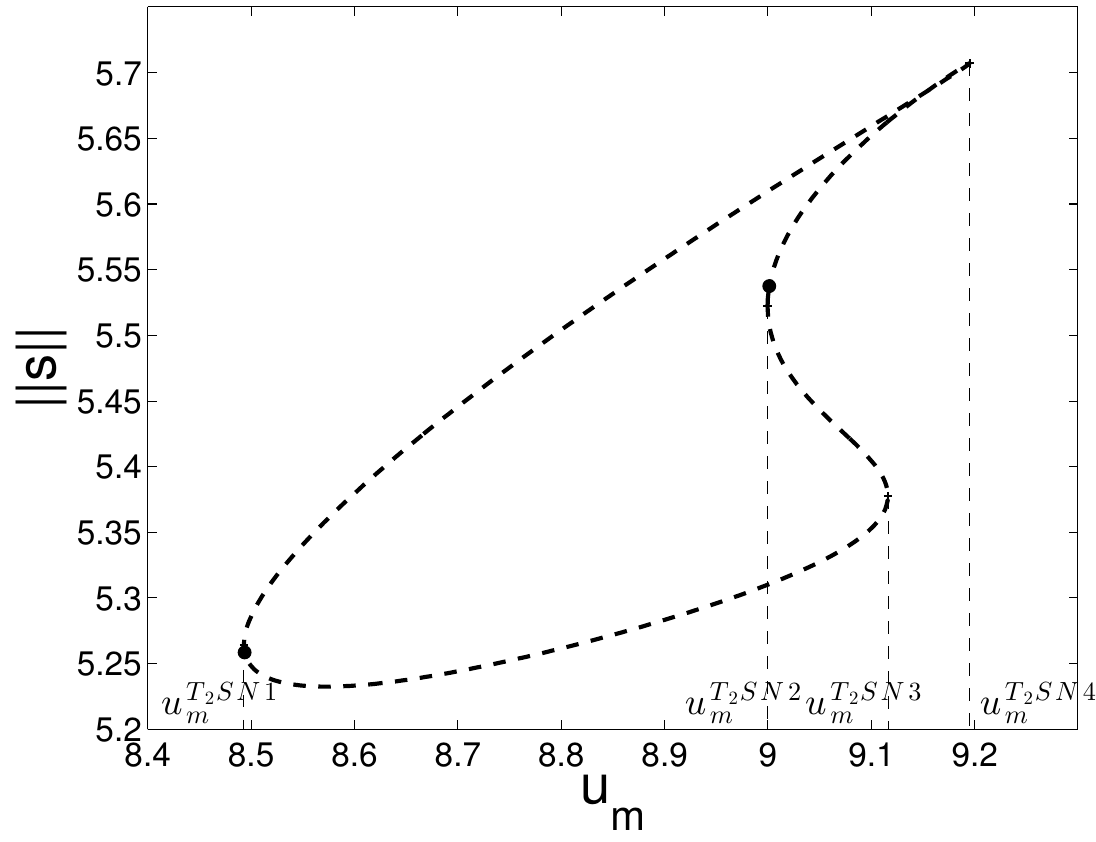}\hfill
\includegraphics[width=0.45\hsize]{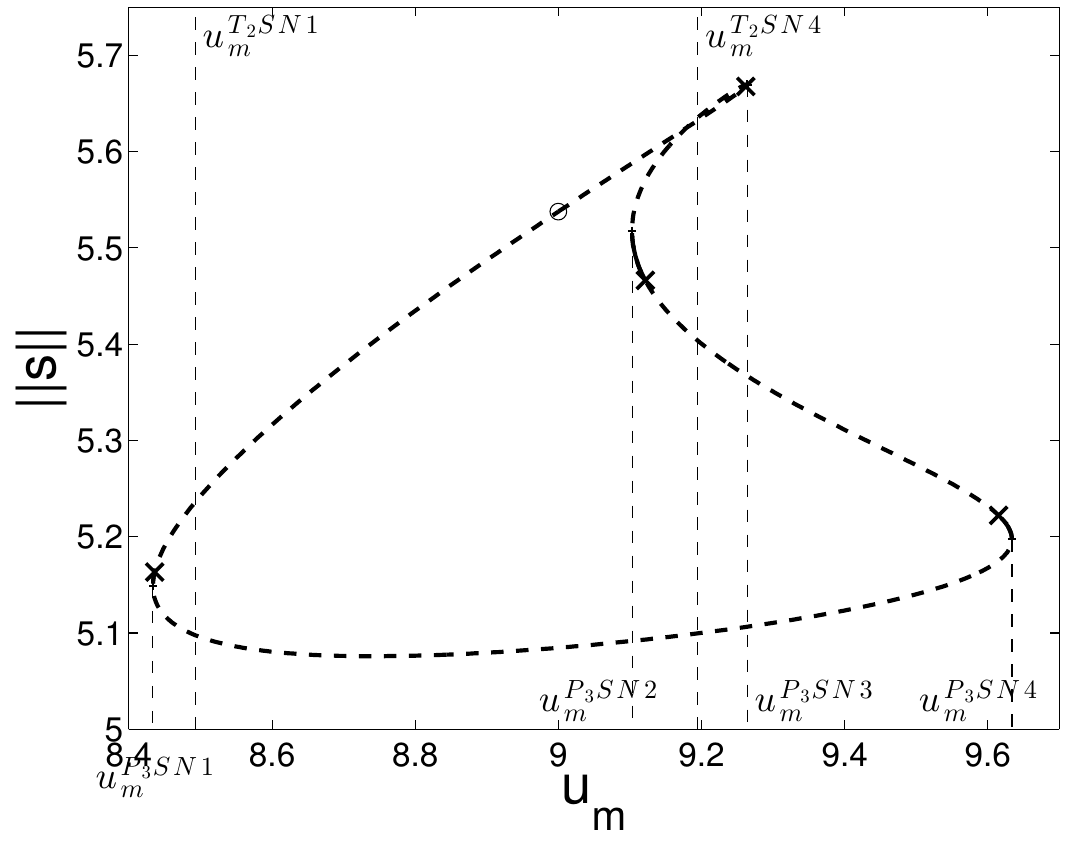}\\
\hfill(a)\hfill(b)\hfill\\
\caption{(a) Branch of transport solutions with $c=\pm1/2$ with $u_m$ as parameter bifurcation, the other parameters being fixed: $\gamma=13.153$, $a=0,65$. The velocity field $u_0$ symmetric. (b) Branch of 3-periodic orbit with  the same bifurcation parameter $u_m$ and the same fixed parameters.\\
Dashed lines indicate unstable solutions in both bifurcation diagrams.}\mylab{fig:c1/2}
\end{figure}
\begin{figure}[tb]
\centering
\includegraphics[width=0.8\hsize]{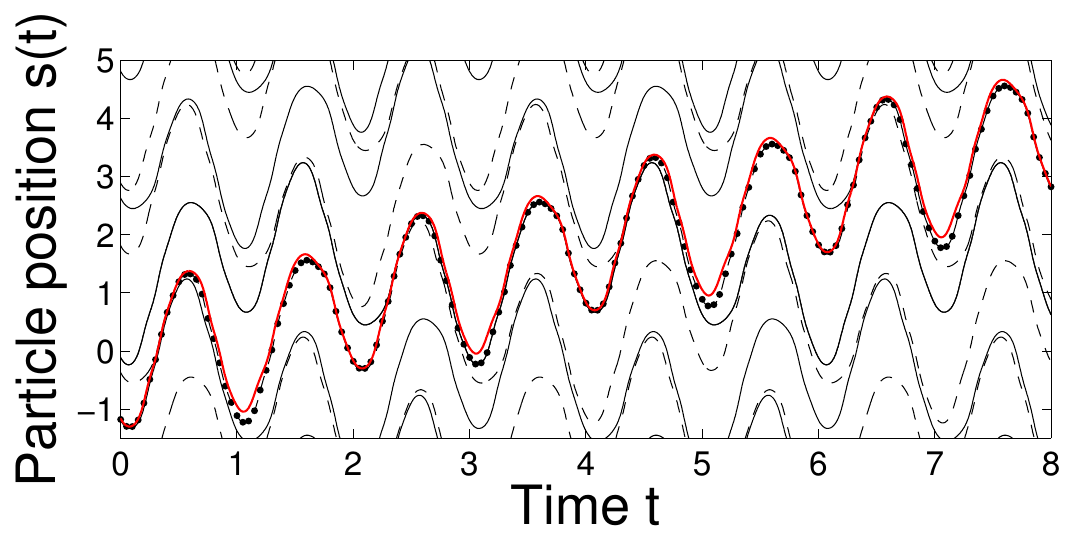}
\caption{Temporal evolution of solutions represented in \myfig{c1/2} by a circle at $u_m=9$. The plain and dashed black lines are 3-periodic solutions translated in space and in time. The trajectory with dashed line $s_{3d}$ is deduced from $s_{3p}$ (plain black line) by: $s_{3d}(t)=s_{3p}(t+2)+1$. The red plain line and the dotted trajectories are the stable and an unstable transport solution, respectively.
}
\mylab{fig:c1/2t}
\end{figure}

%
In order to introduce the transport solutions, one tracks the 1-periodic solutions by expanding the mean velocity $u_m$ at $\gamma=10$ and $a=0.65$ (\myfig{symcu}). The branches $s_0$ and $s_m$ exchange their stability via the asymmetric branch $s_a$ as in the bifurcation diagram \myfig{symgia65cu}. However this last branch $s_a$ loses its stability by period-doubling bifurcations for $u_m$ about $5$ or $10$. One focuses on the period doubling cascade when  $u_m$ is about $10$. The $s_a$ branch loses its stability at $u_m^{2L}=8.17$ where a 2-periodic branch appears (\myfig{symg10cu}). This last branch has 2 saddle-nodes (at $u_m^{sn1}=10.64$ and at $u_m^{sn2}=8.534$) and connects the branch $s_a$ at $u_m^{2R}=11.18$ which becomes stable. The birth at $u_m^{4L}=8.217$ of the 4-periodic branch is very narrow to the 2-periodic  one and connects the 2-periodic branch at $u_m^{4Rsn}$ closer than $5\cdot10^{-7}\%$ to $u_m^{sn1}$ the upper turning point.  Thus in the short range $[ u_m^{4Rsn};u_m^{sn1}]$ (not visible in the \myfig{symg10cu})  the 2-periodic branch is stable. A second 4-periodic branch emerging from the 2-periodic branch at $u_m^{4R}=11.10$ connects the 2-periodic branch at $u_m^{4Lsn}$ close to its second saddle-node $u_m^{sn2}$: $u_m^{4Lsn}$ is about $2\cdot10^{-8}\%$ superior to $u_m^{sn2}$. Thus, again there is a narrow range $[u_m^{sn2};u_m^{4Lsn}]$ of stability of the 2-periodic branch.
It is noteworthy that the turning points of the 2 and 4 periodic branches end roughly at the same value. Indeed this phenomenon occurs also for the 8-periodic solutions (not shown). These branches emerge very close to the 4-periodic branches and therefore, expect for islands of stability close to saddle-nodes,  a window of instability of periodic solutions occurs for approximatively $8.3< u_m<10.7$. The upper limit of this range is due to the branch $s_m$ which is stable after a turning point (see \myfig{symcu}).
In this region we are able to observe quasi-periodic motion as shown in the previous section.\\
However, the simulation shows a regular and stable transport solution at $u_m=9$ too (\myfig{symtw}). After one time period the particle moves of one spatial period. Thus the transport solution $s_t(t)$ seems to be the sum of an uniform translation at the velocity 1 plus an 1-periodic solution $s_p(t)$:
\begin{equation}
s_t(t) = s_p(t) + t \mylab{eq:twsum}
\end{equation}
with 
\begin{equation}
\forall t, s_p(t+1) =s_p(t)  \mylab{eq:sper}
\end{equation}
In other terms $s_t$ verifies
\begin{equation}
\forall k \in \mathbb{Z}, \forall t \in \mathbb{R}: s_t(t+k) = s_t(t) + k \mylab{eq:twper}
\end{equation}
This transport solution will be called synchronized transport solution and 1-periodic transport solution too.
These phenomenon is known as phase synchronization or phase locking effect \cite{Mate00,BaSa00,AlMa06,SLK06,SER07}. 
In point of view of transport of particle, it may be interpreted as a modulated Stokes's drift which occurs for a traveling wave potential. In our case, only the periodicity of the potential is required and so it generalizes this transport kind. 
In order to track this solution and its properties, for instance its existence domain and its  stability, one carries out the path-following of $s_p$, i.e.  the transport solution $s_t$ in the comoving frame.  If $s_t$ is solution of the ODE (\ref{eq:ode}) then $s_p$ is  solution of the new ODE:
\begin{eqnarray}
\ddot{s}_p+\gamma \dot{s}_p&=&\gamma\left[u_m(1+aw_0(s_p(t)+t))\sin(2\pi t)-1\right] \mylab{eq:tw} 
\end{eqnarray}
The presence of the constant biais $-1$ on the right side of the \myeq{tw} recalls the tilted force in ratchet problem \cite{Reim02}. This force tends to move the particle to the right what corresponds to periodic solution $s_t$ in the laboratory referential. This problem is analogous of current reversal phenomenon in tilted inertial ratchet (see e.g. \cite{MaAl08}). Here we seek periodic solutions in the comoving frame so one aims at stopping the "natural" transport. \\
The continuation of 1-periodic solution $s_p$ solution of \myeq{tw}   expanded by $u_m$ shows a existence domain in a eight-shape delimited by the range $u_m^{tsn1}=8.65$ and $u_m^{tsn2}=10.62$ corresponding to two saddle-nodes (\myfig{symctw}). Then there is always two different solutions and at most one solution is stable. On the left, the first range of stability starts from the turning point and ends at the period doubling bifurcation at $u_m^{2L}=8.676$. The second range of stability is delimited by two period doubling bifurcations at $u_m^{2M1}=8.97$ and $u_m^{2M2}=9.37$. Finally, the third range is very narrow (not visible in \myfig{symctw})  between $u_m^{2R}=10.61$ and the rightmost saddle-node $u_m^{sn2}$. In these stability regions the transport solution is stable. Note, that the existence of a transport solution $s_{t+}$ in the direction to $x$ positive implies the existence a transport solution $s_{t-}$ in the negative direction given by:
\begin{eqnarray}
s_{t-}(t)&=&{\cal S}_0(s_{t+}(t))=-s_{t+}(t+1/2)\\
s_{t-}(t)&=&-s_p(t+1/2)-1/2-t \mylab{eq:tw-sum}
\end{eqnarray}
If for an initial condition $(x_i,v_i,t_i)$ the dynamics tends to the $s_{t+}$ solution then for the initial condition $(-x_i,-v_i,t_i+1/2)$ the dynamics tends to the $s_{t-}$. Therefore, there is four 1-periodic transport solutions as mentioned in \cite{SER07}. In the three stability ranges, there is two stable transport solutions in opposite directions  that attract all the dynamics.\\
Let us define the transport velocity $c$ by
\begin{equation}
c=\lim_{t\rightarrow \infty}\frac{x(t)-x(t_0)}{t-t_0}\mylab{eq:c}
\end{equation}
if this limit exists. Otherwise, we introduce the transport velocity at the time t:
\begin{equation}
c(t)=\frac{x(t)-x(t_0)}{t-t_0}\mylab{eq:c(t)}
\end{equation}
According to the Eqs. (\ref{eq:twsum}) and (\ref{eq:tw-sum}) it is clear that the transport velocity $c$ is locked  to $c=\pm1$ for all solutions of the branch displayed  in \myfig{symctw}.\\
In the instability domains a period doubling cascade occurs for the solution $s_p$  similar to the figures \ref{fig:symu5pd}-a or \ref{fig:symg10cu}. \myfig{symctw} displays the 2-periodic branches which connect the four doubling period bifurcations. An example of the resulting transport in laboratory referential corresponding to the 2-periodic solution $s_p$ is shown in \myfig{symtwpd}. Two periods are required in order to shift exactly the particle of two pores. Thus, the transport velocity remains equal to one. In a more general way, all $T$-periodic solutions of the \myfig{symctw} (in co-moving frame) implies the existence of a transport solution with $|c|=1$.
The period doubling cascade implies the existence of a parameters range where all  periodic orbits $s_p$ are unstable and then the transport solution $s_t$ too. In this region, the dynamics  switch alternatively between transport solution to the right direction and transport to the left. An example in given in \myfig{symtwt} for $u_m=8.8$ with two different initial conditions. During a few tens of periods the dynamics are close to the transport solution with $c=1$ or $-1$ and the change of the direction needs only one or two time periods. We do not observe dynamics close to periodic or quasi-periodic motion (laboratory frame) which are unstable too. The discrete dynamics at every entire time $n$ is similar to a random walk. The velocity transport remains close to zero as shown the simulation for $u_m=8.8$ with different initial conditions (\myfig{symtwt}). Therefore there is a brutal transient of the dynamics from a transport at $c=\pm1$ (stable periodic $s_p$ solution) to a zero mean drift $c=0$ (unstable bound solutions $s_p$). Thus the discontinuity of the drift velocity is related with a period doubling and it is known as a {\em chaotic transition} or a {\em crisis} \cite{Mate00, SER07}.
\\
The existence domain of synchronized transport solution is represented in the plane $(u_m,\gamma)$ in \myfig{twrange}. The stable region of transport splits into many branches for $\gamma$ than $8$ (not shown) and it is very reminiscent of the figures  in \cite{SER07}. Indeed in this latter paper, the figures result from a time integration so that only transport solution having attractive manifold appear. It is the advantage of the continuation approach which allows displaying all branches even unstable.
The transport solution does not exist for $\gamma$ superior to $13$ which is still a small value for the experiment described in the introduction. Beyond this existence domain there is  transport events resulting in a complex dynamics. Because of the parity symmetry the mean transport is zero so that we study this dynamics for an asymmetric geometry.\\


The bifurcation diagram \myfig{symctw} corroborates that synchronized transport is "born out of tangent bifurcation" \cite{SER07}. Usually, the end of periodic transport is interpreted in literature as a way of crisis \cite{GOY82,Mate00,SER07}. The continuation of the transport branch shows even this crisis that  the  solution still exists but it is unstable. Then, the synchronized transport emerges and ends at a tangent bifurcation. This bifurcation being generic, it is surprising that such a periodic transport emerges from bound chaotic dynamics such shown in \myfig{symu5quasiper}. In fact, the 2-periodic branch plays a key role in the emergence of the synchronized transport solution.  Let us represent in \myfig{gluing} the 2-periodic solution at the emergence of the transport solution $u_m=u_m^{tsn2}$ (the right turning point of the \myfig{symtw}). Among the three 2-periodic orbits, one takes the orbit presenting the largest shift between both maxima. This $s_2^{(0)}$ solution is closed from the turning point and belongs the branch between both turning points $(u_m^{sn1},u_m^{sn2})$ of \myfig{symtw}.
In the same graph, we represent the $s_2^{(1)}$ solution obtained from  $s_1^{(1)}$ by a space and a time period shifts, i.e. $s_2^{(1)}(t)=s_0(t+1)+1$. This solution exists due to the discrete spatial and temporal translations symmetries. \myfig{gluing} points out that trajectory of two consecutive orbits $s_2^{(n)}(t)$ and $s_2^{(n+1)}(t)$ are close around the time $t=0.25+2k,k\in\mathbb{Z}.$ At this time the velocity (the slope) is maximal and the particle position is roughly in the middle. Because the slope of both curves are almost equal, both trajectories in the 3-dimensional phase space $(x,v,t)$  are close together.
If the 2-periodic solutions is not stable, the dynamics in the vicinity of  $s_2^{(0)}$ may switch  to the $s_2^{(1)}$ at $t$ about $0.25.$ The representation of the transport solution corroborates this fact. 
The synchronized transport solution $s_t$ is close to $s_2^{(0)}$  for $t$ between $[0;0.25]$. At $t$ about $0.25$, the values of $s_t$ are close to both curves $s_2^{(0)}$ and $s_2^{(1)}$. Then, $s_t$ moves slowly away from $s_2^{(0)}$ and brings closer to $s_2^{(1)}$. The trajectories of periodic orbits diverge after passing the narrow region of pore at $x=0$. The $s_t$  solution follows the  $s_2^{(1)}$ orbit and stays close to $s_2^{(1)}$ till $t\simeq1+0.25$.  Because of the spatio-temporal symmetry a similar scenario allows the trajectory to switch to $s_2^{(2)}$.
The "bridge" between two consecutive 2-periodic solutions being spaced of length one, thus one obtains a transport velocity of exactly one unit.\\
Such a key role of the array of 2-periodic orbits in the transport solution existence is observed  for  $u_m^{tsn1}\leq u_m\leq u_m^{tsn2}$. The transport solution is close to the 2-periodic solution which is along the branch joining both turning points $u_m^{sn1}$ and $u_m^{sn2}$ in bifurcation diagram \myfig{symg10cu}. Moreover both turning points of this last branch corresponds roughly to the domain existence boundaries of synchronizes transport solution. This fact may be explained by plotting the solution of the unique 2-period branch when $u_m<u_m^{tsn2}$ or $u_m>u_m^{tsn1}$. The difference between maxima or minima of this 2-periodic orbit is small (\myfig{noglu}) contrary to the solution displayed in \myfig{gluing}.
Indeed this difference has to approach one in order to obtain a neighborhood between consecutive orbits. Therefore, the existence of a common neighborhood  of two consecutive 2-periodic orbits seems to be the key of the synchronized transport existence.
Thus, the whole bound chaotic attractor is not relevant. 
This remark is in agreement with Barbi and Salerno \cite{BaSa00} who point out  phase locking phenomenon without chaotic transition.\\
In a more general way, any synchronized transport solution has the discrete velocity drift $c_q=1/q, q\in \mathbb{Z}$ may be emerged from of an array of $m$-periodic orbits with $m\geq q+1$. Note that for larger characteristic value of $u_m$ the velocity may be rational $c_{pq}=p/q$ with $p,q\in\mathbb{Z}$. However, a  characteristic velocity much larger than 10 is not in agreement with the Stokes approximation so that one considers only velocity $c_q$ inferior to one. 
Such a stable dynamics for $c=1/2$ was found for $u_m=9$ and $\gamma=13.153$. The continuation of this transport solution by varying the velocity $u_m$ is shown in \myfig{c1/2}a.  The temporal evolution of the stable and an unstable transport solution corroborates the relation $s_t(x,t+2p)=s_t(x+p,t)$ in \myfig{c1/2t}. In the same figure, we represent  3-periodic solutions which are translated of one spatial period and {\bf two} temporal periods, i.e. the solutions $s(x+p,t+2p)$, $p\in\mathbb{Z}$. The unstable transport solution is very close to the periodic solution during two periods and then switch to another orbits. As for $c=1$, two consecutive orbits are close where the velocity is maximal (middle of the pore) and the acceleration is small (linear variation). It is in this latter region that the switching occurs. Note that the stable solution is not very close to the orbit, nevertheless  the switching occurs in the same region pointing out the same mechanism. Moreover, the existence domain of transport solution (\myfig{c1/2}a) and the one of  3-periodic solutions which have a large maxima difference (\myfig{c1/2}b) are related. The existence domain is roughly the range between two saddle nodes of the 3-periodic orbit. Beyond the right saddle-node, both 3-periodic branches display orbits which have  a small difference between maxima resulting in a fairly spaced orbits array as in \myfig{noglu}. That explains that there is no transport solution.\\
Finally, note that the stability domain of the transport solution is quasi-discrete so that one observes this solution for a specific parameter set like a resonance phenomenon. In the domain where the transport solution is unstable an intermittent dynamics may occur.
In this latter case, a direct time integration of the governing does not display clearly a transport solution.
\\
 
In conclusion, the symmetric $s_0$ and $s_m$ 1-periodic solutions exchange their stability via a pitchfork bifurcation. For drag about 10 the asymmetric solution $s_a$ presents a cascade of period doubling leading to chaotic but almost bound dynamics. In this parameter domain a synchronized transport known as a phase locked transport  is found.  The velocity of the transport is locked to plus or minus one.  This synchronized transport may be stable and becomes unstable via periodic doubling. The transport dynamics then switch from $c=+1$ to $c=-1$. Tangent bifurcations bound its domain of existence.
This tangent bifurcation occurs in a chaotic region  however the emergence of transport solution is favored simply by  the array of 2-periodic orbits obtaining by the spatial and temporal symmetry translations. Two consecutive orbits having locally a common vicinity.  This vicinity happens when there is a difference between extrema of the periodic orbit is about one. Therefore such transport solution is characteristic of period doubling phenomenon in the presence of a spatial and temporal periodicities. The asymmetry of the transport solution is due to a spontaneous symmetry breaking of the $2$-periodic solutions.
Another transport solution with $c=\pm1/2$ was found too. Its existence is related to the 3-periodic orbits in the same way as transport solution with $c=\pm1$ is related to 2-periodic orbits.\\
Beyond and close to the existence onset of synchronized transport an intermittent transport is observed. These latter dynamics are studied in the asymmetric velocity flow for  which a preferential direction over all initial conditions  is possible.
%

\section{Asymmetric pore}%
\mylab{sec:asym}
\begin{figure}[tbp]
\centering
\includegraphics[width=0.45\hsize]{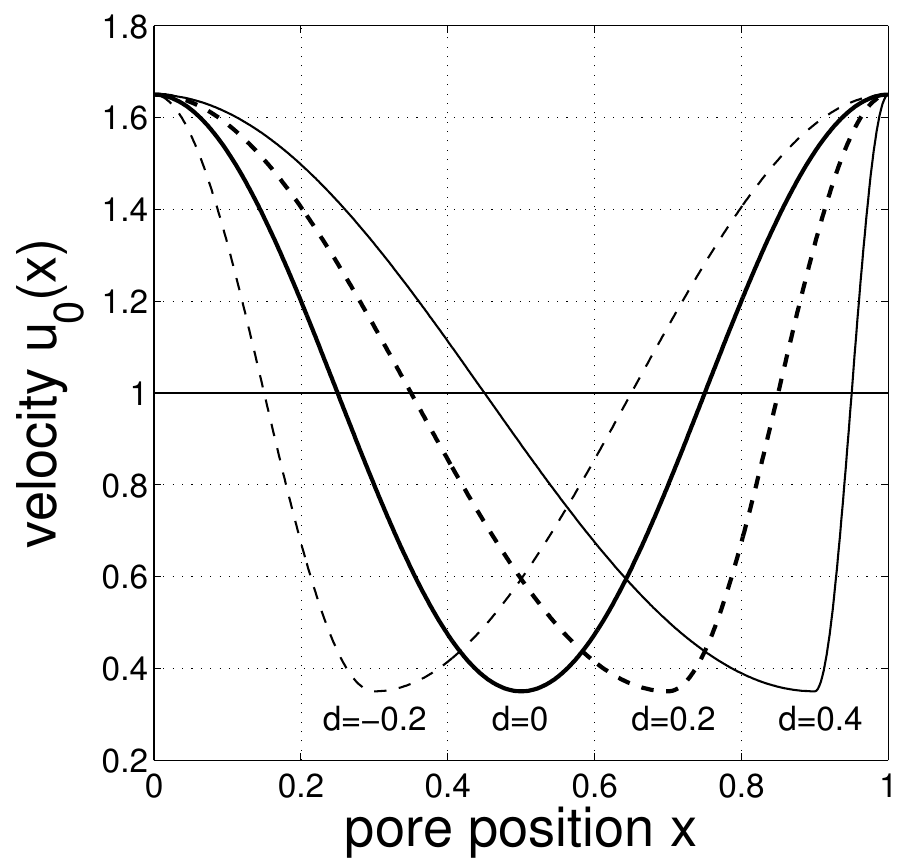}
\includegraphics[width=0.45\hsize]{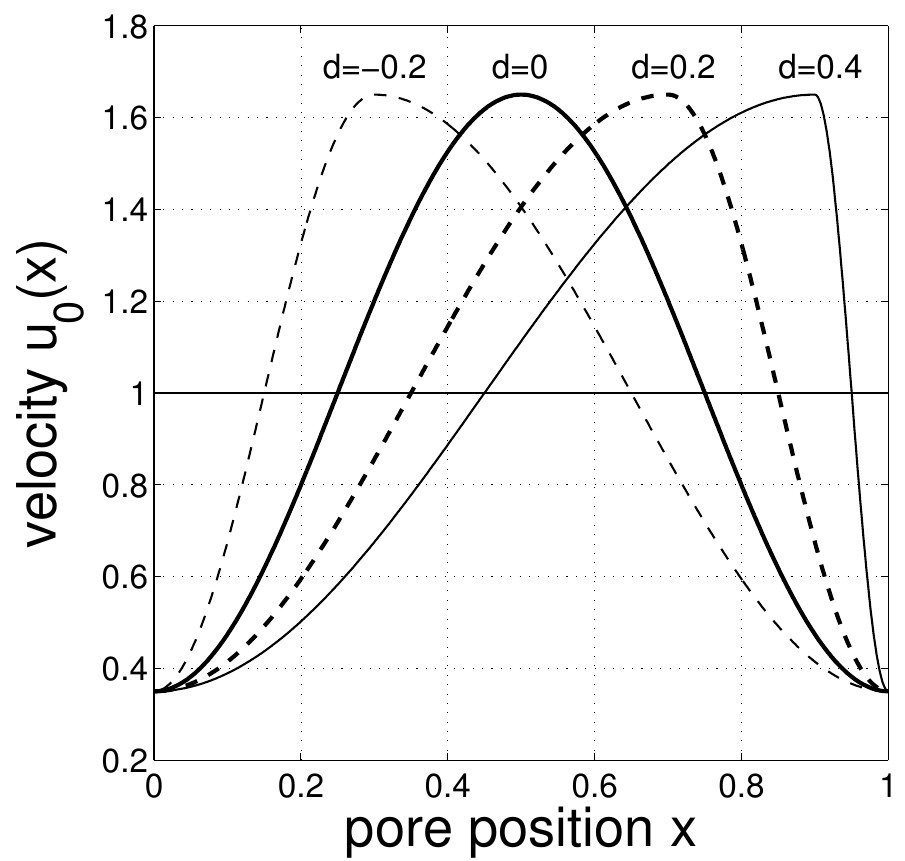}\\
(a)\hfill (b)
\caption{Velocity profiles of the driving flow $u_0(x)$ for $u_m=1$ and $a=+0.65$ (left panel a) and $a=-0.65$ (right panel b). The asymmetry $d$ parameter takes the value $-0.2$, $0$, $0.2$ and $0.4$. }\mylab{fig:velo-asym}
\end{figure}
\subsection{Introduction}
In this section, one breaks the parity-symmetry of the flow, i.e., $u_0(-x)\neq u_0(x)$ and then the existence of a transport solution to one direction does not imply the existence of a symmetrical transport solution in an opposite direction. Thus, an effective transport of particles  is possible: the mean of velocity transport on all initial conditions is non zero. In a first part, we discuss the survey of the synchronized transport. In the second part, we show the existence of intermittent transport. We distinguish a chaotic intermittent transport for $\gamma$ about 10 and quasi-periodic intermittent transport for large $\gamma$.\\
In order to have a dissymmetry parameter $d$,   an asymmetrical flow $u_0$ is constructed by the concatenation of two functions $\cos$ with different scales:
\begin{equation}
u_0(x) = u_m
\left( 1 + a
	\left[ 
	\cos\left(\pi\frac{\bar{x}}{{1/2+d}}\right)I_l(x)+\cos\left(\pi\frac{\bar{x}-1}{1/2-d}\right)(1-I_l(x)) 
	\right] 
\right)
\mylab{eq:voasym}
\end{equation}
with
\begin{eqnarray}
\bar{x}&\equiv& x\mod(1)\\
I_l(x) &=&\left\{\begin{array}{ll}
			1&\mbox{ if } \bar{x}<1/2+d\\
			0&\mbox{ otherwise }
			\end{array}
		\right.
\end{eqnarray}
The parameter $d$ of asymmetry varies in the range $]-1/2,1/2[$. The resulting velocity field is represented on the \myfig{velo-asym}. For $d=0$ the velocity profile is symmetric as in \myeq{u0sym}. When $d$ is positive the minimum of velocity field is shifted to the right: in the region $x\in[0;1/2+d]$  the velocity gradient is smaller  and in the remaining region $[1/2+d;1]$ the velocity gradient is larger. The case $d$ negative is deduced from the case $d$ positive by the transformation $x\rightarrow -x$. Thus, in particular, a drift solution for $d>0$ implies a drift solution in opposite direction for $d<0$. Moreover, contrary to the symmetric case (\myfig{velo-asym}), negative velocity contrast $a$ leads to a new velocity profile. However, one can show that if $s_+(t)$ is solution for a velocity contrast $a$ then $s_-(t)=-s(t+1/2)+1/2+d$ is solution too for the opposition velocity contrast $-a$. For instance, if $s_+$ is a transport solution for $a$, then $s_-$ is a transport direction for $-a$ in opposite direction. Therefore, one will study the influence of the parameter $d$ in the range $[0,1/2[$ and for positive velocity contrast.\\ 
 Note, that the parameter $d$ measures the dissymmetry in the sense that an increasing $d$ implies a stronger contrast of the velocity gradients. The parameter $d$ being positive the negative gradient of $u_0$ is smaller in absolute value than the positive gradient.  The advantage of this formulation compared to the usual profile in the literature as $v_0(x)=\sin(2\pi x) + d \sin(4\pi x)$ in \cite{KRHM00} is, on one hand, to avoid two maxima of the velocity field which does not occur in our pore geometry and, on another hand, to control the dissymmetry of the velocity flow by tuning the parameter $d$ without change the other parameters as $u_m$ and $a$.
The flow $u_0(x)$ remains derivable for all $|d|<0.5$ as the Stokes flow requires it. Note, however that for $d=0.4$ the gradient maximum is about 30. This large value implies a large acceleration what is not consistent with the Stokes approximation. Therefore, in the bifurcation diagrams of this section only the values of $d$ inferior to $0.4$ will be relevant for the transport in micropump.%
\begin{figure}[tbp]
\includegraphics[width=0.43\hsize]{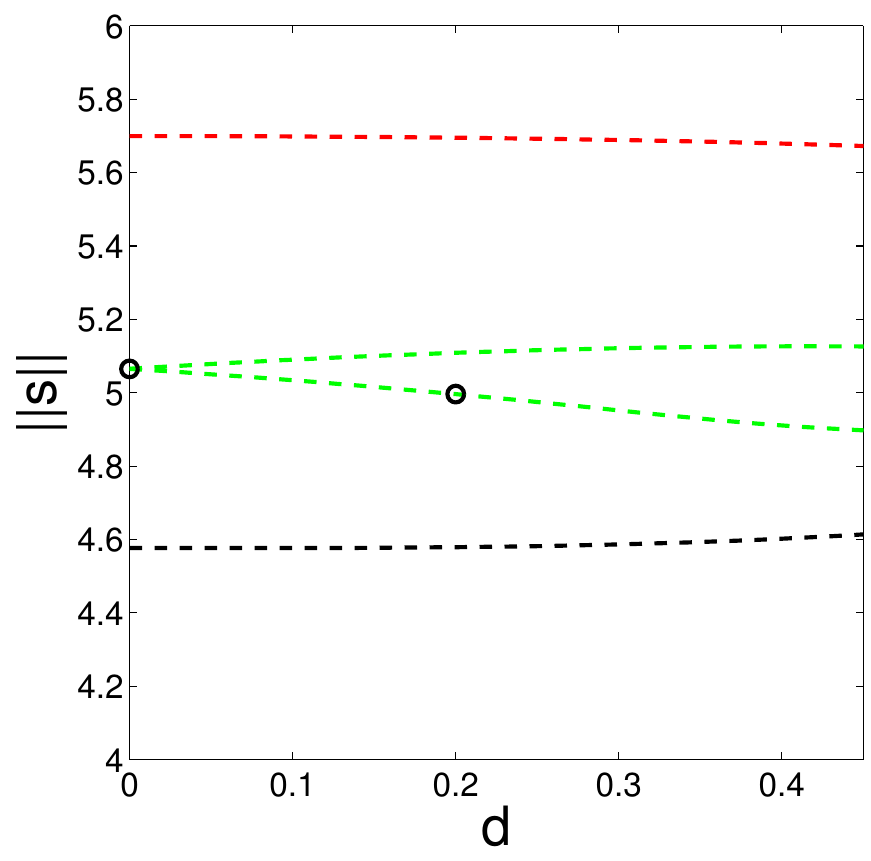}
\includegraphics[width=0.45\hsize]{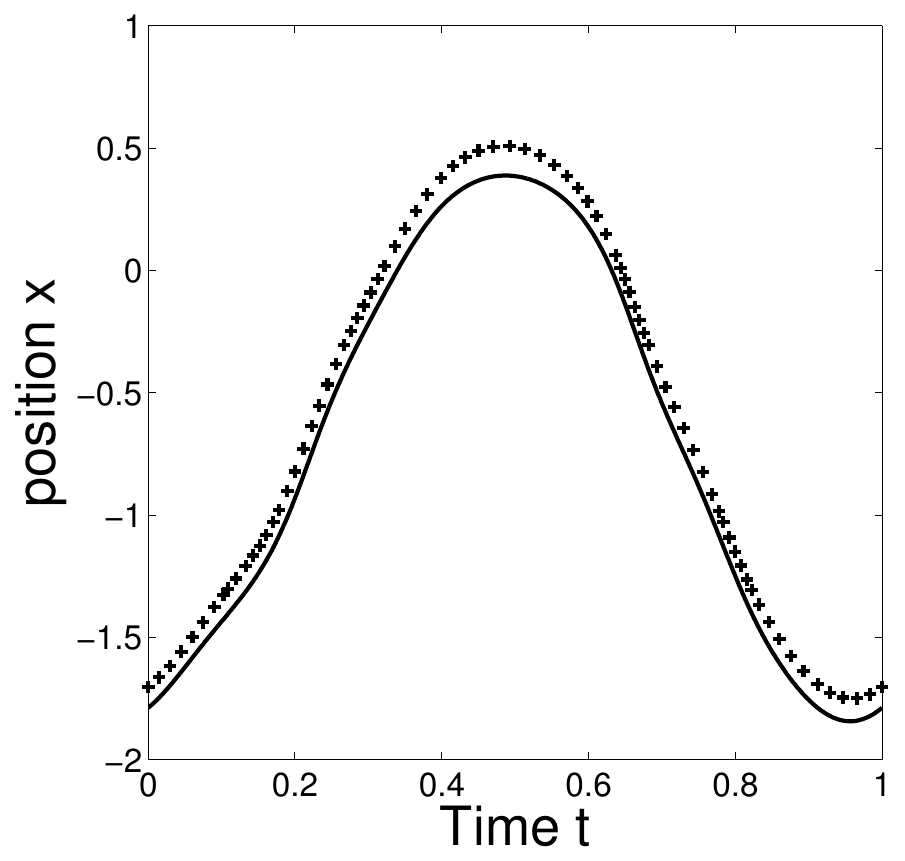}\\
(a)\hfill(b)
\caption{(a) Continuation of the 1-periodic orbits starting in \myfig{symg10cu} at $u_m=9$ and spanned by the asymmetry parameter $d$. Shown are (red) $s_0$, (black) $s_m$ and (green) $s_a$ branches. (b) Time evolution of solution indicated by circles in panel (a) of the branch $s_a$ at $d=0$ and $d=0.2$.}\mylab{fig:perasy}
\end{figure}
\begin{figure}[tbp]
\centering
\includegraphics[width=0.43\hsize]{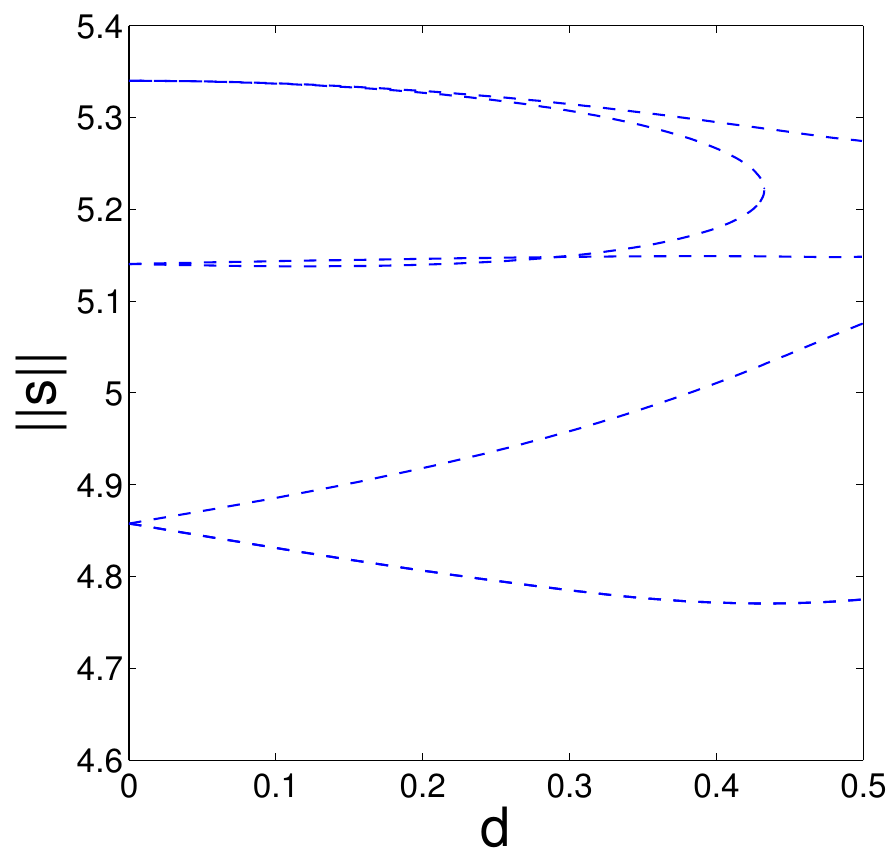}
\caption{(a) Continuation of the 2-periodic solutions starting in \myfig{symg10cu} at $u_m=9$ and spanned by the asymmetry parameter $d$. Each of the three $s_a$ branches splits into two branches when $d\neq0$ leading to 6 branches. Two of them vanish at a turning point.}\mylab{fig:p2cd}
\end{figure}
\subsection{Synchronized transport solutions}
\mylab{sec:amtw}\\
\begin{figure}[tbp]
\includegraphics[width=0.7\hsize]{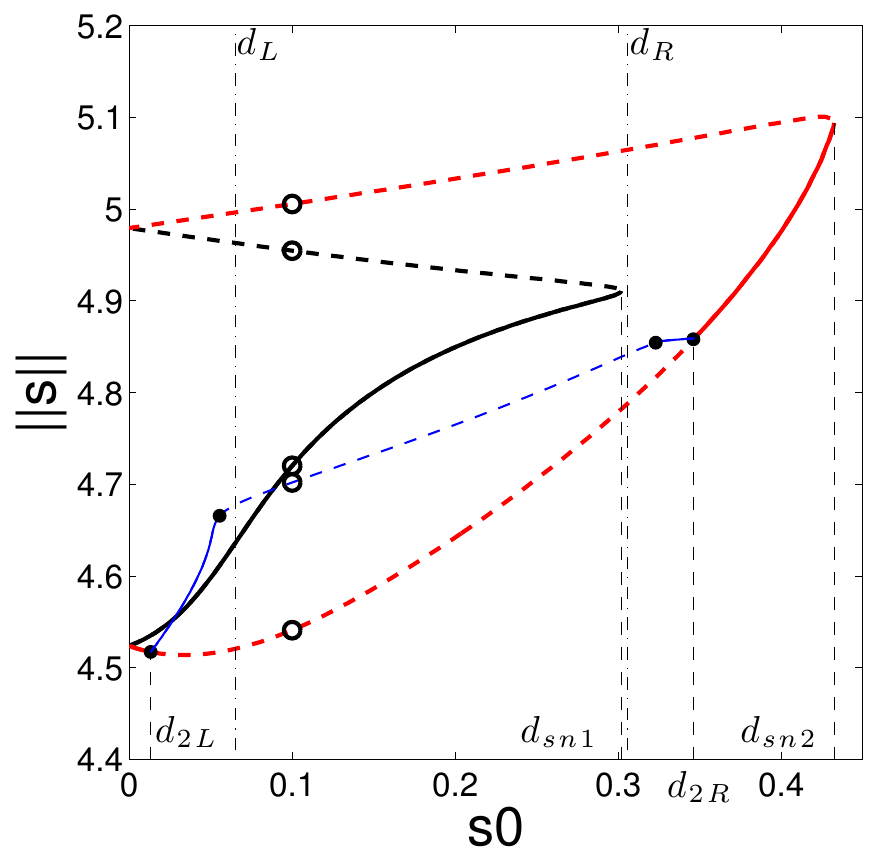}
\caption{Continuation of the synchronized transport solutions starting from \myfig{symtw} at $u_m=9$ ($\gamma=10$, $a=0.65$ and $d=0$) and spanned by $d$. The [black] red branch corresponds to a transport to the [left] right. The blue line displays the 2-periodic transport solution to the right. Solid [dashed] lines indicate stable [unstable] solutions. Dots designate period doubling bifurcations.}\mylab{fig:twu9cd}
\end{figure}

\begin{figure}[tbp]
\centering
\includegraphics[width=0.43\hsize]{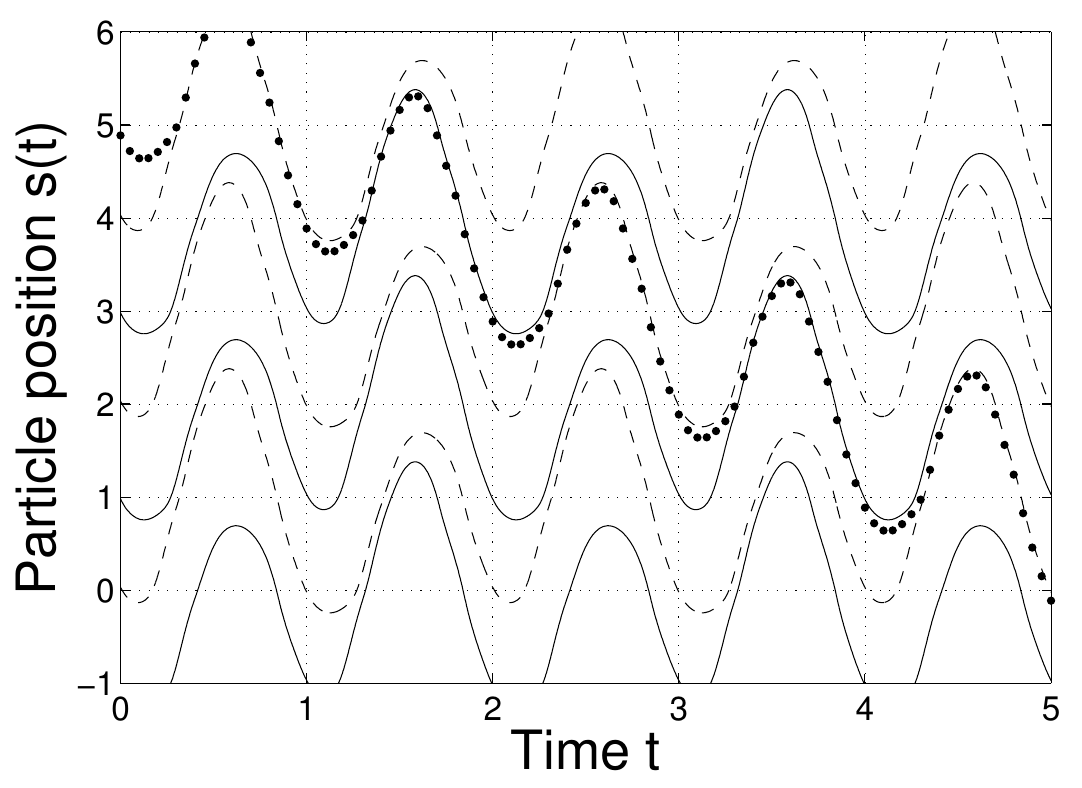}
\includegraphics[width=0.43\hsize]{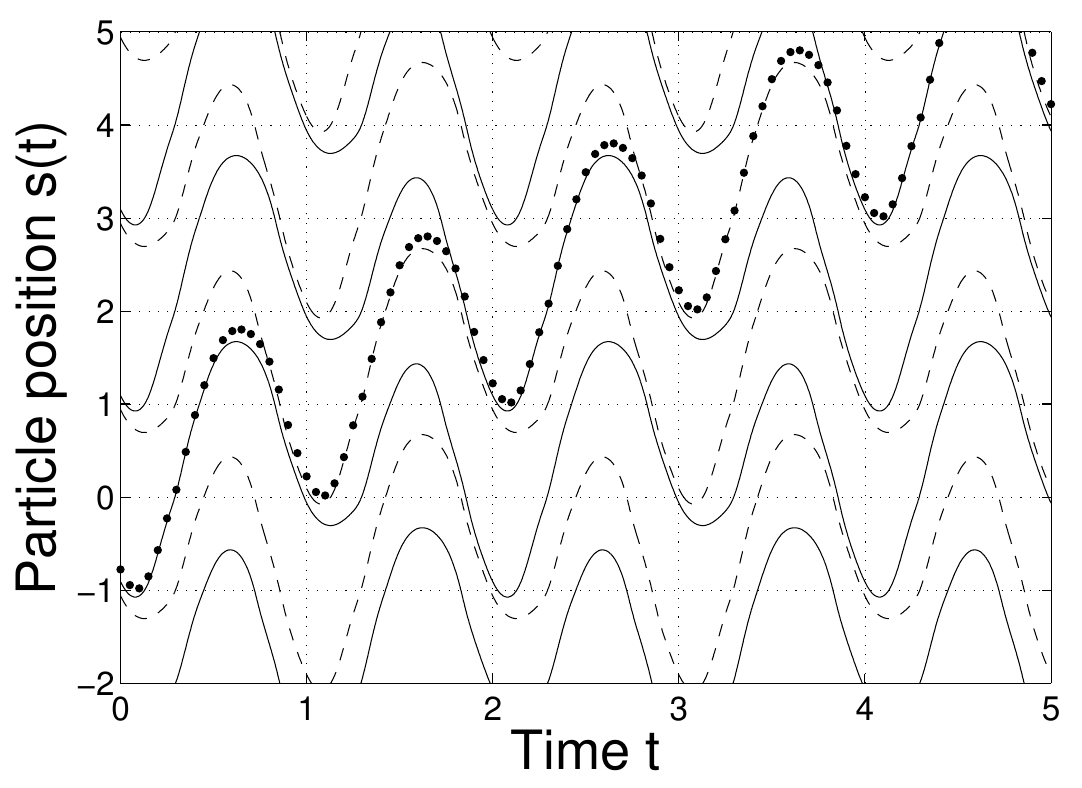}\\
(a) \hfill (b)
\caption{Dotted are time evolutions of transport (a) to the left (b) to the right  close to the onset of their respective existence onset $d_{sn1}$ and $d_{sn2}$. Plain and dashed lines are 2-periodic orbits chosen for the same parameters as the transport solution (circles in  \myfig{p2cd}). The different 2-periodic orbits are shifted one spatial period and one temporal period.}\mylab{fig:gluingd01}
\end{figure}
One studies the influence of the parameter $d$ on the existence of the synchronized transport solutions.
Thus, one focusses on the ranges where all periodic solutions are unstable as in \myfig{symg10cu}.
The asymmetry of the velocity profile has a poor qualitative influence on the periodic solutions.  The bifurcation diagram \myfig{perasy}-a shows that 1-periodic solutions exist for all values of asymmetry and there is no bifurcation, for instance they remain all unstable. Note that when $d=0$ the asymmetric $s_a$ branch contains two solutions deduced from the ${\cal S}_0$ transformation. Then if $d$ is non-zero it splits into two distinct branches. The trajectory of different branches are only slightly modified as shows \myfig{perasy}-b. The bifurcation of 2- and 4-periodic solutions reveals a similar scenario. The  \myfig{p2cd} shows that all 2-periodic solutions are unstable. There is a turning point where two branches end but it occurs for $d>0.4$ which is not relevant in our framework. Then, one may conclude that all periodic solutions remain unstable for relevant values of asymmetry $d$. This conjecture is corroborated by the time-integration of the particle motion \myeq{ode}.\\
One expects that the transport solutions found in the symmetric case remains when $d$ is non-zero. Indeed the transport solutions already break the reflection symmetry ${\cal S}_0$ and thus generically the solutions exists, at least, in a neighborhood of $d=0$. As for the branch $s_a$, the transport solution $s_t$ splits into two solutions $s_{t+}$ and $s_{t-}$ corresponding respectively to the transport to the right (positive $x$) and to the left (negative $x$). One starts from the phase point $u_m=9$, $a=0.65$ and $\gamma=10$ of the symmetric case (circle in \myfig{symctw}) case and according to the \myfig{symctw} there is one stable transport solution and one unstable.
Each transport solution $s_{t+}$ and $s_{t-}$ is defined by respectively Eqs. (\ref{eq:twsum}) and (\ref{eq:tw-sum}) of the symmetric case:
\begin{eqnarray}
s_{t+}(t)&=&s_{p+}(t)+t\\
s_{t-}(t)&=&s_{p-}(t)-t
\end{eqnarray}
where $s_{p+}$ and $s_{p-}$ are 1-periodic functions.
The  $s_{t+}$ and $s_{t-}$ branches are  different when $d$ is non-zero. Then, the continuation of the periodic solutions $s_{p+}$ and $s_{p-}$ requires two different ODEs:
\begin{eqnarray}
\ddot{s}_{p+}&=&\gamma\left(u_m(1+aw_0(s_{p+}(t)+t,d))\sin(2\pi t)-1-\dot{s}_{p+}\right)\\
\ddot{s}_{p-}&=&\gamma\left(u_m(1+aw_0(s_{p-}(t)-t,d))\sin(2\pi t)+1-\dot{s}_{p-}\right)
\end{eqnarray} 
There is four branches starting from both transport solutions of the symmetric case. Stable and unstable transport solutions of each direction meet in a saddle-node bifurcation exchanging their stability (\myfig{twu9cd}). The left transport ceases at $d_{sn1}=0.302$ and the right transport at $d_{sn2}=0.433$ (\myfig{twu9cd}). 
The right transport branch loses its stability in the range $[d_{2L}=0.0133;d_{2R}0.346]$ via two period doubling bifurcations. Again there is a cascade of bifurcations leading to $2^n$-periodic transport solutions. The transport velocity  of these branches remains $c=+1$. Using the time integration, one is found that all these transport to the right become unstable  when $d<d_R\simeq0.306$. Thus $d>d_R$, all dynamics are attracted to a stable transport with $c=+1$.  
In the narrow range $[d_{sn1},d_R]$, the transport to the left is no more stable: the simulation shows that the dynamics switch from left and right transport as a random walk. Because of the asymmetry, generically, the basin of attraction of opposite transport direction are different and then there is mean transport direction after a long time integration. The velocity transport is no more locked and depends continuously on the parameters, here it decreases with $d$ from $1$ to $-1$. The scenario of this transport reversal: transport solution "destroyed" by way of crisis and birth of transport solution at a tangent bifurcation is widely described in deterministic inertia ratchet literature for instance in \cite{SKPK03,SER07}. The continuation diagrams allows to interpret this transport reversal as a consequence of the shift of existence domains of right and left transports. This shift is itself due to the asymmetry of the pore ratchet ($d\neq 0$). That the tangent bifurcation occurs in the chaotic region of the right transport is a coincidence. There is many other scenarios as we show in the next bifurcation diagram \myfig{twu9cg}.
For $d_L<d<d_{sn1}$ only the transport to the left ($c=-1$) is stable and the simulation shows that it attracts all dynamics.
Finally if $d<d_L$, the right transport becomes again stable and both transport solutions, to the left and to the right, are stable. Then the transport direction depends on the initial conditions.  If one considers the initial condition as a random variable, generically, one expects a probability different to go to the left or to the right.\\
The influence of a lattice of shifted 2-periodic orbits is still relevant in the emergence of the transport solution. Among the six 2-periodic branches of continuation diagram in \myfig{p2cd}, we chose the solution for which two consecutive shifted orbits are locally close. The \myfig{gluingd01} shows that the transport is in the vicinity of the 2-periodic orbit during one period and switches to another shifted 2-periodic orbit as explained for the symmetric case. The transport velocity is locked either to $+1$ or $-1$.\\
%
\begin{figure}[tbp]
\includegraphics[width=0.6\hsize]{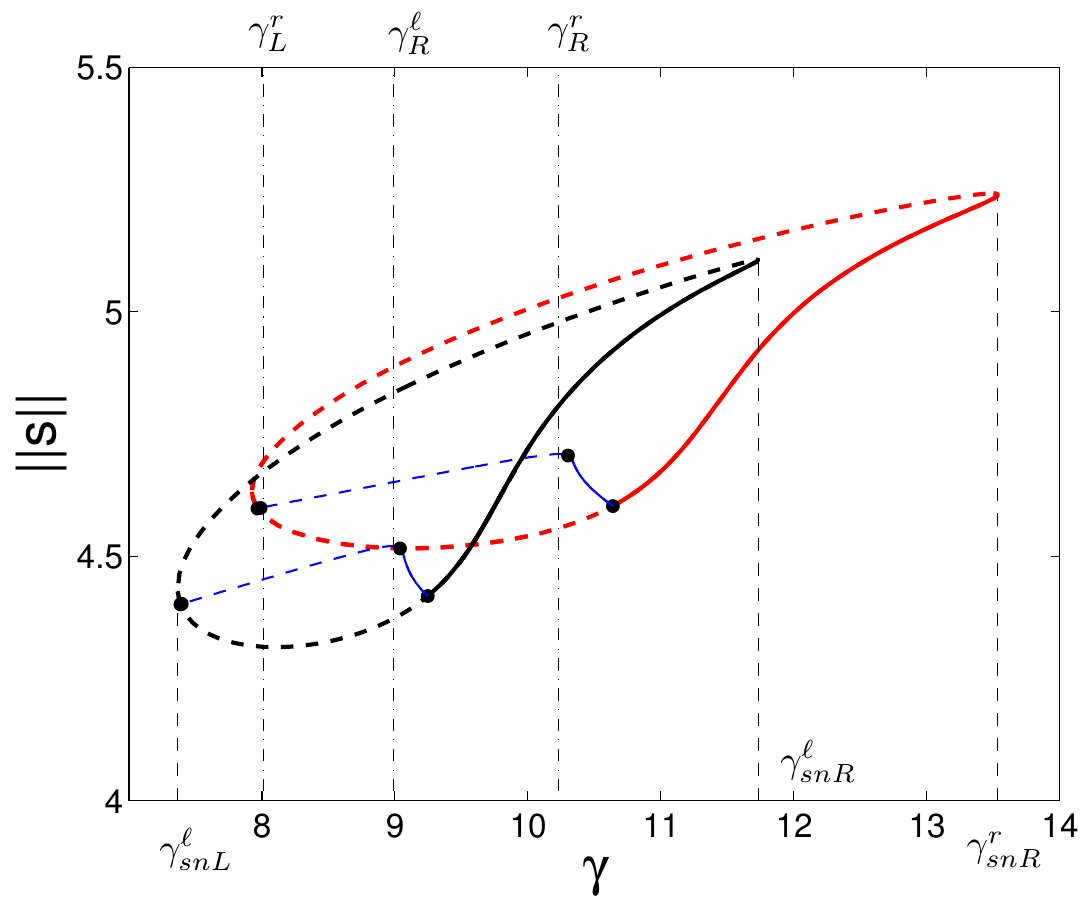}
\caption{Continuation of the left and right synchronized transport solutions starting in \myfig{twu9cd} at $d=0.1$ ($a=0.65$ and $u_m=9$) and spanned by $\gamma$. The color, stability and bifurcation codes as in \myfig{twu9cd}.}\mylab{fig:twu9cg}
\end{figure}
\begin{figure}[tbp]
\includegraphics[width=0.6\hsize]{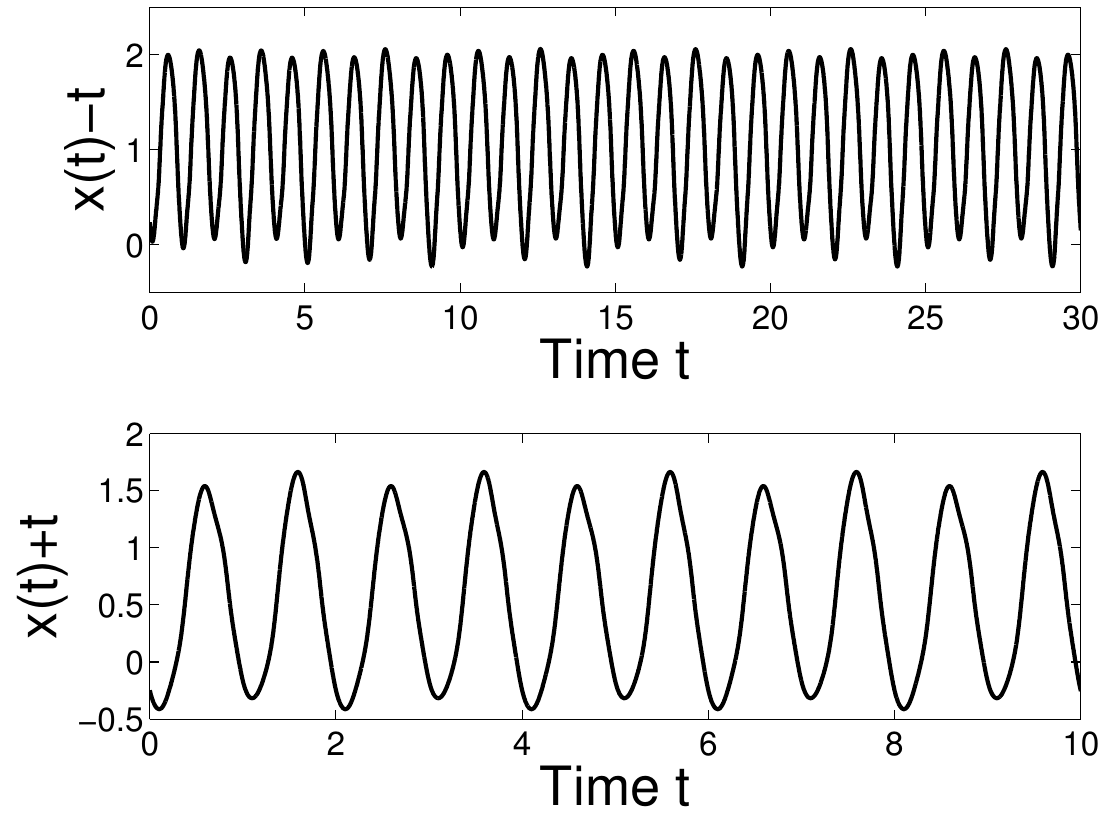}
\caption{Solutions of synchronized solutions of (upper panel) the branch $s_{t+}$ and (lower panel) the branch $s_{t_-}$. Shown is the particle position in the comoving frame [(lower panel) $x(t)=-t$] (upper panel) $x(t)=t$ at [$\gamma=9.1$] $\gamma=10.25$. The  $s_{t+}$ is dynamics is quasi-periodic and $s_{t-}$ is 2-periodic.}\mylab{fig:xtw}
\end{figure}
%
%
\begin{figure}[tb]
\includegraphics[width=0.45\hsize]{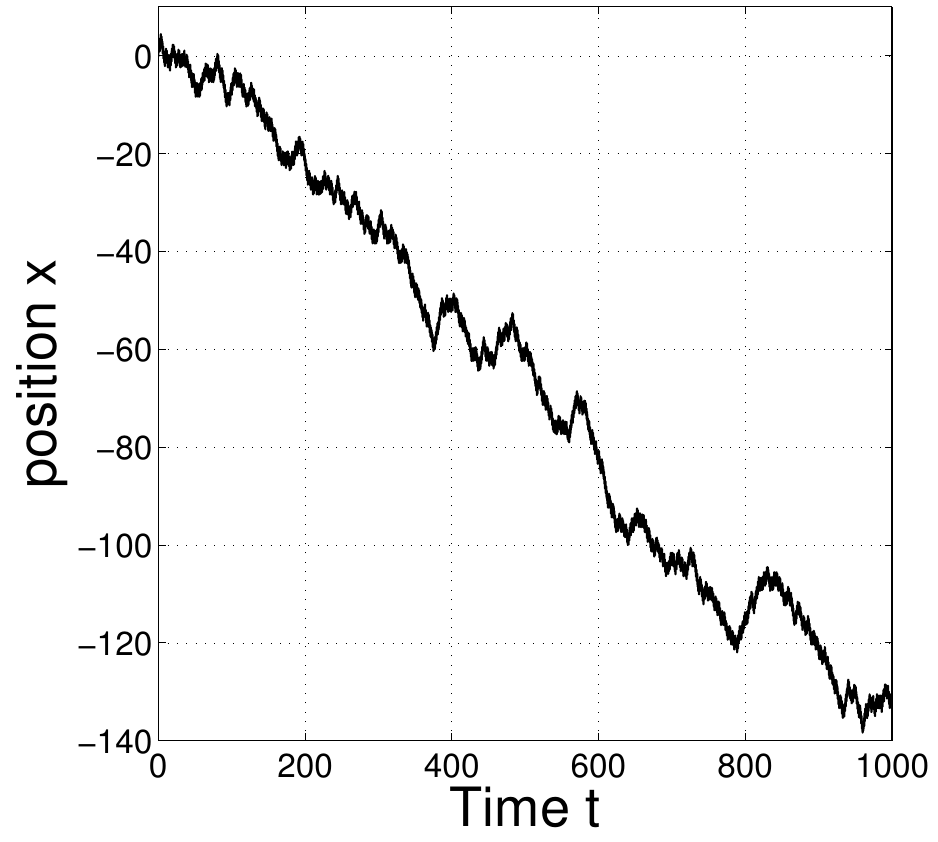}
\includegraphics[width=0.45\hsize]{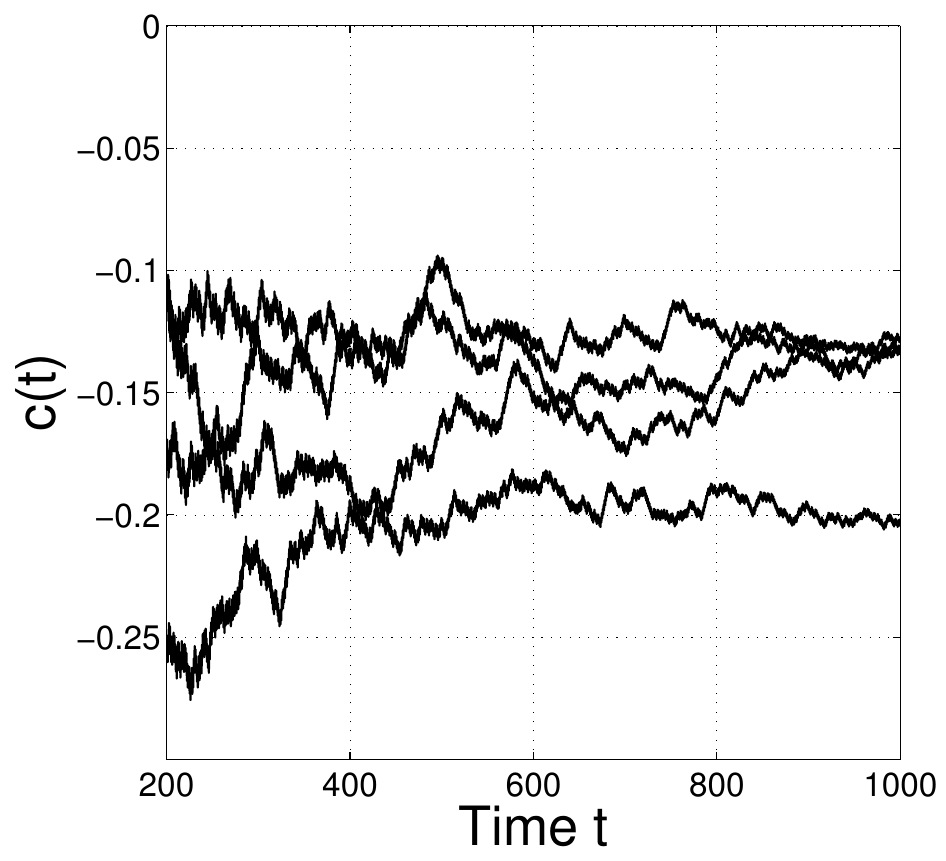}\\
(a)\hfill(b)
\caption{(a) Time evolution of the particle position at $\gamma=8.8$ with the parameters of the bifurcation diagram in \myfig{twu9cg} ( $u_m=9,a=0.65,d=0.1$). (b) Time evolution of the transport velocity $c(t)$ for four different initial values.}\mylab{fig:simutwg8,8}
\end{figure}
Thus  by setting the parameter $d$, i.e. the geometry of the pore, the transport direction may be reversed. However in the experiment a switch of transport direction is observed only by modifying the pumping frequency, is it possible in our model? The parameter $d$ depending only on the pore geometry, it must be fixed. In contrast, a change of the pumping period implies a change of the dimensionless drag $\gamma$. Thus one expands the transport branches by the parameter $\gamma$ and one starts from the transport solution found in \myfig{twu9cd} at $d=0.1$. Each branch of transport solutions expanded by $\gamma$ forms a loop and emerges/ends by a tangent bifurcation (\myfig{twu9cg}) as for the symmetric case.  The left transport exists in the range $[\gamma^\ell_{snL}=7.93;\gamma^\ell_{snR}=13.5349]$ and the right transport for $\gamma\in[\gamma^r_{snL}=7.37;\gamma^r_{snR}=11.72]$. Moreover, for the right transport there is two period doubling bifurcations at $\gamma=7.96$ and $10.64$. Both bifurcations are connected by a 2-periodic transport solution (\myfig{twu9cg}). This branch is unstable via two doubling period bifurcations (\myfig{twu9cg}). Indeed a cascade of period doubling bifurcations occurs. All these branches are transport solutions with the velocity $c=+1$ as explained in the previous paragraph. 
For example, the time integration at $\gamma=10.25$ close to the boundary $\gamma_R^r$ shows a stable transport with $c=+1$. In the comoving frame $x(t)=t$, the dynamics does not seem to be periodic because a cascade of  doubling periodic bifurcations occurred for this parameter value (\myfig{xtw}). There is a narrow range for which this dynamics is stable.
 The time integration of the ODE \myeq{ode} shows that for $\gamma<\gamma_R^r=10.23$ and $\gamma>\gamma_L^r=8.01$ all periodic branches of $2^n$ periods are unstable. This range is slightly shorter that the instability range of the 2-periodic branch (\myfig{twu9cg}).  The left most limit $\gamma_L^r$ is close to the turning point, then there is an island of stability of the right transport in the range $[\gamma_{snL}^r;\gamma_L^r]$. A similar scenario occurs for the transport to the left: the left transport with $c=-1$ is unstable in the range $[\gamma_L^\ell=7.94;\gamma_R^\ell=8.99]$ due to a period doubling cascade. This range is slightly shorter that the range of the 2-periodic solution instability (\myfig{twu9cg}). An example of the transport dynamics is showed in \myfig{xtw} when $\gamma=9.1<\gamma^\ell_R$. The solution in the comoving frame  $x(t)=-t $ is a stable 2-periodic transport solution and attracts all the dynamics. Finally, close to the leftmost saddle node there is an island of stability  of the left transport.\\
The bifurcation diagram \myfig{twu9cg} allows to answer positively of our question  since there is distinct ranges of $\gamma$ values corresponding to stable transport solutions  in different direction. In the range $[\gamma^\ell_{snR};\gamma^r_{snR}]$ only the transport to the right is stable and the time integration shows that the dynamics is attracted by this solution. It is an absolute transport to the right. In the same way, in the range $[\gamma^r_{L};\gamma^\ell_{R}]$ it is an absolute transport to the left.\\
Between these ranges ($[\gamma^r_R;\gamma^\ell_{snR}]$) there is two stable transports in opposite directions. Depending on the initial conditions the dynamics will be attracted by one of these two stable solutions. Because of the dissymmetry of the geometry,  the attraction basin of each solution is different and an effective transport occurs.  If one takes the mean velocity over the initial conditions as in \cite{Mate00} the mean drift velocity varies continuously  from 1 to -1 when $\gamma$ is decreasing from $\gamma^\ell_{snR}$ to $\gamma^r_R$. Such transport reversal was already interpreted as a consequence of bistability  in \cite{BaSa00,SER07}.\\
There is  a competition between left and right transport solutions in the range $[\gamma^r_L;\gamma_R^\ell]$ too. This time both transport solutions are unstable. The numerical simulation shows dynamics which alternate the transport directions (\myfig{simutwg8,8}-a). Moreover, the switch between the directions lasts less than one period, i.e. one does not observe events of quasi-periodic motions. For example, at $\gamma=8.8$, the drift is negative and its velocity is in the range $c=-0.12$ and $-0.2$ (\myfig{simutwg8,8}-b).
In the remaining range $[\gamma^\ell_{snL};\gamma^\ell_{snL}]$ only the transport in the left direction exists but it is unstable. Then, there is a competition between bound attractors as quasi-periodic solutions and the transport solution. \myfig{simutwg7,5}-a displays episodes  of transport close to the synchronized transport solution following by quasi-periodic motion episodes (tens of periods). One does not observe transport to the right. Therefore the velocity transport is negative. For example for $\gamma=7.5$,  the drift velocity $c(t)$ varies between $-0.15 $ to $-0.2$ depending on the initial conditions  (\myfig{simutwg7,5}-b). The mean transport velocity is no more  locked and it is presented in literature as a non phase locked transport \cite{SER07}.
However, our analysis shows that synchronized transports exist and generate this dynamics.\\
\\
To conclude, there is synchronized transport solutions involving 2-periodic orbits as in the symmetric case. The existence domain of transport solution is slightly modify for a non-zero dissymmetry $d$. However, the transport may exist in only one direction and may attract the whole dynamic ensuring an absolute transport to this direction. As for the experiment it is possible to change the transport direction only by varying the frequency of the pumping and not the pore lattice geometry.
\begin{figure}[tb]
\includegraphics[width=0.45\hsize]{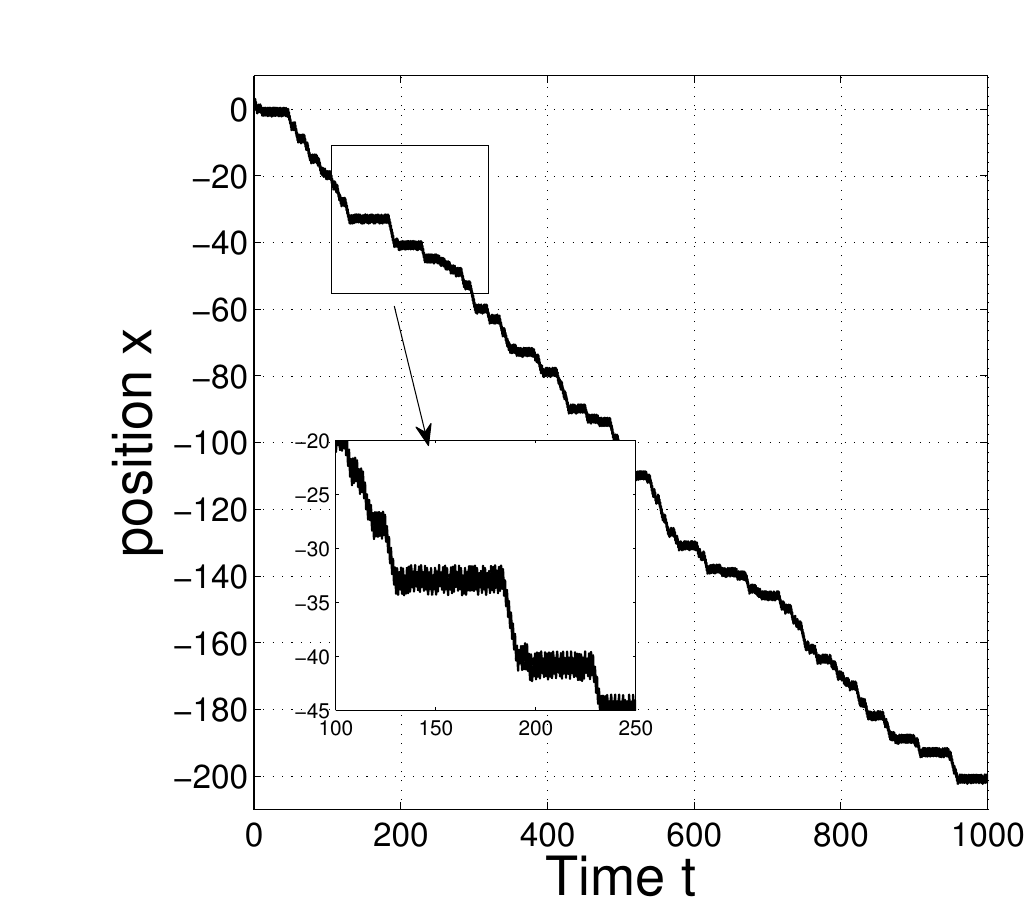}
\includegraphics[width=0.45\hsize]{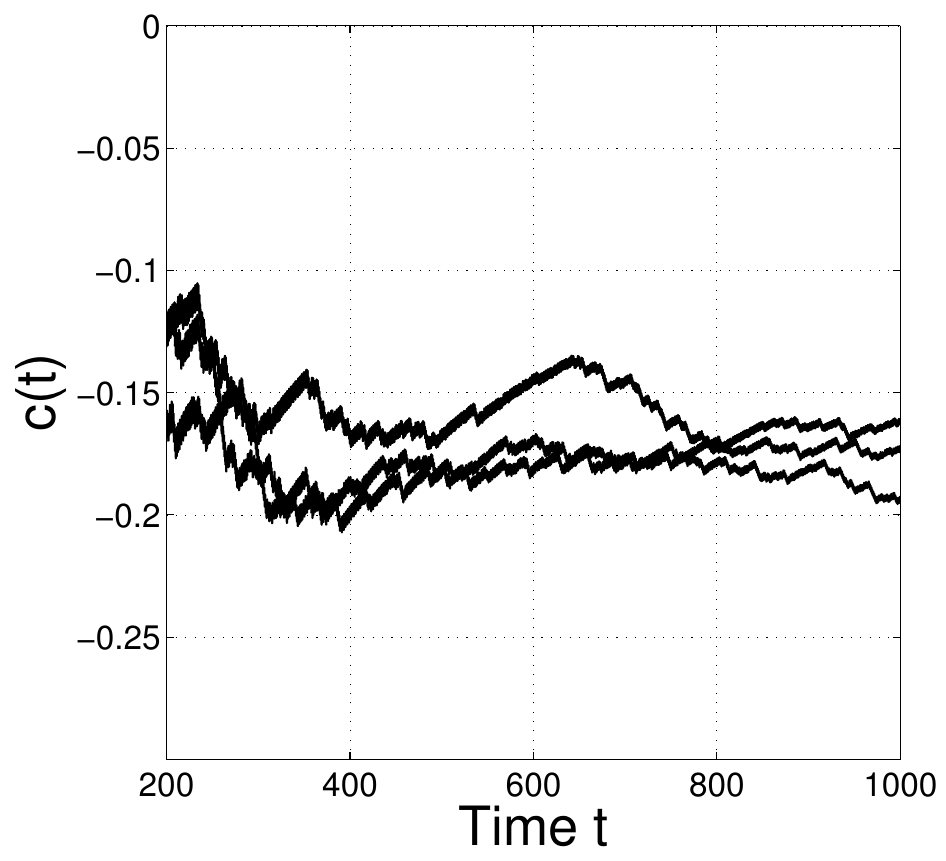}\\
(a)\hfill(b)
\caption{(a) Time evolution of the particle position at $\gamma=7.5$ with the parameters of the bifurcation diagram \myfig{twu9cg} ( $u_m=9,a=0.65,d=0.1$). We observe long periods of oscillations in a pore following by transport episodes leading to a slow drift to the left. (b) Time evolution of the transport velocity $c(t)$ for four different initial values.}\mylab{fig:simutwg7,5}
\end{figure}
\subsection{Chaotic intermittent transport ($\gamma$ about 10)}
\begin{figure}[tb]
\centering
\includegraphics[width=0.65\hsize]{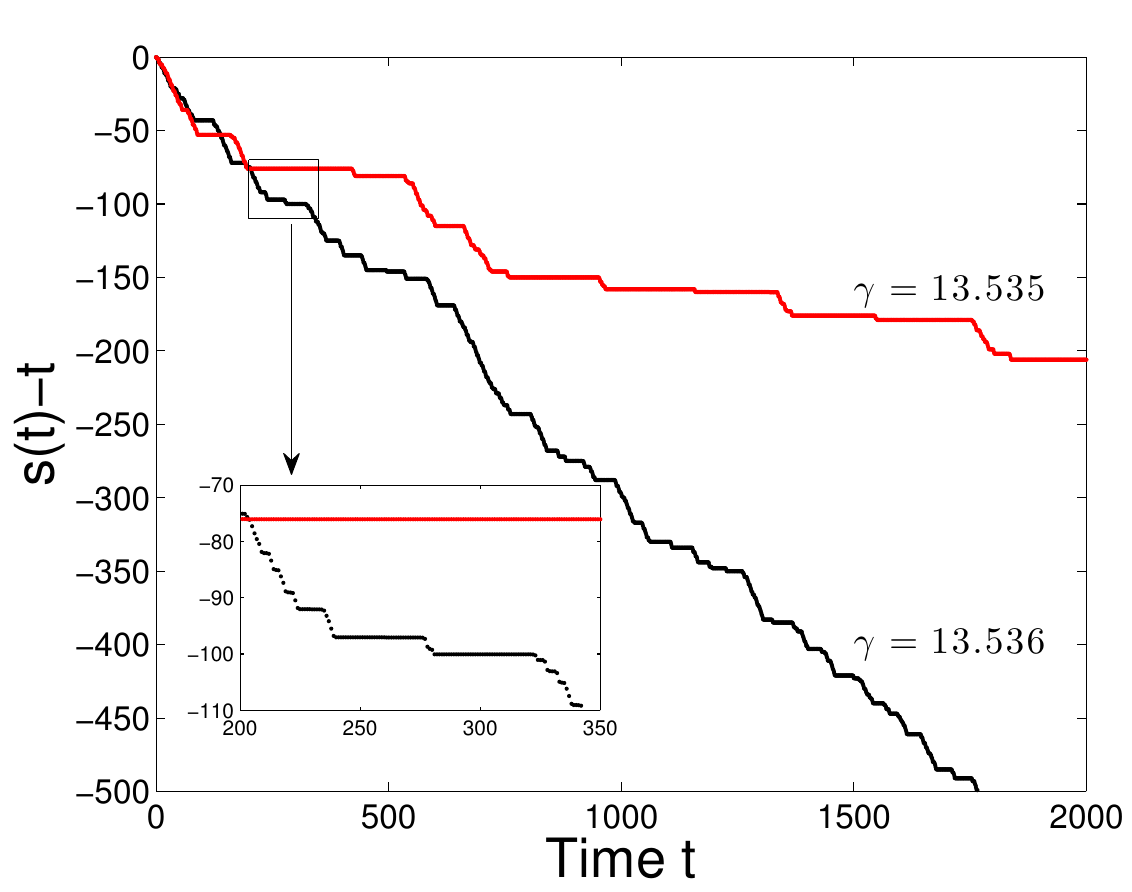}
\caption{Particle position at entire times $n$ in the comoving frame ($c=+1$) for two different values of $\gamma$ close to the tangent bifurcation of synchronized transport. The dynamics displays an irregular staircase for which the length increases and diverges when $\gamma$ approching the tangent bifurcation.}\mylab{fig:intercomov}
\end{figure}
\begin{figure}[tb]
\includegraphics[width=0.45\hsize]{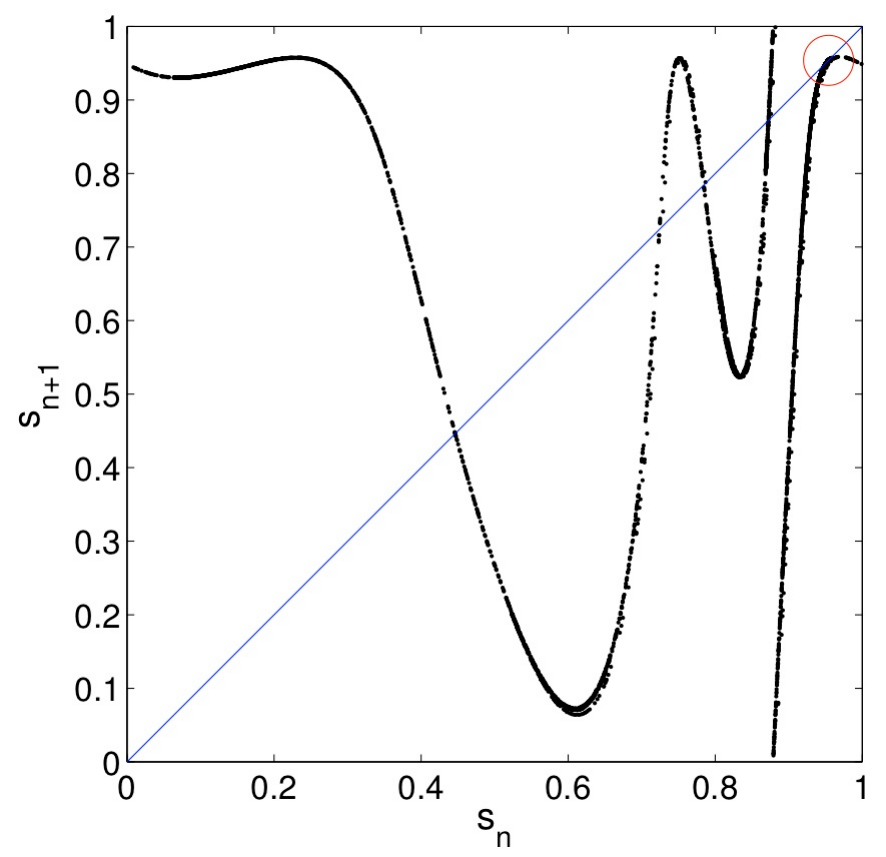}
\includegraphics[width=0.45\hsize]{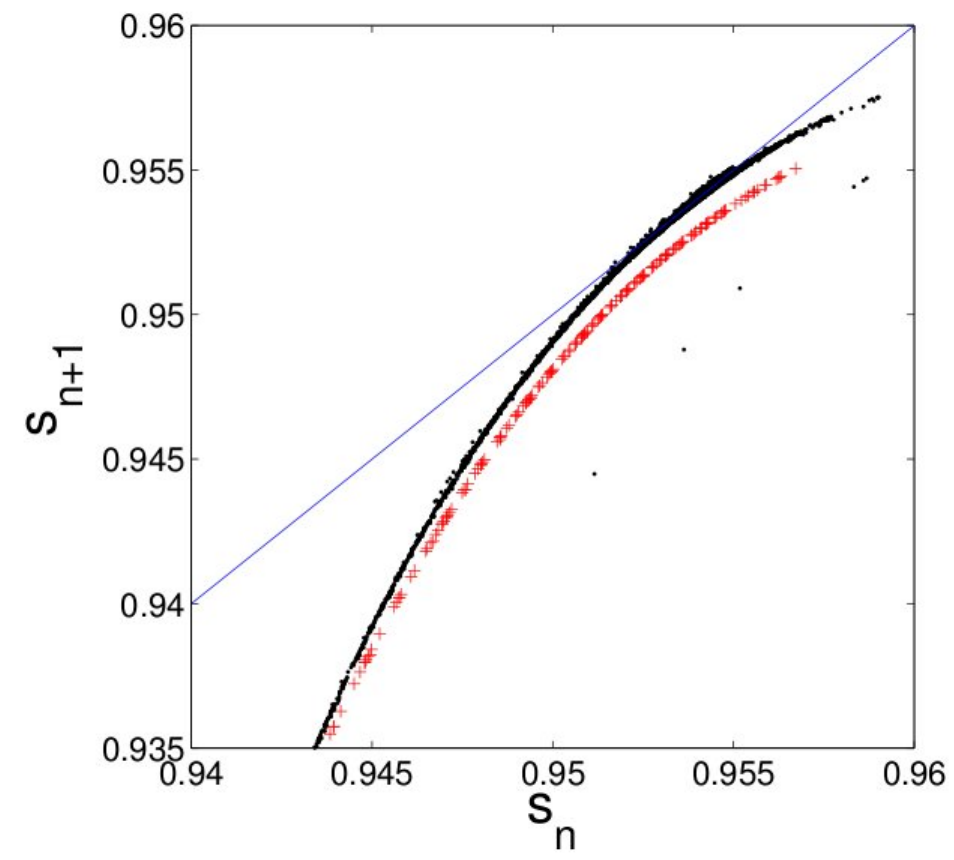}\\
(a)\hfill (b)
\caption{(a) Poincar\'e return map of the particle position $s_n$ at entire times $n$ of the dynamics close to the tangent bifurcation $\gamma^r_{snR}$ of the right transport (bifurcation diagram \myfig{twu9cg}). (b) Magnification of the dynamics close to the tangent bifurcation (circle in (a) panel) for two values of $\gamma$: $\gamma=13.535$ (red '+' ) and $\gamma=13.53499$ (black dots).}\mylab{fig:typ1bifcomov}
\end{figure}

\begin{figure}[tbp]
\includegraphics[width=0.5\hsize]{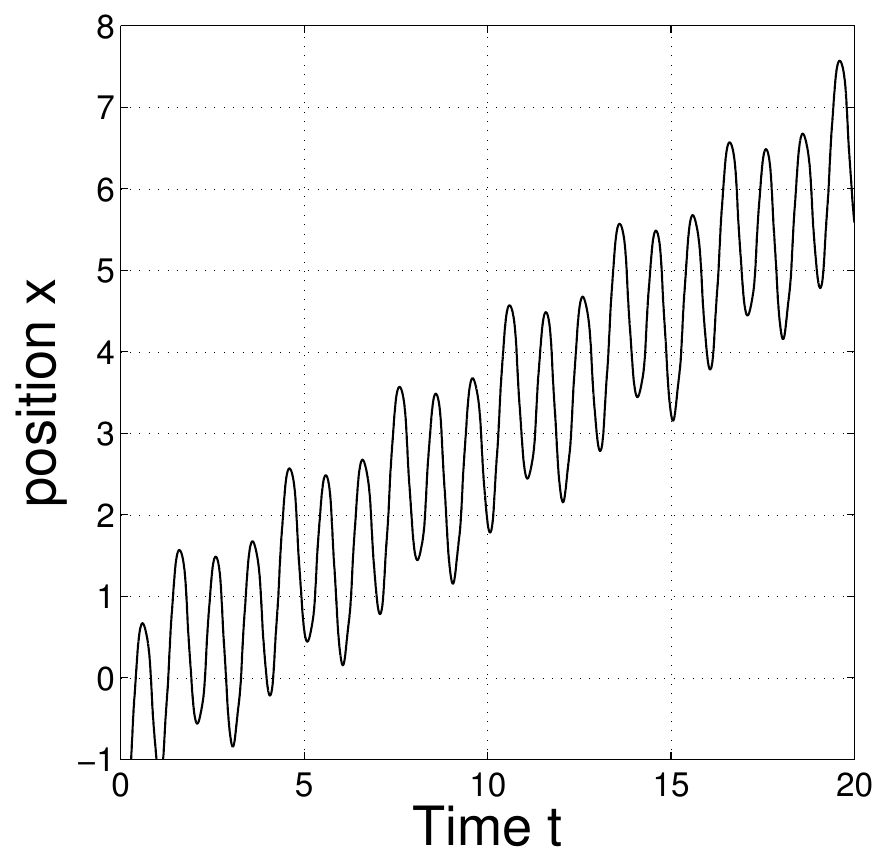}
\caption{Synchronized transport solution with  $c=1/3$ drift velocity obtained at $\gamma=14.2$ with the parameters of the bifurcation diagram \myfig{twu9cg} ( $u_m=9,a=0.65,d=0.1$). After three temporal periods the particles moves exactly of one pore length.}\mylab{fig:twdc13}
\end{figure}
\begin{figure}[tb]
\centering
\includegraphics[width=0.45\hsize]{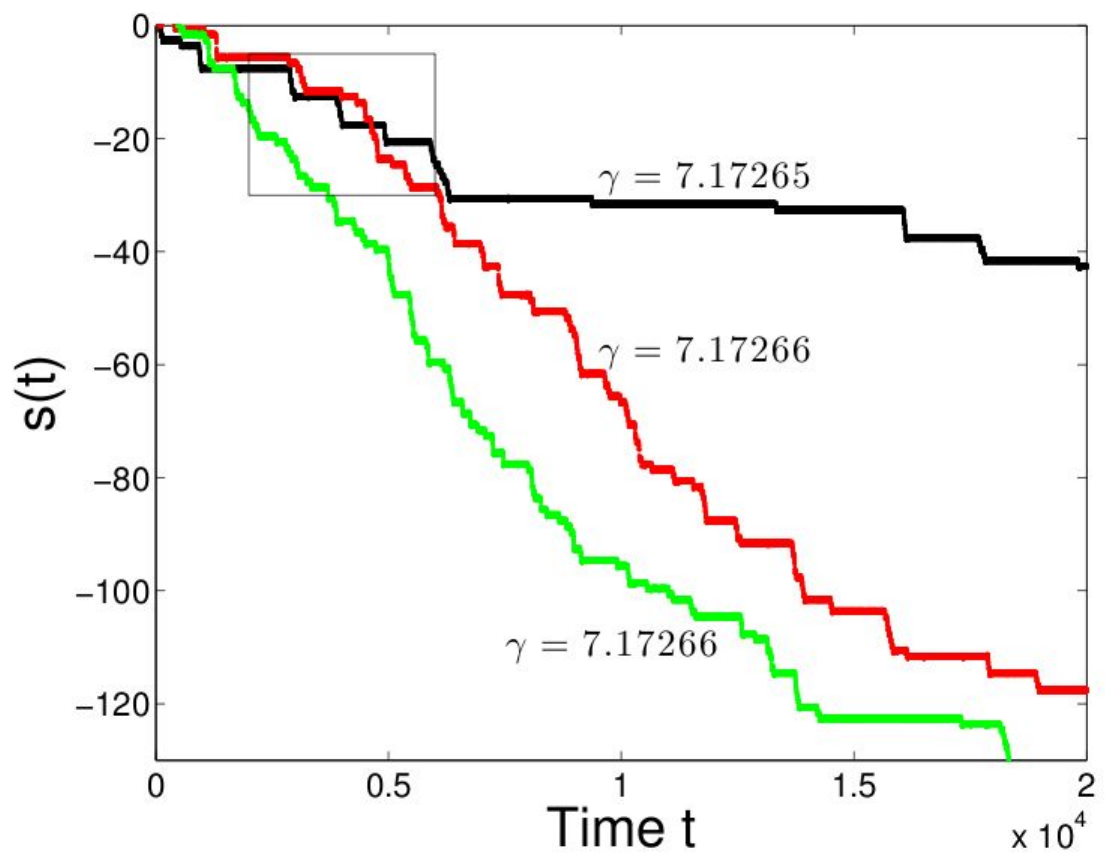}
\includegraphics[width=0.45\hsize]{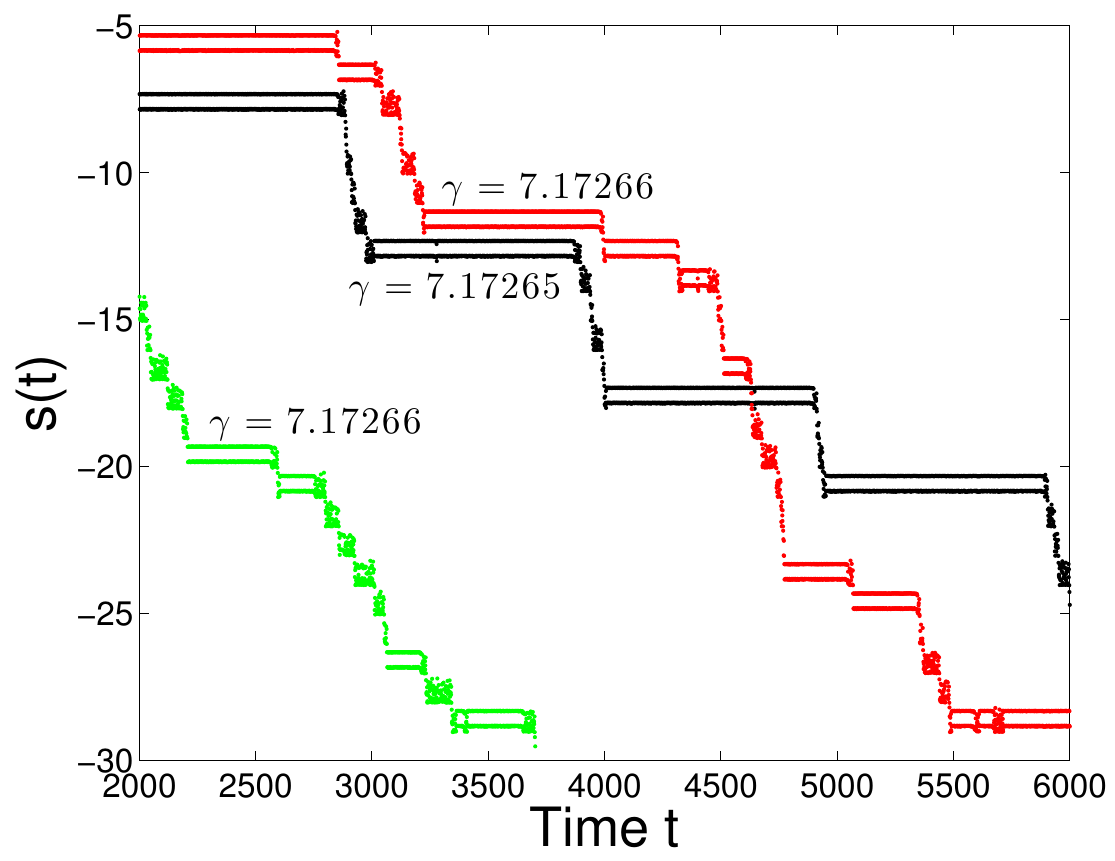}
\caption{Particle position at entire times $n$ close to the onset of the intermittent drift ($\gamma\simeq7.1726$). Other parameters are as in bifurcation diagram \myfig{twu9cg} $u_m=9,a=0.65,d=0.1$. The drift velocity vanishes rapidly approaching the onset (\myfig{twu9cg}).}\mylab{fig:twinterchaotic}
\end{figure}
\begin{figure}[tb]
\includegraphics[width=0.45\hsize]{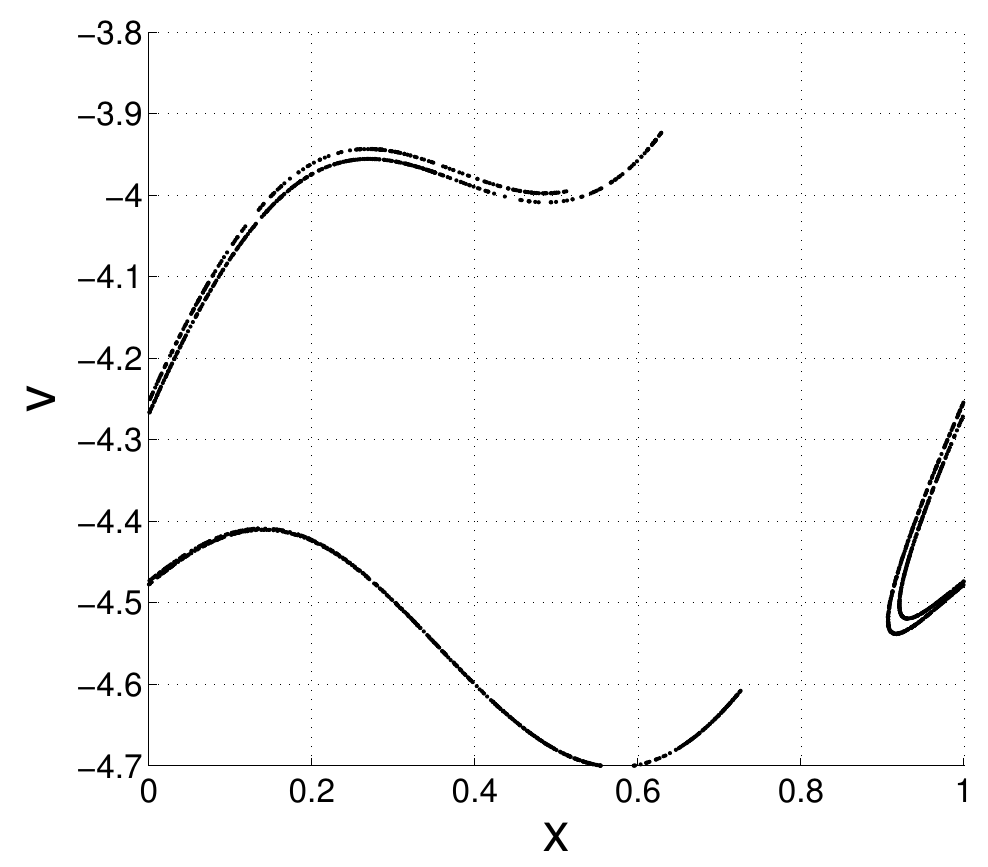}
\includegraphics[width=0.45\hsize]{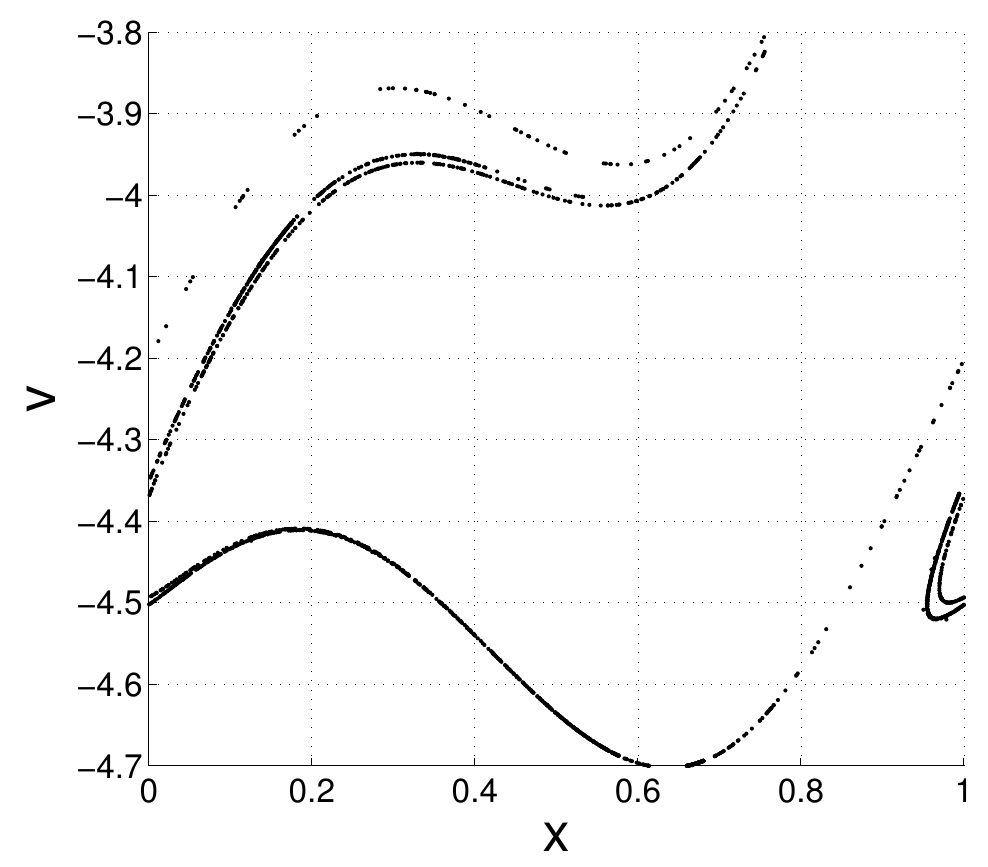}\\
(a)\hfill (b)
\caption{(a) Poincar\'e $(x,v)$ map  of the strange attractor before the drift  onset ($\gamma=7.172$).  The $x$ axis is  the congruence modulo 1 of the particle position $x$. Remaining parameters are the same as in \myfig{twu9cg}. (b) Poincar\'e $(x,v)$ map  of the intermittent drift shown in \myfig{twinterchaotic} after the onset ($\gamma=7.175$). There is supplementary points allowing the connection between the shifted attractors.}\mylab{fig:poincareg7,17}
\end{figure}
Beyond the existence domain of synchronized transport there is a complex transport. The tangent bifurcations at $\gamma_{snR}^r$ suggests the presence of an intermittent bifurcation in the comoving frame. The time integration of the solution in the comoving frame $s(t)-t$ for $\gamma>\gamma_{snR}^r$ shows long plateaux corresponding to a transport with velocity exactly equal to plus one. These episodes of transport are interrupted by a bounded dynamics in the laboratory frame resulting in a descending staircase in the comoving frame(  \myfig{intercomov}). The length of plateaux are irregular but are statistically longer approaching the tangent bifurcation $\gamma_{snR}^r$. It is characteristic of an intermittent bifurcation and such a staircase is reminiscent of  the Devil's staircase. The return map (\myfig{typ1bifcomov}) of the position suggests  that is an intermittent bifurcation of type I described by Manneville and Pomeau \cite{MaPo80}: Before the existence of the synchronized transport solution the dynamics is nearly tangent and the time spent in the vicinity diverges. When the dynamics escapes from this point a chaotic behavior takes place. Such a behavior agrees with the curve shape   in the return map (\myfig{typ1bifcomov}) since it is not monotonous and the slope may be large. Thus two close points are after an iteration far away from each other.  When one moves away from the critical point ($\gamma$ larger) the episodes of transport and bounded dynamics are equivalent. At $\gamma$ about $14.2$  a stable synchronized transport with $c=+1/3$, i.e. a 3-periodic transport solution (\myfig{twdc13}), attracts all the dynamics so that it is not possible to observe the end of the complex transport solution. \\

The dynamics for $\gamma$ smaller than the onset $\gamma_{snL}^\ell$ of synchronized left transport solution shows switching between transport to the left and bound dynamics as for the intermittent bifurcation to the right. However, one does not observe a divergence of duration of the transport solution approaching the turning point $\gamma_{snL}^\ell$ on the contrary the dynamics is similar to the one found in \myfig{simutwg7,5} when the (unstable) synchronized solution exists.
Moreover the transport episodes are no more regular as the synchronized transport.
Thus there is not an intermittent bifurcation as the turning point as it occurs for the right transport. Note that the time integration does not show the existence of a synchronized transport solution and then the continuation method is needed to detect the tangent bifurcation which is the birth of the synchronized transport.
One found out the onset  $\gamma \simeq 7.172$ of the intermittent transport. The simulation close to this point displays a vanishing drift  (\myfig{twinterchaotic}). The  time evolution in the laboratory frame of the particle position  evokes again a Devil's staircase. The step does not correspond to a periodic state but rather nearly 2-periodic solutions. The dynamics becomes chaotic but still bound just before or after the jump between plateaux. The duration close to the bound attractor increases when approaching the critical value of $\gamma$.  Therefore we believe the existence of an intermittent bifurcation from a bound attractor (\myfig{poincareg7,17}-a). After the transition, there is additionally points in the Poincar\'e map which connect the array of attractors allowing the drift  (\myfig{poincareg7,17}-b).

To conclude, complex transport emerges from the chaotic attractors via an intermittent transition resulting in a slow drift. Such a dynamics was called in \cite{SER07} aperiodic non phase locked transport. As for this latter paper, their existence domain bound includes  the synchronized transport domain. Our analysis allows to distinguish non phase locked transport resulting from the competition between synchronized transport solutions as described in previous section and the present intermittent transport. Contrary to the literature for which both non phase locked transports are not discerned,  we are able to determine the birth of synchronized transport. 
In the next section, we show that such a intermittent bifurcation exists but involving simply the one-periodic solutions for large drag.
%
%

\subsection{Quasi-periodic intermittent transport (large $\gamma$)}
\mylab{sec:aint}
%
\begin{figure}[tb]
\includegraphics[width=0.6\hsize]{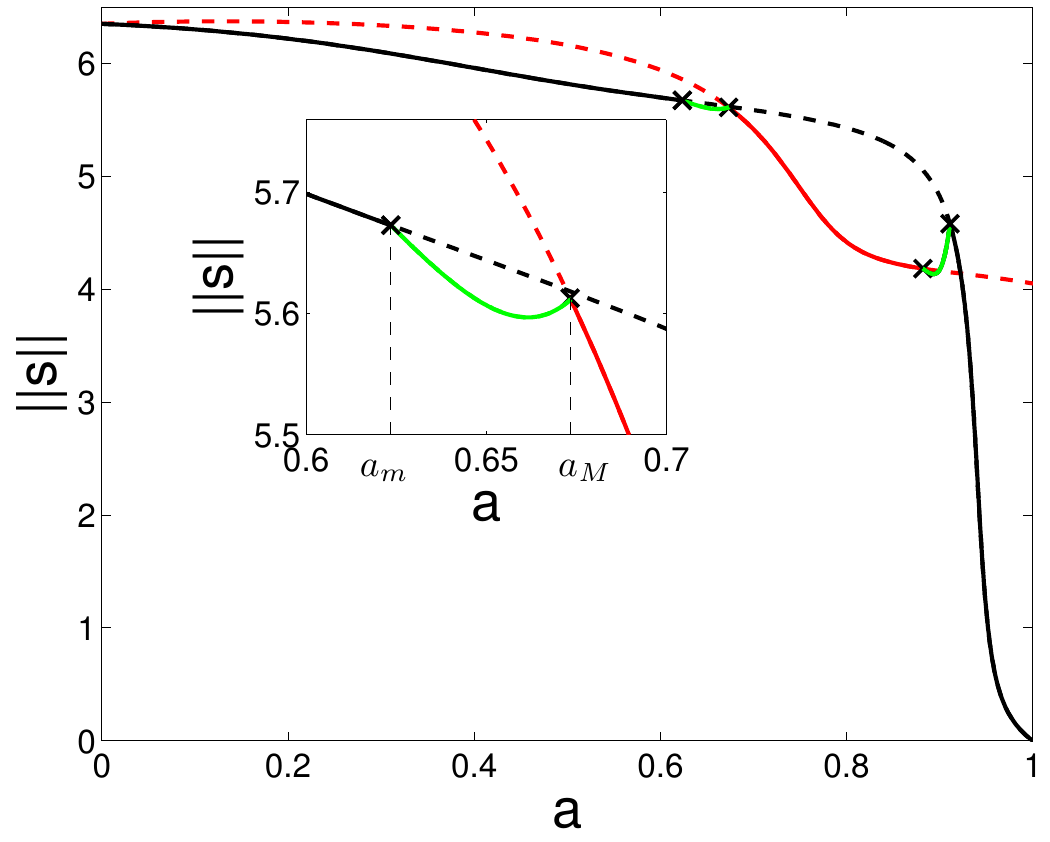}
\caption{Continuation of one-periodic orbits for the symmetric velocity profile $v_0(x)$ \myeq{u0sym} by varying  $a$. Parameters $\gamma=100$, $u_m=9$. Red line is the $s_0$ branch, balck line is the $s_m$ branch and the green line is the $s_a$ branch. Dashed lines indicate unstable solution.}\mylab{fig:asymu9}
\end{figure}
\begin{figure}[tb]
\includegraphics[width=0.29\hsize]{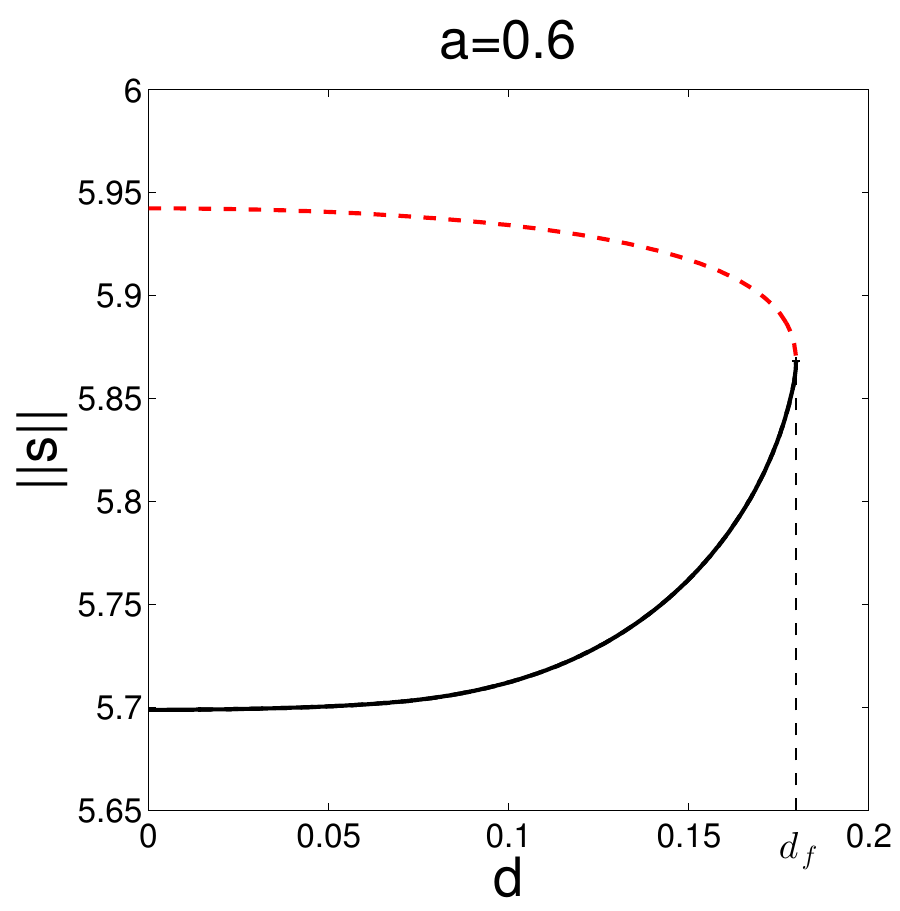}
\includegraphics[width=0.29\hsize]{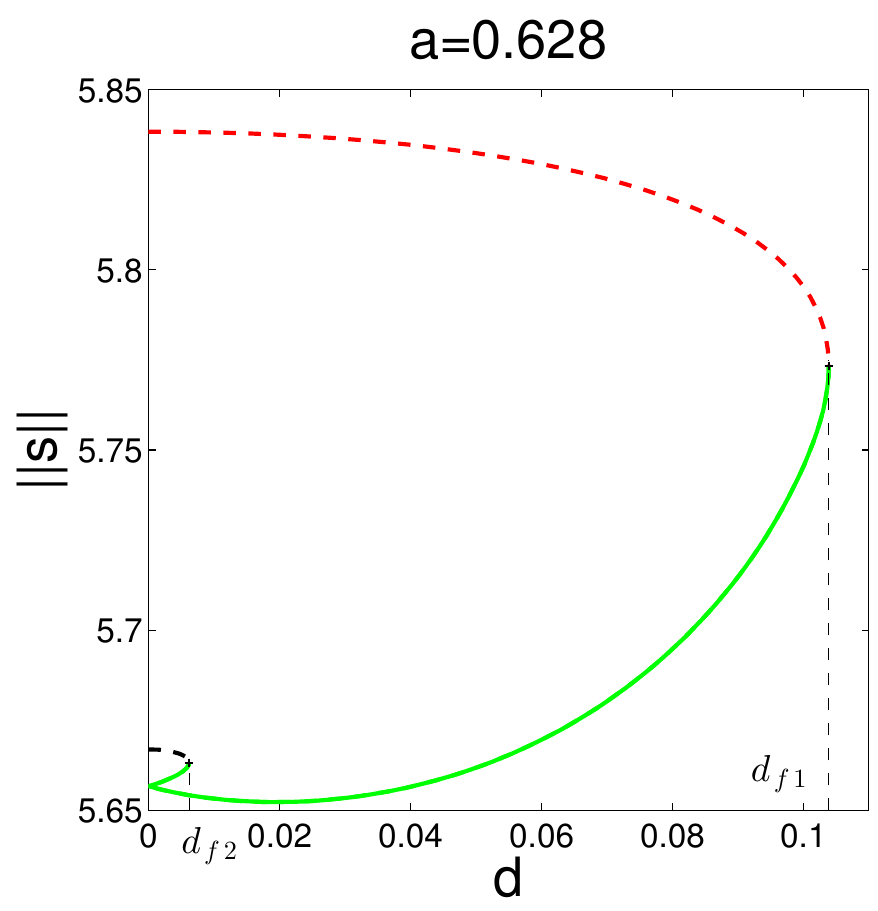}
\includegraphics[width=0.29\hsize]{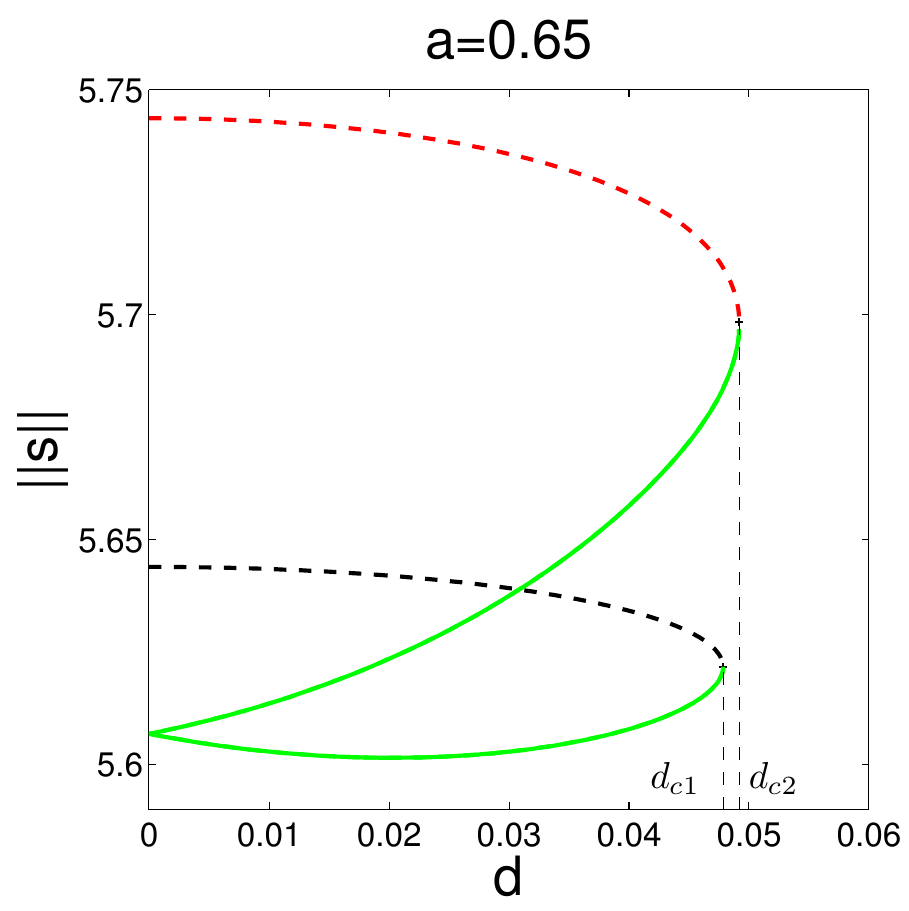}\\
\includegraphics[width=0.29\hsize]{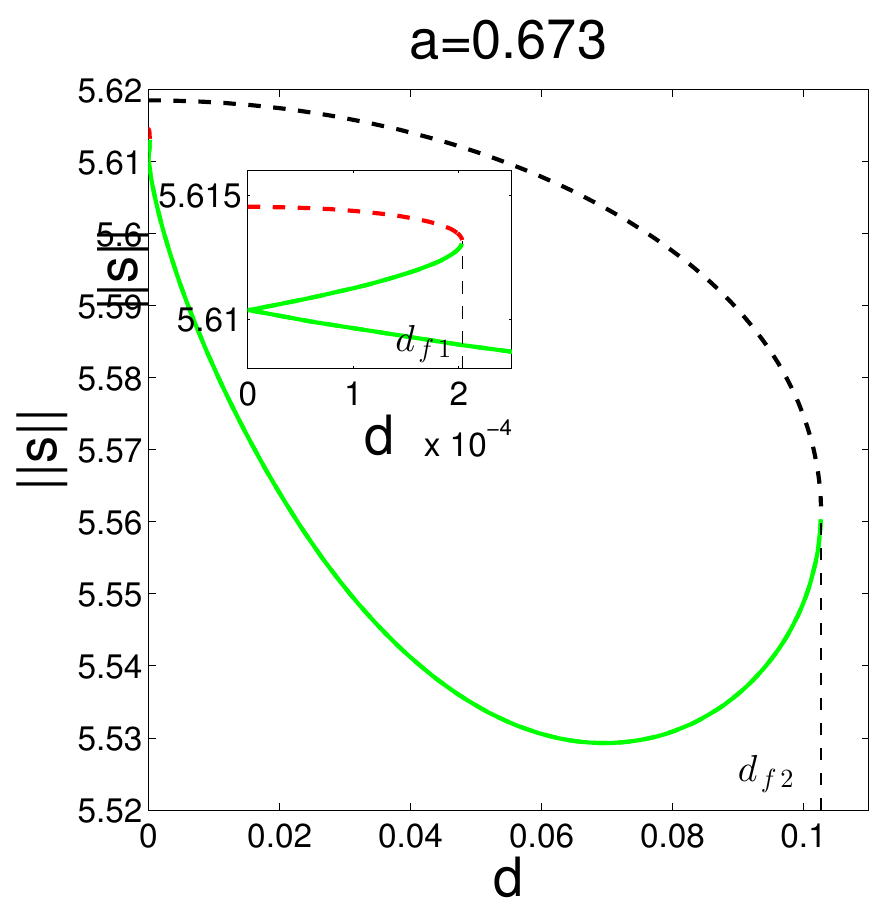}
\includegraphics[width=0.29\hsize]{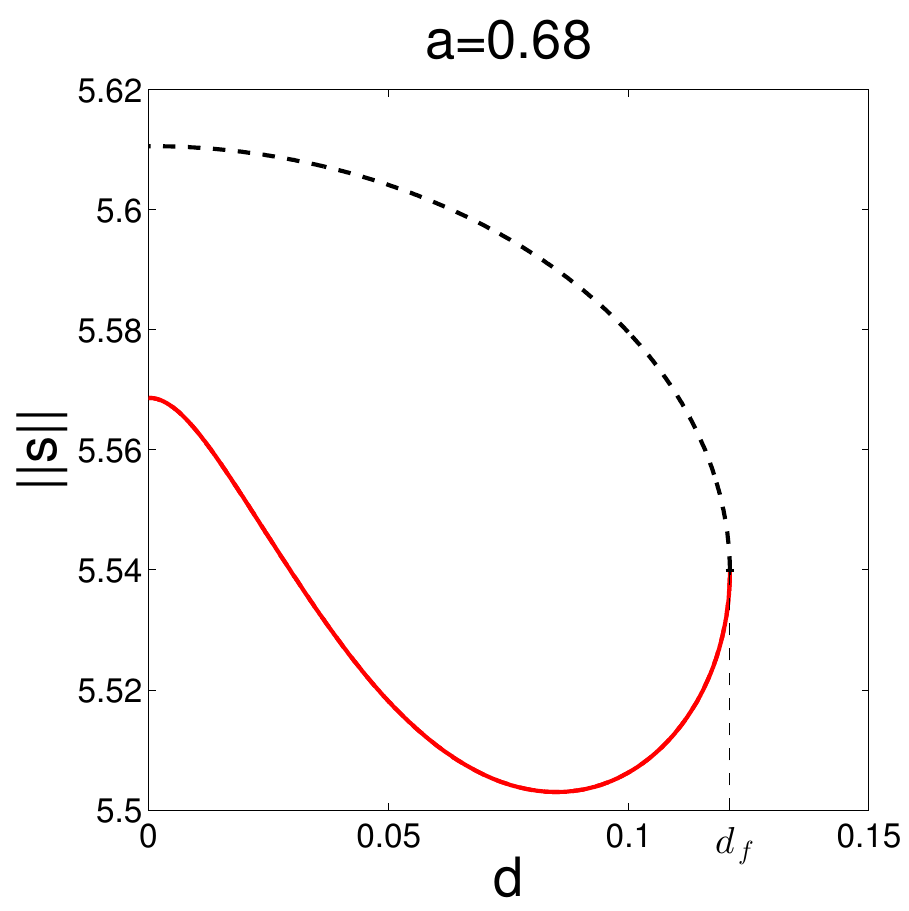}
\includegraphics[width=0.29\hsize]{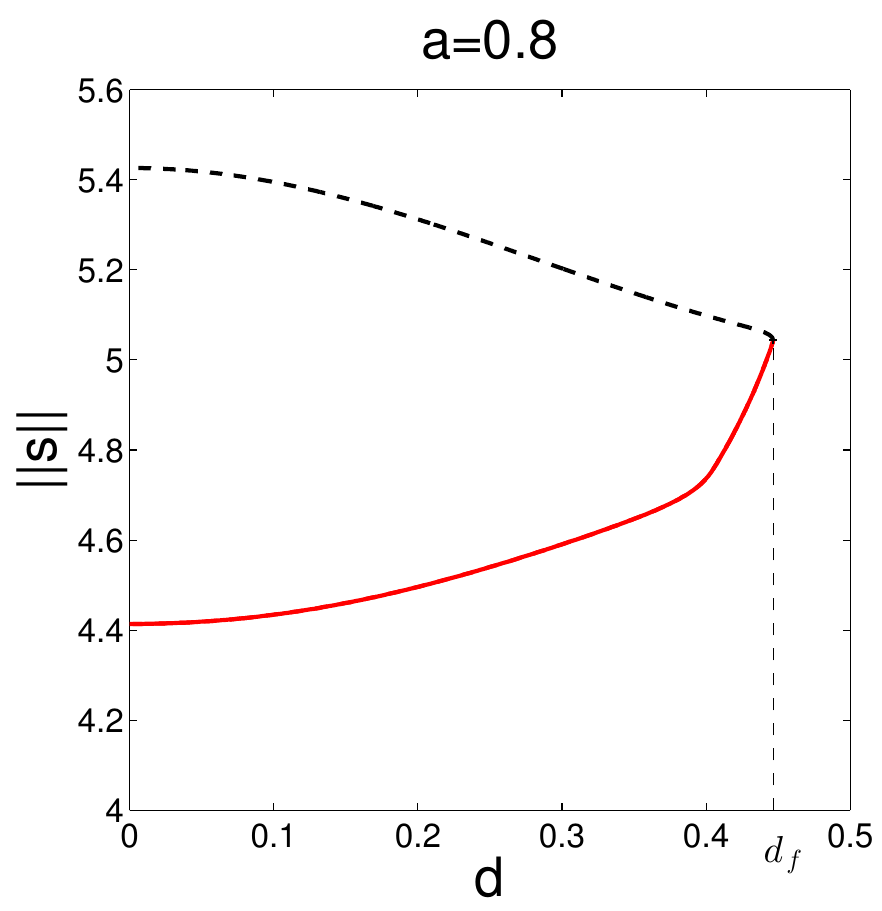}
\caption{Continuations of one-periodic orbits by varying $d$  for different values of $a$. Other parameters being fixed: $\gamma=100$, $u_m=9$. The color code is as in \myfig{asymu9}}\mylab{fig:g100um9cd}
\end{figure}
\begin{figure}[tb]
\includegraphics[width=0.6\hsize]{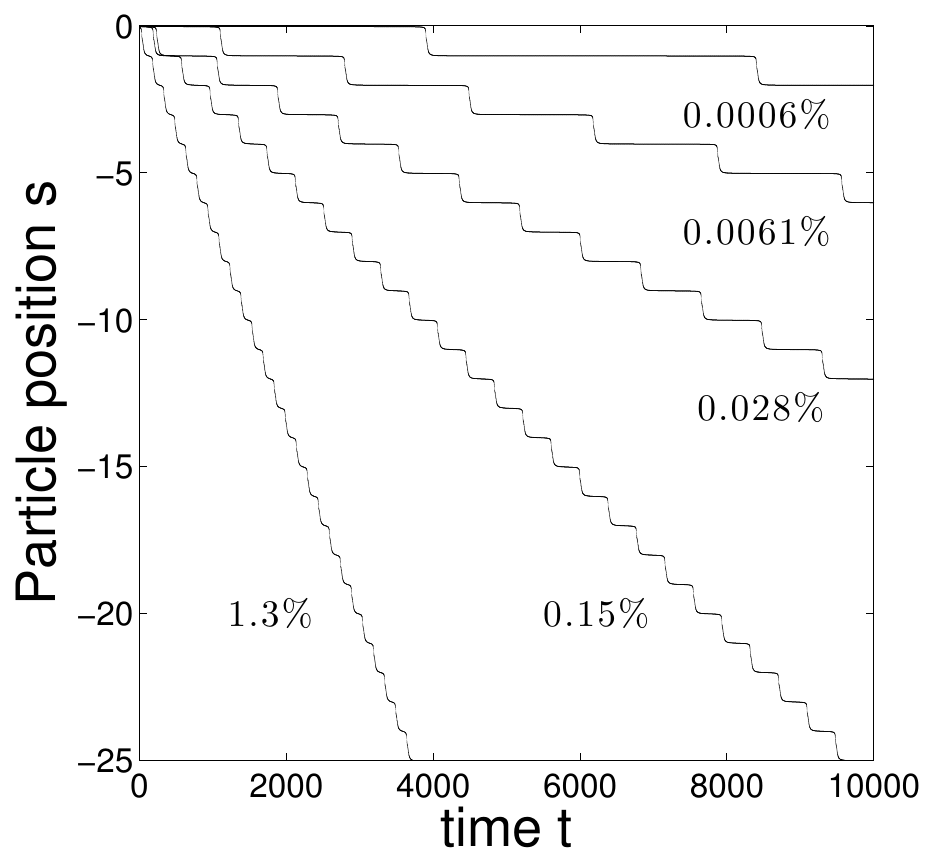}
\caption{Poincar\'e map in the $(s,t)$ plane of the dynamics at $a=0.6$ and different values of $d$ superior to the onset $d_f$ of \myfig{g100um9cd}. The percentage of the relative $d$ variation: $(d-d_f)/d_f$ is indicated below each plot. Remaining parameters: $\gamma=100$, $u_m=9$.}\mylab{fig:g100um9a06xt}
\end{figure}
\begin{figure}[tb]
\includegraphics[width=0.45\hsize]{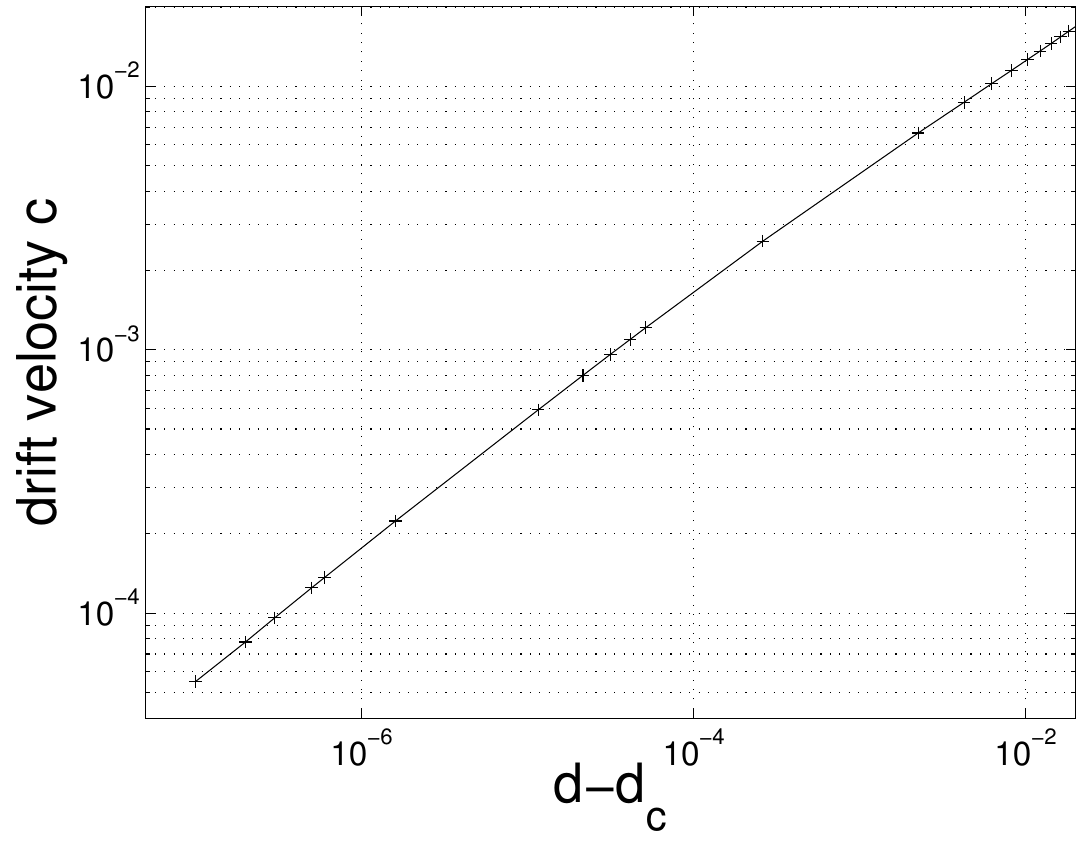}
\includegraphics[width=0.45\hsize]{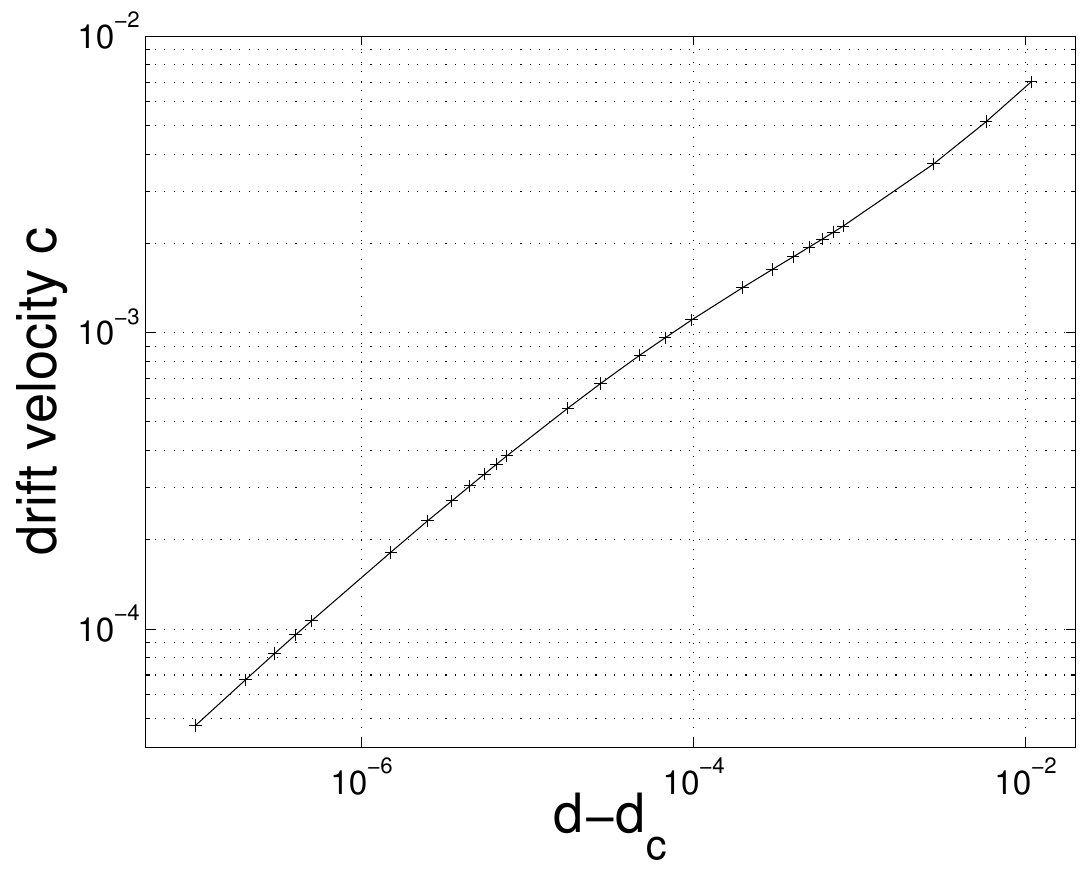}\\
(a)\hfill(b)\\
\caption{Logarithmic scale plots of the dependence of $d-d_f$ on the drift velocity for (a) $a=0.6$ and (b) $a=0.65$. $d_f$ is the saddle-node onset in the bifurcation diagrams of \myfig{g100um9cd}. Remaining parameters: $u_m=9$, $\gamma=100$.}\mylab{fig:powerlaw}
\end{figure}
\begin{figure}[tb]
\includegraphics[width=0.45\hsize]{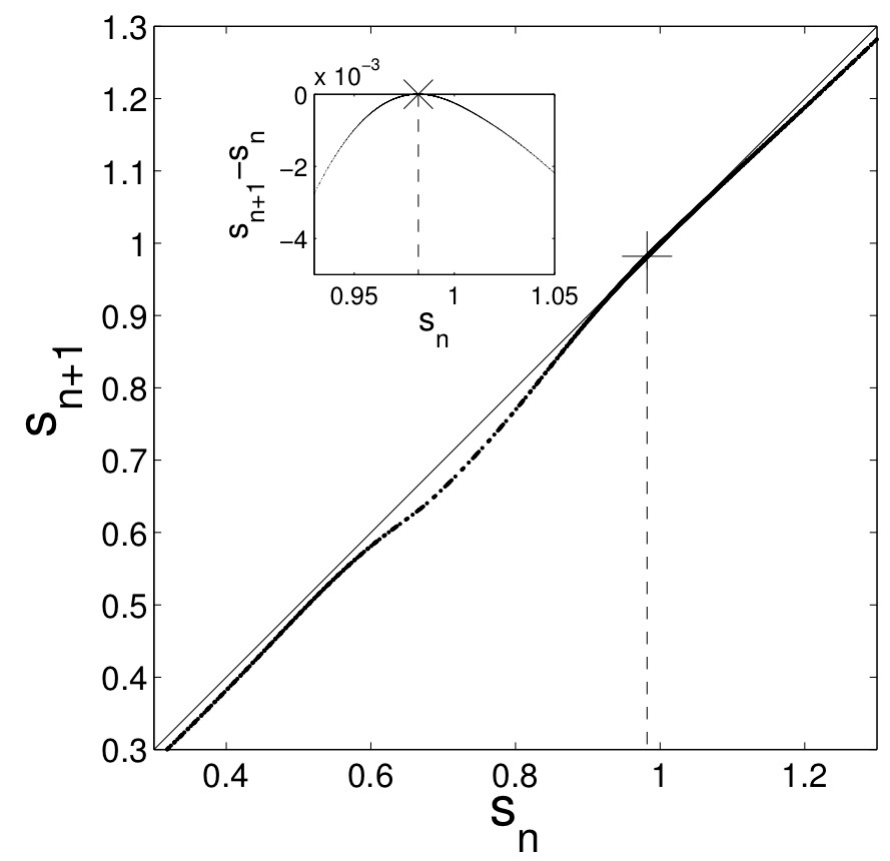}
\includegraphics[width=0.45\hsize]{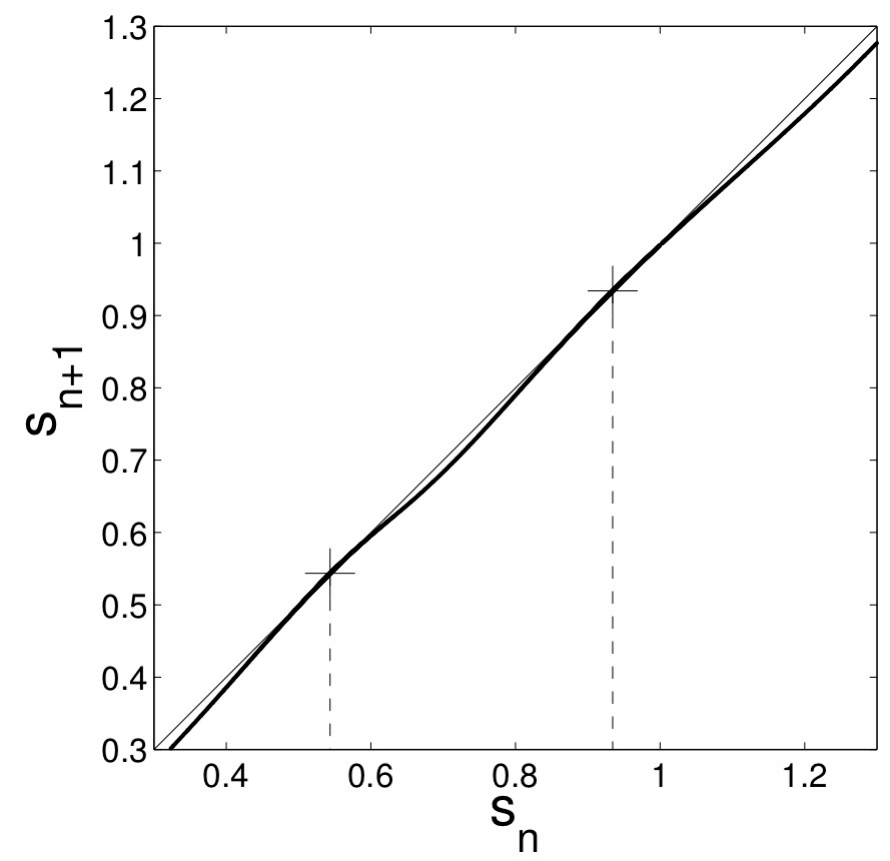}\\
(a)\hfill(b)\\
\caption{Return map of the particle position $s_n$ at entire times $n$ of the dynamics close to the intermittent bifurcation $d_f$ for (a) $a=0.6$ and (b) $a0.65$. The '+' symbol indicates the fixed point at $d_f$. For the panel (b) there is two fixed points according to \myfig{g100um9cd} panel $a=0.65$. The curve passes very close to this point but it does not cross the bisectrix.}\mylab{fig:typ1bif}
\end{figure}
\begin{figure}[tb]
\includegraphics[width=0.45\hsize]{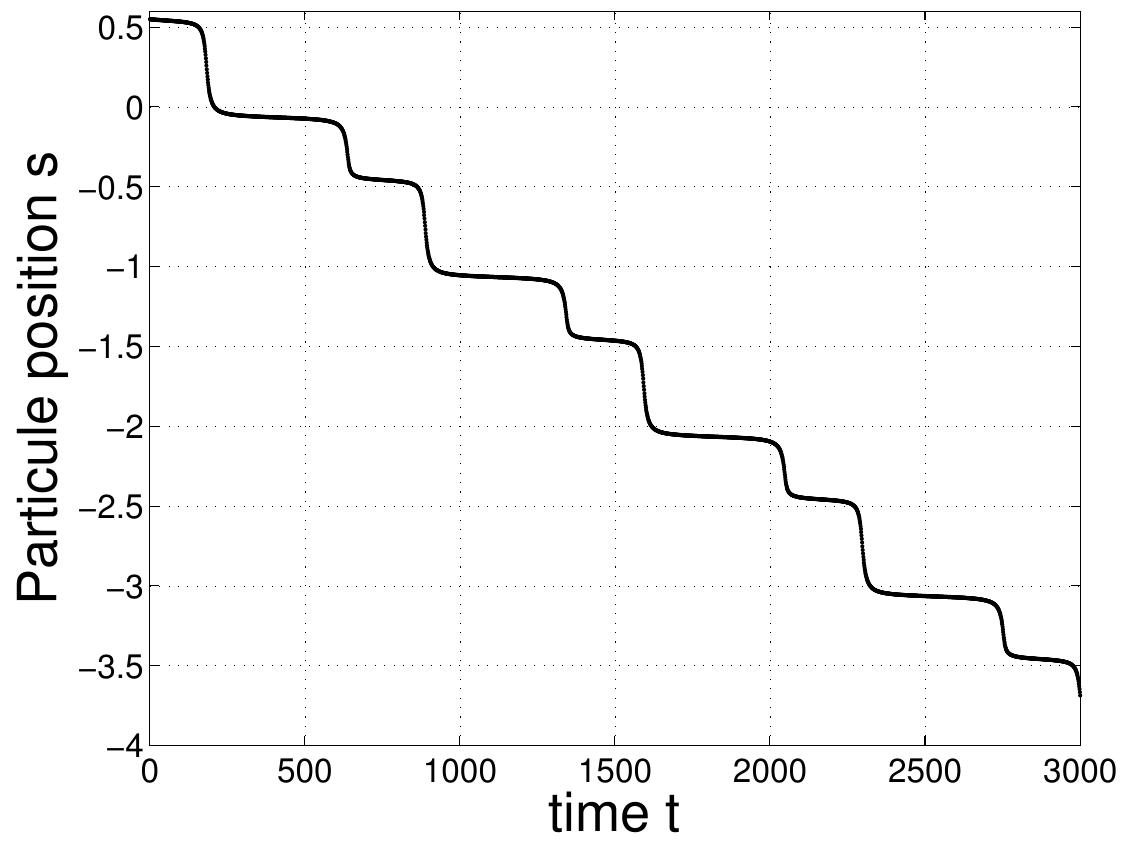}
\caption{Poincar\'e map in the $(s,t)$ plane and $d=0.0494$ beyond the onset $d_{c1}$ in the bifurcation diagram \myfig{g100um9cd}-c. Each plateau is close to a fixed point at the onset. Parameters $a=0.65$, $u_m=9$, $\gamma=100$. }\mylab{fig:g100um9a065xt}
\end{figure}
%
%
%
\begin{figure}[tb]
\includegraphics[width=0.7\hsize]{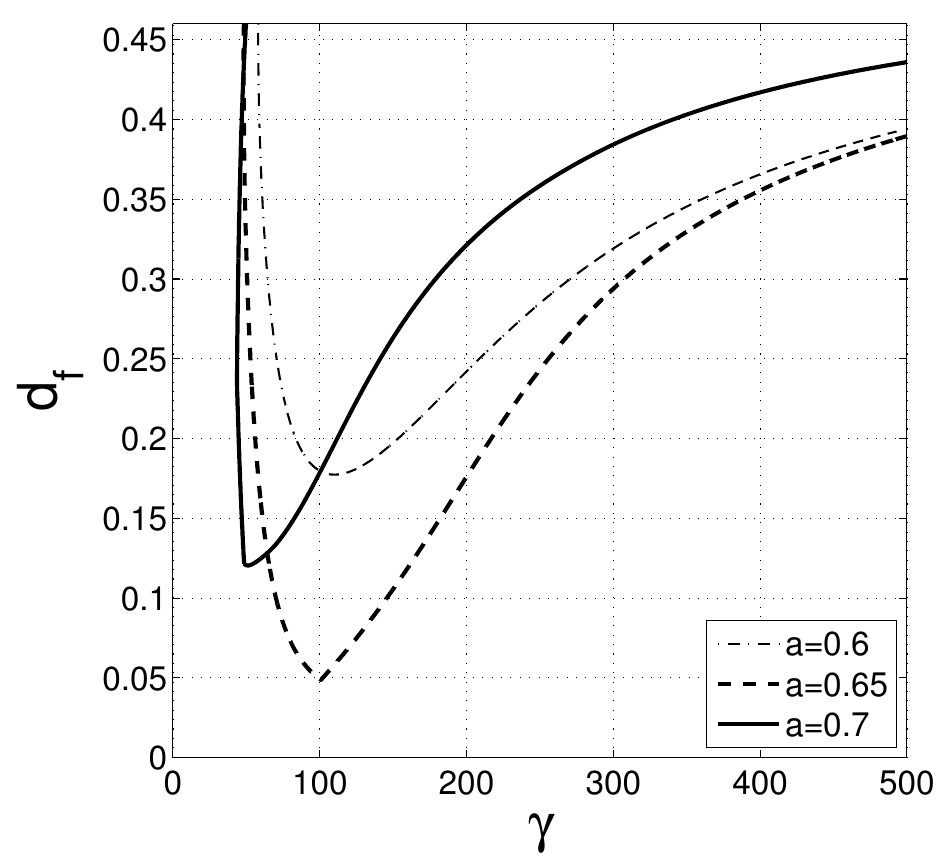}
\caption{Loci $(\gamma,d_f)$ of intermittent bifurcation at $u_m=9$ for three velocity contrasts  (fine dashdotted line)  $a=0.60$; (dashed line) $0.65$ and (plain line) $0.70$. The upper domain bound by each curve is the intermittent transport domain.}\mylab{fig:loci-gd}
\end{figure}

%

%
For the symmetric case no transport solution may exist when the drag $\gamma$ is large (about 100) because of the existence of a stable 1-periodic solution as in the bifurcation diagram of Figs. \ref{fig:symgia65cu}. However, when the velocity profile is asymmetric, the existence of such a "symmetric" branch solution is no more guaranteed.
By varying $d$, one shows, in this section, that one-periodic oscillations solutions  $s_0$, $s_m$ and also $s_a$ may vanish for large drag $\gamma$.\\
\\
The diagram bifurcation (\myfig{asymu9}) of the 1-periodic solutions in the axisymmetric case ($d=0$) is expanded using the velocity contrast $a$ parameter, the drag $\gamma$ being setting to $100$ and  the characteristic velocity $u_m$ to $9$.  
The  1-periodic solutions $s_0$, $s_m$ exchange their stability via two pitchfork bifurcations connecting the asymmetric branch $s_a$  (\myfig{asymu9}). It is a similar scenario as it occurs in the figures \ref{fig:symgia65cu} and \ref{fig:symcu} of the \mysec{sym}. The $s_a$ branch exists in the small ranges  around $a=0.65$ ($[a_m;a_M]$ inset of \myfig{asymu9}) and around $a=0.9$.\\
Now, let us study the existence of these solutions for an asymmetric velocity profile, i.e., when $d$ varies in the range [0;1/2]. It is found that if $a$ is smaller than a critical value $a_c$ about $0.507$ then both branches $s_0$ and $s_m$ remain for all values of $d$ and their stability is unchanged.
For larger value $a$ than the critical one $a_c$ both branches annihilate in a fold bifurcation at $d=d_f<0.5$.
According to numerical simulation, the fold bifurcation seems to start at $d_f=0.5$ when $a=a_c$.
When $a$ is  superior to $a_c$, the fold bifurcation corresponds to the vanishing of 1-periodic solutions  as shows the bifurcation diagram spanned by $d$ in \myfig{g100um9cd}-a for $a=0.6>a_c$.
In this case $d_f=0.1797384$. The time integration of the motion equation \myeq{ode} for $d>d_f$ and close to the turning point does not show a periodic motion rather a slow drift to the left consisting of  two distinct evolution phases that take place on distinct timescales: the particle oscillates in the vicinity of the vanished periodic solution at the onset and then drift to the next pore to the left. The Poincar\'e map in the $(s,t)$ plane (\myfig{g100um9a06xt}) displays a regular descending staircase for different values of $d$: the plateau corresponding to oscillations close to the onset. The plateaux being longer when $d$ approaches $d_f$.
The computation of the drift velocity $c$ as a function of $d-d_f$ indicates  a power law dependence: $c\%(d-d_f)^{1/2}$ (\myfig{powerlaw}-a). This dependence may be interpreted as type-I intermittent bifurcation at the turning point described by  Pomeau and Manneville \cite{MaPo80,PoMa80}. To corroborate that one is plotted the first return map of the discrete dynamics $s_{n+1}=f(s_n)$ of the particle position at different time $t=n$. The function $f$ has a  parabolic trajectory curve which before the bifurcation ($d<d_f$) intersects the bisectrix into two fixed-points (periodic orbits $s_0$ and $s_m$), then for $d=d_c$ tangents the bisectrix and finally for $d>d_c$ do no intersect the bisectrix. The \myfig{typ1bif} shows the return map for $d>d_f$ where the parabolic curve is clearly visible in the intset. Such an intermittent bifurcation is already mentioned for deterministic inertia ratchet for current reversal \cite{SKPK03} with a similar second order evolution equation. However, the intermittent bifurcation occurs for smaller friction (smaller $\gamma$) and the dynamics far from the dynamics is chaotic as it was found in the previous section. This last point contrasts with the regularity of the drift.\\
Let us focus on the understanding of this drift regularity. First we study the discrete dynamical system
\begin{equation}
s_{n+1}=f(s_n)=s_n+g(s_n)
\end{equation}
Here, we consider this system as one dimensional problem and then we will discuss the influence of the second phase parameter the velocity $v_n$. According to \myfig{typ1bif} one has the following properties
\begin{enumerate}
\item $g$ is 1-periodic,
\item $g$ is small,
\item $|g'|$ is not large,.
\item $g$ is strictly negative
\end{enumerate}
The three first assumptions are in agreement with the physical assumptions of the drift problem.  The periodicity is deduced from the spatial periodicity. Due to the large drag the motion is quasi-advective and $x_{n+1}-x_{n} $ should be small then second assumption is checked. Although $g$ is small, its derivates could be large. Due to the Stokes flow the acceleration cannot be large and then $|g'|$ is not large.  The fourth assumption is related to the fact that there is no 1-periodic solution.\\
If one calls $s_0$ the point closest to the bisectrix, if after $N$ iterations $s_N$ is the smallest point such that $s_N\geq s_0+1$ then one needs either $N$ or $N-1$ iterations to be larger than $s_0+2$ (See appendix \ref{a:it} for a proof). We show moreover that this behaviour does not depend on the initial condition. That explains the regularity and also the stability since to descend one staircase step, $N$ or $N-1$ iterations are needed for all dynamics. The simulation corroborates quite well this result: for $d=0.18000$ the number of iteration varies between 378-379 and for $d=0.17975$ the number of iteration is either 1692 or 1693.  Approaching the critical point the variation of number of iteration may be great: when $d=0.1797385$ the period is approximatively 18000 but this number vary of $\pm500$ iterations. That can be explained by the 2-dimensional aspect of the dynamics. In fact approaching the critical point the step length vanishes and the presence of the velocity in the iteration implies a small error comparing to the 1D model.  This error leads to an important number of iterations since the step length vanishes.\\
Therefore, the regularity of the dynamics does not necessarily imply the existence of a periodic solution. Even if it exists an unstable periodic orbits the proof of its existence is a complex question \cite{FGZ93,CKP96}. Usually, one focuses only on  periodic shadowing orbits which are orbit which stay "close" to the unstable periodic orbit. In this sense, we show that for large drag the dynamics is intermittent and quasi-periodic. This regular staircase contrasts with the "Devil's stair case" found in \myfig{intercomov} and in \myfig{twinterchaotic}.\par
This intermittent bifurcation still exists when the third branch $s_a$ is present. Indeed when $a_m<a<a_M$ and $d=0$, the branch $s_a$ emerges from $s_0$ (\myfig{asymu9}). In the bifurcation diagram spanned by $d$ (\myfig{g100um9cd}b) the branch starting from $s_a$ splits into two branches which connect the branch $s_0$ and $s_m$ in two fold bifurcations. The state $s_m$ being close to $s_a$ when $d=0$ then they connected at a turning point $d_{f_2}$ close to zero in the bifurcation diagram spanned by $d$ (\myfig{g100um9cd}b). This fold bifurcation is not an intermittent one because of the presence of the second stable branch $s_a$ (\myfig{g100um9cd}b). In contrast, the fold bifurcation connecting this branch $s_a$ and $s_0$  leads to an intermittent drift.
 When $a$ increases this branch becomes larger and in particular $d_{f_2}$ increases. On the other hand,  the critical value $d_{f1}=d_f$ decreases.  Then, for $a=0.65$, we have $d_{f1}\simeq d_{f2}\simeq 0.05$ (\myfig{g100um9cd}-c).
 Again the intermittent bifurcation occurs for the rightmost turning point, for instance $d_{f2}$ in \myfig{g100um9cd}-c, with the characteristic power law $c\%(d-d_f)^{1/2}$ (\myfig{powerlaw}-b). However the second fold bifurcation influences qualitatively the drift: the Poincar\'e map of the dynamics (\myfig{g100um9a065xt}) displays  a regular staircase with two different steps. Each plateau corresponds to the oscillations close to the fold bifurcation.
 When $a$ is close to $a_M$ the periodic states $s_0$ and $s_a$ collapse and so e.g. at $a=0.673$ the branch connecting these solutions is very small (\myfig{g100um9cd}-d).
Beyond the value $a_M$ the $s_a$ branch vanishes and the saddle-node bifurcation occurs at $d_f$ which increases with $a$ (see Figs.\ref{fig:g100um9cd} at $a=0.68$ and $a=0.8$).
%
\par 
The numerical simulations shows that the intermittent drift attracts all the dynamics and it is stable even far away from the onset. The transport does not change its direction and so the drift velocity $c$ remains negative. The function $|c(d)|$ increases with the asymmetry parameter $d$. However, the maximal value is not larger than $0.1$, e.g. for  $a=0.65$, $u_m=9$ and $d=0.4$, one obtains $c(0.4)=-0.0654$. It is still small comparing to the synchronized transport. 
 It is noteworthy that the drift remains for large drag, at least $\gamma=500$ as shows the loci of bifurcation points in $(\gamma,d_f)$ plane (\myfig{loci-gd}). Beyond $500$  the critical dissymmetry parameter $d_f(\gamma)$  is larger than  $0.4$ and then this result is less relevant for our model. The curves corresponding to $a=0.65$ and $a=0.7$ have a sharp corner are due to the competition between both fold bifurcations as it occurs in \myfig{g100um9cd}. 
 For smaller drag, the intermittent transport ceases at about $\gamma=50$ depending on the other parameters as the velocity contrast. For $a=0.7$ the domain curve displays a turning point for $\gamma\simeq41$. This fold bifurcation is due to the fusion/annihilation of two saddle-nodes. The first saddle-node is coming from the $s_0$ and $s_m$ branches of \myfig{g100um9cd}. The second saddle-node is due to 1-periodic branches starting from $d=0.5$. We do not study these branches because they occur for non realistic parameter range. Then large asymmetry is not always benefit to the the ratchet effect.
Anyway, in the experiment, the particles have a large drag at least about $50$ then the intermittent transport  is relevant for this framework.\\
One finds transport solution for smaller characteristic velocity too. The transport direction remains the same. The existence domain range is a smaller than $u_m=9$. For instance for $u_m=5$ the range of intermittent bifurcation is  $\gamma\in[80;120]$ (\myfig{lociu5-gd}). For $\gamma=120$ the critical asymmetry parameter $d$ is larger than 0.4, thus the lager value are not relevant.
\begin{figure}[tb]
\includegraphics[width=0.7\hsize]{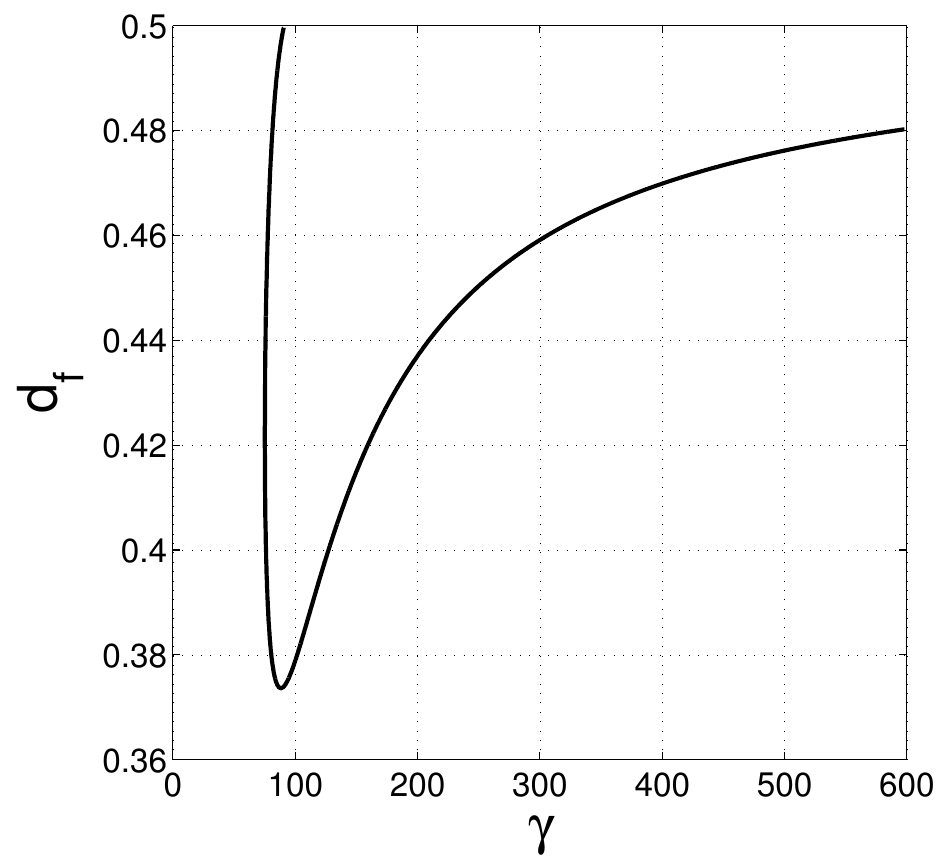}
\caption{Loci $(\gamma,d_f)$ of intermittent bifurcation at $u_m=5$ and $a=0.65$. The upper domain bound by the curve is the intermittent transport domain.}\mylab{fig:lociu5-gd}
\end{figure}
\\
\\

The intermittent transport is stable and exists in a large range of parameters for $u_m$ superior to $5$, large drags about $100$ and for large enough velocity contrast about $0.5$. The transport is slow about 0.1, i.e. at least 10 temporal periods are needed to drift of one pore. A noteworthy aspect of this transport solution is that the drift is always to the left direction (negative $x$). Thus, for $a>0$ and $d>0$ or $a<0$ and $d>0$, the possible transport intermittent is always to the left direction. According to the \myfig{velo-asym} the drift  direction corresponds to the sense of the slow increase of the velocity amplitude. Then, only the parameters of the velocity profile $u_0$ determines the direction of the transport. Since the velocity of the driving flow is related to the pore shape, this latter determines the transport direction.  This fact contrasts to the synchronized transport solutions  for which the direction of transport is strongly dependent on  frequency,  mean velocity, size of particles etc... 
The chaotic intermittent transport found close to the synchronized transport solutions are not observed for large $\gamma$. Indeed, for large drag it does not exist n-periodic solution with $n>1$, then this intermittent transport involving complex periodic solutions does not exist. Physically it is due to the quasi-advective motion.

\section{Conclusion}
\mylab{sec:conc}\\
In this paper we have studied the process of the deterministic transport of particles in a pore lattice filled of  viscous liquid. This liquid is pumped back and forth. To explore the particle dynamics in the parameter space, we have formulated the transport as a bifurcation problem. We focus on a generic problem related to such an experiment described in \cite{KRHM00}.
A simple model problem is considered in which the only force acting on the particle is the linear Stokes' drag in a quasi-static creeping flow. Moreover the problem is one dimensional then the particle cannot  rotate and the interaction with the pore wall are neglected. Therefore the only nonlinearity is due to the spatial flow variations. %
\par
For large particles drag, $\gamma$ about $100$, their motion is quasi-advective and they get the periodicity of the pumping. For a symmetric field flow and a small velocity contrast $a$ there is two types of oscillation: one stable $s_m$ around the minimum flow velocity and the second one unstable $s_0$ around the maximum. These solutions are invariant by  the symmetric transformation ${\cal S}_0$ or ${\cal S}_m$ and then there are called symmetric solutions. Taking a larger velocity contrast, there is another 1-periodic solution $s_a$ emerging from one of symmetric solution branches via a pitchfork bifurcation. This new solution breaks spontaneously the symmetry ${\cal S}_0$ or ${\cal S}_m$. In its existence domain, it is always stable. Therefore for all parameters values, one of these three solutions is stable and attracts all the dynamics forbidding any transport. It is no more true when one breaks the symmetry of the velocity profile, i.e. $d>0$, which $d$ measures the difference between the two extrema of spatial velocity gradient. All these solutions may end at a fold bifurcation. The numerical simulations show that it is an intermittent bifurcation of type I \cite{PoMa80} leading to a slow and regular transport. Close to the onset there is two phases in the transport: many oscillations in the vicinity of the onset solution following by a drift to another pore. One shows that the quasi-periodicity of the intermittence is due to the large friction (overdamped ratchet) and the creeping flow. Such a transport does not seem to be described in the literature of ratchet problems.  An interesting aspect of this transport is that its existence domain corresponds to a parameter range  relevant for the micro-pump experiment. The simulation shows that the drift velocity increases with the asymmetry velocity profile $d$.  For realistic flow velocity profiles, the drift velocity is smaller than $0.1$, i.e. one pore drift needs 10 temporal periods. The simulation shows that the drift direction is determined by the kind of asymmetry: in the sense of small positive velocity gradient and large negative  gradient. Therefore, for this transport the asymmetry of the pore geometry  is the crucial parameter which determines its existence and its velocity. \par
The experiment displayed sometimes a faster transport and furthermore the direction of the drift may change only by tuning either the frequency or the pumping pressure,  the pore geometry being unchanged. Thus, the intermittent transport cannot explain all these phenomena.
By taking smaller friction values $\gamma$ about $10-20$, the asymmetric branch $s_a$ loses its stability due to a cascade of period doubling. All periodic solutions become unstable in a window of parameters and the resulting dynamics  is quasi-periodic or chaotic and bound.
There is an intermittent bifurcation from this strange attractor which leads to slow drift.
However this transition does not lead to the synchronized transport which have a locked transport velocity to one, i.e. the spatial periodicity over the time periodicity. This latter dynamics exists in a range slightly smaller than the intermittent transport and emerges/vanishes at a tangent bifurcation. In the comoving frame, this bifurcation may lead to an  intermittent behavior: regular transport interrupted by chaotic oscillations.
However, we show that the synchronized transport solution is due to the presence of 2-periodic solutions. The spatial and temporal periodicities imply the existence of a lattice of 2-periodic  shifted  orbits. If the extrema of one orbits have a significant difference, two consecutive orbits may be locally closed to each other.  The trajectory of the synchronized transport is during one period close to a 2-periodic orbit and switches to the neighbor orbit. Therefore the drift velocity is locked to one. 
This mechanism possess similarity with heteroclinic connection \cite{Kirkal10} since the unstable manifold of a orbit is also the stable manifold of the next 2-periodic solution. This latter condition is sufficient for the existence of a synchronized transport. However it differs from heteroclinic cycles because the dynamics does not approach asymptotically the periodic orbit.\\
There is other transport solution with $c=1/q$, $q\in\mathbb{Z}$ too. For instance one finds drift velocities $c=1/2$ and $c=1/3$. For $c=1/2$ we show the transport solution is driven during two period by a 3-periodic orbit and then switches to the  orbit in the next pore shifted by two temporal periods. Therefore the synchronized transport does not emerge from a chaotic dynamics but rather from a array of $m$-periodic orbits. Indeed the chaotic dynamics is a consequence of a cascade of period-doubling, -tripling\dots 
Moreover the synchronized transport solution appears to be quite generic of ratchet since only spatial and temporal translation symmetries are required. The other mechanisms are due to non-linear phenomena: The existence of the $s_a$ branch results from a pitchfork bifurcation (parity symmetry breaking) and then via a period doubling it is possible to obtain the array of $2$-periodic shifted orbits.\\
All synchronized solutions have a range for which they are stable. If it is the unique stable branch then it attracts all the dynamics. The branch loses its stability via a doubling period cascade. In the comoving frame the dynamics may be quasi-periodic or chaotic but the drift velocity remains lock to one.\par
As shows \mysec{sym} the synchronized transport solution exists from the parity symmetry of ratchet and the asymmetric case appears as a perturbation of this case since the transport is related to bifurcations from the $s_a$ branch. Moreover the solution are quantitatively slightly modified but the relevant role of the asymmetry is that the left-right transports have different existence domains. Thus the current reversal is a generic phenomenon for slightly asymmetric ratchet because of both opposite transports exist in different parameter ranges.
There is many scenarios for the current reversal depending on relative positions of branches of solutions.\par
All these phenomena as synchronized transport or transport reversal are described in deterministic ratchet problems. However, the transition involving the so-called crisis is evoked for different transitions in our case and the emergence of the transport is related only to the stable transport. Indeed, in literature only stable solutions are studied even in the detailed analysis of \cite{SER07}. In contrast, our study shows that the unstable branches are required to understand the birth of transport solution and the transport direction reversal. it allows to point out the relevant role of the periodic orbits.
\\
\par

Our 1-D model does not take into account the pore boundary in the drag force. However, for the experiment design, it may be relevant for instance in the narrow region of the pore. In this case the drag force should be smaller than in our model because of the presence of the wall slows down particle drag. Then, the synchronized may appear for smaller particles as predicted in our model for which a drag force about 10 or 20 corresponds to particle sizes about 5 $\mu m$. Moreover, the dependence of the drag coefficient $\gamma$ as a function of the position $x$ adds nonlinearity in the evolution equation and then in ODE point of view, spontaneous symmetry breaking as period doubling may still exist.\\
A more realistic simulation requires obviously a 3D simulation for which the particle motion does not stay on the axis. Because of particle rotation and particle/pore interactions the problem is much more complicated. But this complexity is still consistent with non linear mechanism which may lead to the presented transport solutions.\\
\par
Even this work is deterministic, it allows to point out the possible role of noise in the emergence of synchronized transport. Indeed, if it exists a array of 2-periodic orbits  as described in the \mysec{synchro transport}, the particle trajectory may switch from a 2-periodic orbit to another close orbit thank the noise even if the transport solution does not exist. Because of the asymmetry in the phase space of the orbits, there is a preferred direction of the particle. It would be interesting to combine non-linear analysis and stochastic analysis for slightly noisy ratchet problems.

\section{Appendices}
\subsection{proof of the boundness of the  velocity and acceleration.}
 \mylab{sec:bound}\\
%
Let us consider the general form of the ODE  \myeq{ode} governing the particle motion:
\begin{equation}
\ddot{x}+\gamma \dot{x}=\gamma v_0(x(t),t)
\end{equation}
where $|v_0|$ is bound by a constant $M$.\\
If $\dot{x}>M$ then $\ddot{x}<0$. Therefore it is easy to show that if $\dot{x}(t_0)\leq M$ at the time $t_0$ then $\forall t\geq t_0: \dot{x}(t)\leq M$. Moreover if $\dot{x}(t_0) > M$ then $\dot{x}$ decreases. And if  $\dot{x}\leq M$ and one comes back to the previous case. So, for $t\leq t_0$:  $\dot{x}(t) \leq max(\dot{x}(t_0),M)$\\
One may apply a similar reasonment for the case $\dot{x}<M$.
So one concludes that  $|\dot{x}(t)|$ is bounded.\\
The triangle inequality implies that $ |\ddot{x}(t)|\leq 2\gamma max(|\dot{x}(t_0)|,M)$.\par
Let us remark that it is the acceleration may be large large for large drag $\gamma$. The reason is that there is no assumption of the acceleration of the  field flow. Anyway, for a finite $\gamma$ the damping forbids any velocity particle discontinuity.
\subsection{Quasi-periodic intermittent transport.}\label{a:it}
The intermittent transport for large drag $\gamma$ displays a very regular aspect. Let us prove that the one-dimensional  map
\begin{equation}
x_{n+1}=f(x_n)=x_n+g(x_n)
\end{equation}
with the following assumtpions
\begin{enumerate}
\item $g$ is 1-periodic,
\item $g$ is small,
\item $|g'|$ is not large,
\item $g$ is strictly negative,
\end{enumerate}
allow to understand this behavior. 
 The 2-dimensional aspect of the dynamics is discussed later.\\
We construct the map $x_n$ starting from $x_0$. Because of $g$ is negative $x_n$ is decreasing. We call  $N$ the smallest iteration number such that $x_N\leq x_0-1$ (\myfig{schemxn}). The assumption 3 implies that $f$ is  a strictly increasing function. Then $x_0-1<x_{N-1}$ implies that $x_{1}-1<x_{N}$.\\
\begin{figure}[tb]
\includegraphics[width=0.7\hsize]{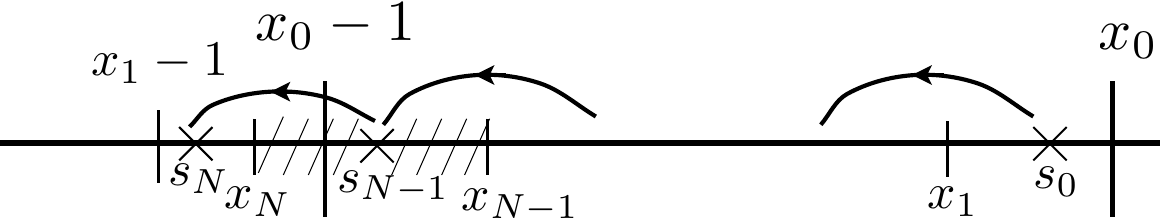}
\caption{Scheme of the $(x_n)$ and $(s_n)$ maps. The hatched segment  is the range of possible values of $s_{N-1}$. If $x_0-1<s_{N-1}$ as in this scheme then $x_1-1<s_N<x_N$. Otherwise $x_2-1<s_N<x_1-1$. }\mylab{fig:schemxn}
\end{figure}
Now consider the map $s_n$ starting from $x_1<s_0\leq x_0$. We seek the minimal iteration number in order such $s_M<x_0-1$. 
Because of $f$ is strictly decreasing then $x_{n+1}<s_n\leq x_n$ for all $n$. For instance $x_{N}<s_{N-1}\leq x_{N-1}$. The range $[x_N,x_{N-1}]$ contains the point $x_0-1$ (\myfig{schemxn}) and there is two cases depending if $s_{N-1}$ is inferior to $x_0-1$ or not. If $s_{N-1}\leq x_0-1$ then $M=N-1$ is the wanted iteration number. Otherwise $x_1-1< s_{N}\leq x_N\leq x_0-1$ and then $M=N$ (case considered for \myfig{schemxn}).\\
Therefore the map $x_n$ modulo 1 passes  in the interval $]x_1,x_0]$ every $N-1$ or $N$ iterations. Because $x_0$ is unspecified all dynamics have this quasi-periodic behavior. Furthermore, if $x_0$ is near the tangent bifurcation then $]x_1,x_0]$ is small and then the dynamics seems numerically periodic. \par
The previous proof does not yield if one considers the two-dimensional discrete dynamical system to take into account  the velocity. According to the numerical simulations, it acts as a small perturbation to the previous process. Approaching the bifurcation point, the intervals $[x_n,x_{n+1}]$ vanish  then a perturbation of this map even small may have a large influence of the  number of iterations. It is was it observed when $d$ is very close to $d_c$. However the variation of iterations is smaller compared the mean iteration number. 

%
\acknowledgments
We acknowledge support by the DFG [SFB 486, B13].


\clearpage
\bibliographystyle{siam}
\bibliography{%
/Users/Philippe/DOC/BIBLIO/biball%
}
\end{document}